\documentclass[prd,twocolumn,showpacs,superscriptaddress,nofootinbib,floatfix,showkeys,10pt]{revtex4-2}
\usepackage{graphicx}
\usepackage{amsmath}
\usepackage{bm}
\usepackage{yhmath}
\usepackage{mathtools}
\usepackage{wasysym}
\usepackage[colorlinks,citecolor=blue,urlcolor=blue,linkcolor=blue]{hyperref}
\usepackage{subfigure}
\usepackage{color}
\usepackage{cases}
\usepackage{subfigure}
\usepackage{times}
\usepackage{dcolumn,booktabs,bm}
\usepackage{slashed}
\usepackage{amsfonts,amssymb,stmaryrd,latexsym,amsmath}
\usepackage{textcomp}
\usepackage{multirow}
\usepackage{cancel}
\usepackage{array}
\usepackage{orcidlink}
\usepackage{enumitem}

\newcommand{\clabel}[2][]{#2}
\newcommand{\change}[1]{#1}

\newcommand{\etal}{\textit{et al}. }

\begin{document}


\title{Ground state baryons in the flux-tube three-body confinement model using Diffusion Monte Carlo}

\author{Yao Ma\,\orcidlink{0000-0002-5868-1166}}\email{yaoma@pku.edu.cn}
\affiliation{School of Physics, Peking University, Beijing 100871, China}

\author{Lu Meng\,\orcidlink{0000-0001-9791-7138}}\email{lu.meng@rub.de}
\affiliation{Ruhr-Universit\"at Bochum, Fakult\"at f\"ur Physik und
Astronomie, Institut f\"ur Theoretische Physik II, D-44780 Bochum,
Germany }

\author{Yan-Ke Chen\,\orcidlink{0000-0002-9984-163X}}\email{chenyanke@stu.pku.edu.cn}
\affiliation{School of Physics, Peking University, Beijing 100871, China}

\author{Shi-Lin Zhu\,\orcidlink{0000-0002-4055-6906}}\email{zhusl@pku.edu.cn}
\affiliation{School of Physics and Center of High Energy Physics,
Peking University, Beijing 100871, China}

\begin{abstract}

We make a systematical diffusion Monte Carlo (DMC) calculation for all ground state baryons in two confinement scenarios, the pairwise confinement and the three-body flux-tube confinement. With the baryons as an example, we illustrate a feasible procedure to investigate the few-quark states with possible few-body confinement mechanisms, which can be extended to the multiquark states easily. For each baryon, we extract the mass, mean-square radius, charge radius, and the quark distributions. \clabel[Jackknife]{We use the Jackknife resampling method to estimate the statistical uncertainties of masses to be less than 1 MeV.} To determine the baryon charge radii, we include the constituent quark size effect, which is fixed by the experimental and lattice QCD results. Our results show that both two-body and three-body confinement mechanisms can give a good description of the experimental data if the parameters are chosen properly. In the flux-tube confinement, introducing different tension parameters for the baryons and mesons are necessary, specifically, $\sigma_Y= 0.9204 \sigma_{Q\bar{Q}}$. The lesson from the calculation of the nucleon mass with the DMC method is that the improper pre-assignment of the channels may prevent us from obtaining the real ground state. With this experience, we obtain the real ground state (the $\eta_c \eta_c$ threshold with the di-meson configuration) of the $cc\bar{c}\bar{c}$ system  with $J^{PC}=0^{++}$ starting from the diquark-antidiquark spin-color channels alone, which is hard to achieve in the variational method and was not obtained in the previous DMC calculations.

\end{abstract}

\maketitle

\section{Introduction}~\label{sec:intro}

The quark model is widely used to study the hadron mass spectrum. The quarks inside a hadron are modeled as the constituent quarks interacting via the effective potentials. Various quark potential models have been used for the conventional and exotic hadron spectrum over the years~(see Refs.~\cite{Richard:2012xw,Richard:2016eis} for the quark model reviews and see Refs.~\cite{Meng:2022ozq,Chen:2022asf,Liu:2019zoy,Chen:2016spr,Chen:2016qju} for the reviews about the exotic hadrons.). The quark level interactions usually include the color-dependent Coulomb interaction, spin-dependent chromo-magnetic interaction, tensor interaction, and spin-orbit interaction. Basically, the above interactions can be derived from the one-gluon-exchange mechanism~\cite{Godfrey:1985xj}. In addition, there is a confinement term to describe the long-range interaction. Quark potential models behave nicely to describe the conventional hadron spectrum by solving the two-body or three-body problems. For instance, the Coulomb-plus-Linear Cornell potential proposed by Eichten \etal can well reproduce the charmonium and bottomonium spectra~\cite{Eichten:1974af,Eichten:1978tg,Eichten:1979ms}. A relativized quark model constructed by Isgur and his collaborators works successfully for all mesons and baryons~\cite{Godfrey:1985xj,Capstick:1986ter}. After that, Semay and Silvestre-Brac found that if the parameters are chosen correctly, the non-relativistic approach could accurately simulate the spectra from the relativistic one~\cite{Semay:1992xq}, and they built a new non-relativistic potential that works equally well in the meson and baryon sectors~\cite{Semay:1994ht,Silvestre-Brac:1996myf}.

\begin{figure}[htbp]
  \centering
  \includegraphics[width=0.3\textwidth]{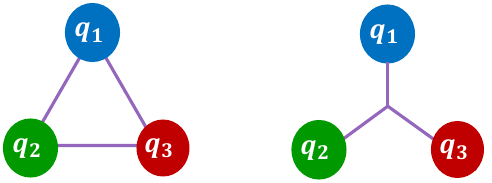} 
  \caption{\label{fig:two_confine} Two confinement scenarios for the baryons. The left and right panels represent the pairwise confinement mechanism ($\Delta$-type) and the three-body confinement mechanism (Y-type) respectively. }
    \setlength{\belowdisplayskip}{1pt}
\end{figure}

However, the confinement mechanism in quantum chromodynamics (QCD) is still elusive and ambiguous, which is usually introduced phenomenologically~\cite{Li:2009zu} or inspired by lattice QCD (LQCD) simulations~\cite{Bali:2000gf}. For the $q\bar{q}$ mesons, the confinement potential is apparently pairwise. But for the $qqq$ baryons, there is no compelling reason to assume the confinement interaction to be a two-body one as shown in the left panel of Fig.~\ref{fig:two_confine}, which is called the $\Delta$-type potential. Actually, another Y-type interaction was suggested by Artru in the string model for the baryons~\cite{Artru:1974zn}. In this scheme, three strings are connected at one point and carry quarks at their ends, as shown in the right panel of Fig.~\ref{fig:two_confine}.  This Y-type confinement mechanism has been investigated in different models, such as the QCD bag model~\cite{Hasenfratz:1980ka}, non-relativistic quark model~\cite{Richard:1983mu,Blask:1990ez}, semi-relativistic quark model~\cite{Carlson:1982xi,Sartor:1985ss,Stancu:1988gb,Stancu:1989iu}, and relativized quark model with chromodynamics~\cite{Capstick:1986ter}. A comparison between the $\Delta$-type and Y-type interactions was made by LQCD, and its fitting results seemed to favor the Y-type one~\cite{Takahashi:2000te,Takahashi:2002bw}. 

Nevertheless, this Y-type potential is in fact difficult to calculate with the conventional variational principle. If one expresses the minimum total length $L_{\mathrm{min}}$ of the flux tubes in terms of the three-body Jacobi coordinates, it will turn into a form containing the square-root as well as some complicated angle-dependence, making the evaluation of the integral in the matrix elements a difficult task~\cite{Dmitrasinovic:2009dy}. And certainly, extending the multi-body confinement to the multiquark states is a tougher work~\cite{Bicudo:2015bra}. Besides, there are other shortcomings in the framework of the variational method. For instance, when the number of particles (equivalently the dimensions of the wave function) is large, the number of the basis increases exponentially, so does the matrix dimension. Thus meeting the 
demand for the storage space and computing resources could become a challenge~\cite{Vary:2009qp}.

A promising alternative of the variational method is the diffusion Monte Carlo (DMC) method. In this formalism, the distribution of the so-called walkers is used to represent the spatial wave function, which will gradually evolve to the ground state over time in principle. In this scheme, the spatial integrals are replaced by summations over walkers. Consequently, there is no extra complexity in handling the Y-type interaction compared with the $\Delta$-type interaction. Meanwhile, the uncertainties of the Monte Carlo methods decrease as $1/\sqrt{N}$, where $N$ is the number of walkers. This behavior is independent on the dimension of the wave function, which is a promising advantage against most deterministic methods that depend exponentially on the dimensions. As for the multi-particle problem, the DMC method is easily parallelized to carry out the high-performance simulations~\cite{Kosztin:1996fh,krogel2012population}. The DMC has been well used in the simulations of the molecular physics~\cite{suhm1991quantum}, solid physics~\cite{Foulkes:2001zz} and nuclear physics~\cite{Carlson:2014vla}. 

In hadronic physics, the DMC method has been used in quark models in several works. Bai \etal calculated the $bb\bar{b}\bar{b}$ tetraquark ground state energy in a non-relativistic model with the flux-tube confinement potential~\cite{Bai:2016int}. A $0^{++}$ bound state with a mass of 18.69 GeV was predicted, which is 100 MeV below the $\eta_b\eta_b$ threshold. The $0^{++}$ $bb\bar{b}\bar{b}$ system was also calculated by Gordillo \etal using a similar DMC method~\cite{Gordillo:2020sgc}, with the pairwise confinement interaction. They gave a different mass of 19.199 GeV, which is 300-400 MeV above the $\eta_b\eta_b$ and $\Upsilon(1S)\Upsilon(1S)$ thresholds. In addition to the $bb\bar{b}\bar{b}$ system, they also calculated other fully-heavy tetraquarks in Ref.~\cite{Gordillo:2020sgc}. The predicted masses are all above the corresponding meson-meson thresholds, and agree with the results of the variational method in Ref.~\cite{Wang:2019rdo}. The difference between the fully-heavy tetraquark states of these two DMC calculations could arise from the different interactions, especially the confinement part, the different color configurations assumed (the diquark-antidiquark or di-meson configurations ) or the details of the DMC algorithm. Recently, the DMC method was also used to investigate other multiquark systems~\cite{Gordillo:2021bra,Alcaraz-Pelegrina:2022fsi,Gordillo:2022nnj}. Experimentally, an analysis using data from the CMS detector reported an 3.6$\sigma$ enhancement at 18.4 GeV in the invariant mass distribution of the $\Upsilon(1S)\ell^+\ell^-$ final state~\cite{durgut2018search,durgut2018evidence}, which might be a candidate for the $bb\bar{b}\bar{b}$ tetraquark state. But this enhancement was not seen by LHCb~\cite{LHCb:2018uwm}. Recently, the LHCb~\cite{LHCb:2020bwg}, CMS~\cite{yiCMSResults} and ATLAS~\cite{bouhovaATLAS} collaborations reported the observation of several resonance structures in the di-$J/\psi$ or $J/\psi\psi(2S)$ channels. All of these structures are above the $J/\psi J/\psi$ and $\eta_c\eta_c$ thresholds. 

Theoretically, in the variational method, it was shown that the tetraquark states above the di-meson thresholds in Ref.~\cite{Wang:2019rdo} will become either the scattering states or resonances if more quark-clustering configurations are considered~\cite{Wang:2022yes}. However, the primitive DMC is a method to calculate the ground bound states without quark-clustering beforehand (In order to calculate the excited states~\cite{suhm1991quantum} and resonances~\cite{Klos:2016fdb,Gandolfi:2016bth} with DMC, other nontrivial ingredients should be included.). For the tetraquark system, if there exists no bound state solution, one should expect that the wave function in DMC evolves exactly to the corresponding meson-meson threshold, rather than a state above the threshold as in Ref.~\cite{Gordillo:2020sgc}. Therefore, the doubt arises whether the DMC really avoids the usual quark-clustering and provides an exact estimate of the ground state energy as well as the wave function.

Considering the above issues, a simpler system is needed as a benchmark to test the DMC method and explore the reason why the lowest state cannot be obtained in some cases. Though the DMC method has been used for the nucleons, electrons etc., the multiquark systems are different because of the color confinement. The baryon system is a suitable platform to examine the DMC method, because of the following reasons: (i) the baryon is a bound state, having no ambiguity with the resonance; (ii) the color component is simple. There is only one possible color configuration; (iii) the experiment results are precise; (iv) the flux-tube confinement can be included; (v) although quarks are Fermions, the baryon is equivalent to a boson system, which avoids the notorious sign problem. This can be easily seen from its wave function. The baryon total wave function is the product of the spatial, spin, flavor and color parts:
\begin{align}
|\Psi_{tot}\rangle=|\phi_\mathrm{spatial}\rangle|\chi_\mathrm{spin}\rangle|\chi_\mathrm{flavor}\rangle|\chi_\mathrm{color}\rangle\,.
\end{align}
For the baryons, the only possible color configuration is that any two quarks form a color $\bar{3}_c$ representation and then combine with the third quark in the $3_c$ representation to become a color singlet. In this way, the color wave function is fully anti-symmetric and the color factor $\langle\chi_{\mathrm{color}}|\frac{\lambda_i}{2}\cdot\frac{\lambda_j}{2}|\chi_{\mathrm{color}}\rangle$ for any $(ij)$ quark pair is always $-\frac{2}{3}$. Thus if there are identical particles in the system, the exchanging symmetry renders the remaining $|\phi_\mathrm{spatial}\rangle|\chi_\mathrm{spin}\rangle|\chi_\mathrm{flavor}\rangle$ part to behave like a boson system. Therefore, there is no Fermionic sign problem in the DMC calculation~\cite{toulouse2016introduction}.

The fully-heavy baryon system has been calculated by Gordillo \etal\cite{Gordillo:2020sgc}. However, the baryon systems containing the light quarks are more interesting, especially the nucleon system. Hopefully the discussions about the baryons can also give enlightenment on the understanding of the threshold problem of the tetraquark system and provide some hints for the future explorations.

This paper is arranged as follows. In Sec.~\ref{sec:DMC}, the diffusion Monte Carlo method is introduced, including the single-channel and coupled-channel formalisms, respectively. In Sec.~\ref{sec:Hamiltonian}, the quark model Hamiltonian with different types of the confinement interaction is presented. In Sec.~\ref{sec:results}, the numerical results for all ground state baryons are given. The results among different calculation methods and different confinement potentials are compared. In Sec.~\ref{sec:disc}, we discuss the tetraquark threshold problem, clustering problem, and give some prospects on further applications of the DMC method in the field of hadron physics. Finally, a brief summary is given in Sec.~\ref{sec:sum}. In Appendix~\ref{app:conv-diff-eq}, we give the formalism of the convection–diffusion equation. In Appendix~\ref{app:Rc2_size_contribution}, we illustrate the constituent quark size contribution to the charge radius.

\section{Diffusion Monte Carlo method}~\label{sec:DMC}
\subsection{Imaginary time Schr\"odinger equation}
The DMC algorithm can be introduced from the imaginary time Schr\"odinger equation (in natural units $\hbar=c=1$),
\begin{equation}\label{schrodinger}
-\frac{\partial\Psi(\boldsymbol{R},t)}{\partial t}=[H-E_{R}]\Psi(\boldsymbol{R},t)\,,
\end{equation}
where  $\boldsymbol{R}\equiv(\boldsymbol{r}_1,\boldsymbol{r}_2,...,\boldsymbol{r}_m)$ represents the positions of particles and $E_{R}$ is the shift of energy. The wave function $\Psi(\boldsymbol{R},t)$ can be expanded in terms of a complete set of eigenfunctions $\Phi_{i}(\boldsymbol{R})$ of the Hamiltonian,
\begin{equation}
\Psi(\boldsymbol{R},t)=\sum_{i}c_{i}\Phi_{i}(\boldsymbol{R})e^{-[E_{i}-E_{R}]t}\,.
\end{equation}
When the value of $E_{R}$ is taken properly close to the energy of the ground state $E_0$, the wave function $\Psi(\boldsymbol{R},t)$ will approach the ground state after a long enough evolution time, as long as $c_0$ is not too small~\cite{Boronat1994}. The other components will be suppressed exponentially by the long time evolution.

\subsection{Importance sampling}

In principle, one can construct the DMC algorithm directly from Eq.~\eqref{schrodinger}, see Ref.~\cite{Kosztin:1996fh}. In this naive algorithm, the wave function $\Psi(\boldsymbol{R},t)$ is sampled. However, the algorithm is usually unstable due to the drastic statistic fluctuation. To make it more practical, the importance sampling technique~\cite{Kalos1974} is used. Instead of directly sampling the wave function $\Psi(\boldsymbol{R},t)$, a newly defined function $f(\boldsymbol{R},t)$ is sampled
\begin{equation}\label{f}
f(\boldsymbol{R},t)\equiv\psi_T(\boldsymbol{R})\Psi(\boldsymbol{R},t)\,,
\end{equation}
where $\psi_T(\boldsymbol{R})$ is a time-independent trial function (importance function). The $\psi_T(\boldsymbol{R})$ should be chosen as close to the ground state $\Phi_{0}$ as possible. We will see that the importance sampling will reduce the statistical fluctuation caused by the sharply changing potential in some regions~\cite{Hjorth-Jensen:2017gss}.

The imaginary time Schr\"odinger equation in Eq.(\ref{schrodinger}) can be rewritten in terms of $f(\boldsymbol{R},t)$ as
\begin{align}\label{schrodingerForf}
-\frac{\partial f(\boldsymbol{R},t)}{\partial t}=&-\sum_{i=1}^m \frac{1}{2m_i}\boldsymbol{\nabla}_{\boldsymbol{r}_{i}}^{2}f(\boldsymbol{R},t)\nonumber\\
&+\sum_{i=1}^m\frac{1}{2m_i}\boldsymbol{\nabla}_{\boldsymbol{r}_{i}}(\boldsymbol{F}_i(\boldsymbol{R})f(\boldsymbol{R},t))\nonumber\\
&+[E_{L}(\boldsymbol{R})-E_{R}]f(\boldsymbol{R},t)\nonumber\\
\equiv&(A_1+A_2+A_3)f(\boldsymbol{R},t)\nonumber\\
\equiv&Af(\boldsymbol{R},t)\,,
\end{align}
where $E_{L}(\boldsymbol{R})=\psi_T(\boldsymbol{R})^{-1}\hat{H}\psi_T(\boldsymbol{R})$ is called the local energy and $\boldsymbol{F}_i(\boldsymbol{R})=2\psi_T(\boldsymbol{R})^{-1}\boldsymbol{\nabla}_{\boldsymbol{r}_i}\psi_T(\boldsymbol{R})$ is the drift force acting on particle $i$. One can identify the Eq.~\eqref{schrodingerForf} is the convection–diffusion equation in Appendix~\ref{app:conv-diff-eq}. $A_1$, $A_2$ and $A_3$ terms are the the diffusion term, drift term and source (sink) term respectively.

Considering a small time separation $\Delta t$, the solution of Eq.(\ref{schrodingerForf}) can be approximated to the order of $(\Delta t)^2$~\cite{Boronat1994},
\begin{align}\label{evolution}
f(\boldsymbol{R'},t+\Delta t)=&\int \langle \boldsymbol{R}'|e^{-A\Delta t}|\boldsymbol{R}\rangle f(\boldsymbol{R},t) \mathrm{d}\boldsymbol{R}\nonumber\\
\approx&\int \mathrm{d}\boldsymbol{R_1}\mathrm{d}\boldsymbol{R_2}\mathrm{d}\boldsymbol{R_3}\mathrm{d}\boldsymbol{R_4}\mathrm{d}\boldsymbol{R}\nonumber\\
&\times G_{3}(\boldsymbol{R'},\boldsymbol{R_{1}},\frac{\Delta t}{2})G_{2}(\boldsymbol{R_{1}},\boldsymbol{R_{2}},\frac{\Delta t}{2})\nonumber\\
&\times G_{1}(\boldsymbol{R_{2}},\boldsymbol{R_{3}},\Delta t)\nonumber\\
&\times G_{2}(\boldsymbol{R_{3}},\boldsymbol{R_{4}},\frac{\Delta t}{2})G_{3}(\boldsymbol{R_{4}},\boldsymbol{R},\frac{\Delta t}{2})\nonumber\\
&\times f(\boldsymbol{R},t)\,,
\end{align}
with
\begin{eqnarray}
    G_{1}(\boldsymbol{R'},\boldsymbol{R},t)&=&\prod_{i=1}^m\left(\frac{2\pi t}{m_{i}}\right)^{-3/2}\exp \left[-\frac{m_{i}}{2t}(\boldsymbol{r}'_{i}-\boldsymbol{r}_{i})^{2}\right]\,, \nonumber\\
    G_{2}(\boldsymbol{R'},\boldsymbol{R},t)&=&\prod_{i=1}^m\delta\left(\boldsymbol{r}'_{i}-\boldsymbol{r}_{i}-\frac{\boldsymbol{F}_{i}(\boldsymbol{R})}{2m_{i}}t\right)\,, \nonumber\\
    G_{3}(\boldsymbol{R'},\boldsymbol{R},t)&=&\exp[-(E_{L}(\boldsymbol{R})-E_{R})t]\delta \left(\boldsymbol{R'}-\boldsymbol{R}\right)\,.
    \end{eqnarray}

In DMC algorithm, the function $f(\boldsymbol{R},t)$ is represented by the spatial distribution of a large number of walkers. Each walker $i$ is characterized through its position in the hyperspace $\boldsymbol{R}^{(i)}=(\boldsymbol{r}_1^{(i)},\boldsymbol{r}_2^{(i)},...,\boldsymbol{r}_m^{(i)})$. Any distribution containing the ground state component can be chosen as the initial $f(\boldsymbol{R},0)$ to start the evolution.  To implement the evolution process of $f(\boldsymbol{R},t)$ in Eq.(\ref{evolution}), each walker performs the following steps:
\begin{enumerate}[label=(\alph*)]
    \item Drift. Make a displacement of 
    \begin{equation}
\left(\frac{\boldsymbol{F}_{1}(\boldsymbol{R}^{(i)})}{2m_{1}}\frac{\Delta t}{2},\frac{\boldsymbol{F}_{2}(\boldsymbol{R}^{(i)})}{2m_{2}}\frac{\Delta t}{2},...,\frac{\boldsymbol{F}_{m}(\boldsymbol{R}^{(i)})}{2m_{m}}\frac{\Delta t}{2}\right)\nonumber
\end{equation}
under the drift force.
\item Diffusion. Make a random displacement of $(\boldsymbol{\chi}_1,\boldsymbol{\chi}_2,...,\boldsymbol{\chi}_m)$, where $\boldsymbol{\chi}_j$ is drawn from the 3-dimensional Gaussian distribution $\exp[-m_j\chi_j^2/(2\Delta t)]$.
\item Repeat step (a).
\item Birth–Death process. Replicate the walker $n_{r}$ times with
\begin{equation}\label{birth_death}
n_{r}=\text{Floor}\left[e^{-\left(\frac{E_L(\boldsymbol{R}')+E_L(\boldsymbol{R})}{2}-E_R\right)\Delta t}+u \right]\cdot\frac{N_0}{N}\,,
\end{equation}
where the Floor function only retain the integer part. The random number $u$ uniformly distributed in the interval $[0,1]$ is introduced to make the rounding smoother. $N_0$ is the target total number of walkers, and $N$ is the current total number of walkers. The factor $N_0/N$ is introduced to keep the number of walkers roughly stable. Its effect will be decreased with the damping of the fluctuation and will not change the final distribution of walkers. The stable walker number is $N_0$. The value of $E_R$ is taken as the mixed-energy~\cite{hammond1994monte}
\begin{eqnarray}
\langle E\rangle_{mixed}&=&\frac{\int\psi_T(\boldsymbol{R})\hat{H}\Psi(\boldsymbol{R},t)\mathrm{d}\boldsymbol{R}}{\int\psi_T(\boldsymbol{R})\Psi(\boldsymbol{R},t)\mathrm{d}\boldsymbol{R}}\nonumber\\
&=&\frac{\int E_{L}(\boldsymbol{R})f(\boldsymbol{R},t)\mathrm{d}\boldsymbol{R}}{\int f(\boldsymbol{R},t)\mathrm{d}\boldsymbol{R}}\,.
\end{eqnarray}
As $\Psi(\boldsymbol{R},t)$ approaches the ground state $\Phi_0(\boldsymbol{R})$ during the evolution,  $\langle E\rangle_{mixed}$ gets closer to the ground state energy $E_0$.
\end{enumerate}

The whole procedure above should be repeated enough times until $\Psi(\boldsymbol{R},t)$ evolves to the ground state wave function and $E_{R}$ stabilizes at $E_0$.

One can identify the effect of the importance sampling from Eq.~\eqref{birth_death}. If we neglect the rounding effect and the $N_0/N$ factor, the replicating factor should be 
\begin{eqnarray}
    n_{r}	&=&\exp\left[-\left(\frac{E_{L}(\boldsymbol{R}')+E_{L}(\boldsymbol{R})}{2}-E_{R}\right)\Delta t \right]\nonumber \\
    &=&\exp\left[-\left(\frac{\hat{H}\psi_{T}(\boldsymbol{R}')}{2\psi_{T}(\bm{R}')}+\frac{\hat{H}\psi_{T}(\boldsymbol{R})}{2\psi_{T}(\bm{R})}-E_{R}\right)\Delta t\right].\,    
\end{eqnarray}
Apparently, without importance sampling (taking $\psi_T=1$), the factor reads $n_r=\exp[-(V(\bm{R})/2+V(\bm{R}')/2-E_R)\Delta t]$. The walkers appearing near the divergence of the potential will lead to drastic fluctuations of the population. If one can take $\psi_T=\Phi_0$ ideally, the factor becomes $n_r=\exp[-(E_0-E_R)\Delta t]$, which approaches 1 if we could choose $E_R$ properly. Therefore, the importance sampling reduces the fluctuation of the population. 

In the practical simulation, the $\psi_T$ is unknown beforehand. One choice is the Jastow correlation factor 
\begin{equation}
    \psi_T(\bm{R})=\prod_{i < j}\exp\left(\frac{a_{ij}r_{ij}}{1+\beta_{ij} r_{ij}}\right),
\end{equation}
where $a$ and $\beta$ are adjustable parameters. In this way, the $E_L(\bm{R})$ and $\bm{F}_i(\bm{R})$ can be calculated analytically. In our simulation, we choose a more specific form following Ref.~\cite{Gordillo:2020sgc},
\begin{equation}\label{psi_T}
\psi_T(\boldsymbol{R})=\prod_{i<j} e^{-a_{ij}r_{ij}}\,.
\end{equation}
Here $a_{ij}$ are adjustable constants and their values are set to minimize the fluctuation. 

\subsection{Forward walking technique }\label{forwardwalking}

One can obtain the energy of the ground state and function $f(\bm{R},t)$ using the DMC with importance sampling directly. However, one has to remove the trial wave function $\psi_T(\boldsymbol{R})$ from $f(\boldsymbol{R},t)$ to obtain the square of the ground state wave function $|\Phi_0(\boldsymbol{R})|^2$, as well as the pure expectation value $A_{p}\equiv\frac{\langle\Phi_0|A|\Phi_0\rangle}{\langle\Phi_0|\Phi_0\rangle}$ of the observable $A$. To this end, the forward walking technique~\cite{BARNETT1991} is used. Following Ref.~\cite{Liu1974}, $\Phi_{0}(\boldsymbol{R})/\Psi_{T}(\boldsymbol{R})$ can be obtained from the asymptotic population $P(\boldsymbol{R},t\rightarrow\infty)$ of one walker starting at $\boldsymbol{R}$,
\begin{eqnarray}\label{asymptoticPopulation}
\Phi_{0}(\boldsymbol{R})/\Psi_{T}(\boldsymbol{R})&=\frac{P(\boldsymbol{R},t\rightarrow\infty)}{\langle\Psi_{T}|\Phi_{0}\rangle}\,.
\end{eqnarray}
With this replacement, the ground state wave function $|\Phi_0(\boldsymbol{R})|^2$ becomes
\begin{align}\label{Phi02}
|\Phi_0(\boldsymbol{R})|^2&=\Psi_{T}(\boldsymbol{R})\Phi_{0}(\boldsymbol{R})\cdot\Phi_{0}(\boldsymbol{R})/\Psi_{T}(\boldsymbol{R})\nonumber\\
&=f(\boldsymbol{R})P(\boldsymbol{R})\,,
\end{align}
and the pure expectation value of the observable $A$ becomes
\begin{align}\label{Apure}
A_{p}\equiv\frac{\langle\Phi_{0}|A|\Phi_{0}\rangle}{\langle\Phi_{0}|\Phi_{0}\rangle}&=\frac{\langle\Psi_{T}|A\Phi_{0}/\Psi_{T}|\Phi_{0}\rangle}{\langle\Psi_{T}|\Phi_{0}\rangle}/\frac{\langle\Psi_{T}|\Phi_{0}/\Psi_{T}|\Phi_{0}\rangle}{\langle\Psi_{T}|\Phi_{0}\rangle}\nonumber\\
&=\frac{\sum_{i}A(\boldsymbol{R}^{(i)})P(\boldsymbol{R}^{(i)})}{\sum_{i}P(\boldsymbol{R}^{(i)})}\,,
\end{align}
where $i$ represents walker $i$. 

The value of $P(\boldsymbol{R}^{(i)})$ can be obtained by the following way. When the evolution has already stabilized after a time period $t_s$, the distribution of walkers represents $f(\boldsymbol{R})=\psi_T(\boldsymbol{R})\Phi_0(\boldsymbol{R})$. However, if one focus on a single walker $i$ at $\boldsymbol{R}^{(i)}$, it is equivalent to a function $f'(\boldsymbol{R}')\propto\delta(\boldsymbol{R}'-\boldsymbol{R}^{(i)})$ that has not reached stabilization. So one continues to evolve for a long enough period of time $t_b$ until this $f'(\boldsymbol{R}')$ function has reached stabilization too. At this moment, the number of walkers replicated from the original walker $i$ is the value of $P(\boldsymbol{R}^{(i)})$. The detailed algorithm implementation is described in Ref.~\cite{Casulleras1995}. 

\subsection{Coupled-channel formalism }~\label{sec:couple-channel}

When dealing with multiple the spin-color channels, the above algorithm needs some changes.  One can decompose the wave function $\Psi(\boldsymbol{R},t)$ to
\begin{eqnarray}
\Psi(\boldsymbol{R},t)=\sum_{\alpha}\Psi_\alpha(\boldsymbol{R},t)\chi_\alpha\, ,
\end{eqnarray}
where $\chi_\alpha$ is the wave function of discrete quantum numbers. The space wave function $\Psi_\alpha(\boldsymbol{R},t)$ of channel $\chi_\alpha$ satisfies 
\begin{align}
-\frac{\partial\Psi_{\alpha'}}{\partial t}&=\sum_{\alpha}\hat{H}_{\alpha'\alpha}\Psi_{\alpha}-E_{R}\Psi_{\alpha'}\,.
\end{align}
In the importance sampling technique, Eqs.(\ref{f}-\ref{evolution}) are replaced by
\begin{eqnarray}
f_{\alpha}(\boldsymbol{R},t)&\equiv\psi_T(\boldsymbol{R})\Psi_{\alpha}(\boldsymbol{R},t)\,,
\end{eqnarray}
\begin{align}
-\frac{\partial f_{\alpha'}(\boldsymbol{R},t)}{\partial t}=&-\sum_{i=1}^m \frac{1}{2m_i}\boldsymbol{\nabla}_{\boldsymbol{r}_{i}}^{2}f_{\alpha'}(\boldsymbol{R},t)\nonumber\\
&+\sum_{i=1}^m\frac{1}{2m_i}\boldsymbol{\nabla}_{\boldsymbol{r}_{i}}(\boldsymbol{F}_i(\boldsymbol{R})f_{\alpha'}(\boldsymbol{R},t))\nonumber\\
&+[E_{L}(\boldsymbol{R})-E_{R}]f_{\alpha'}(\boldsymbol{R},t)\nonumber\\
&+\sum_{\alpha}V_{\alpha'\alpha}(\boldsymbol{R})f_{\alpha}(\boldsymbol{R},t)\nonumber\\
=&Af_{\alpha'}(\boldsymbol{R},t)+\sum_{\alpha}V_{\alpha'\alpha}(\boldsymbol{R})f_{\alpha}(\boldsymbol{R},t)\,,
\end{align}
and
\begin{align}\label{evolution_f_alpha}
f_{\alpha'}(\boldsymbol{R'},t+\Delta t)=&\int \mathrm{d}\boldsymbol{R} \langle \boldsymbol{R}'|e^{-A\Delta t}|\boldsymbol{R}\rangle \nonumber\\
&\times\sum_{\alpha}[e^{-\mathbb{V}(\boldsymbol{R})\Delta t}]_{\alpha'\alpha} f_\alpha(\boldsymbol{R},t)\nonumber\\
\approx&\int \mathrm{d}\boldsymbol{R_1}\mathrm{d}\boldsymbol{R_2}\mathrm{d}\boldsymbol{R_3}\mathrm{d}\boldsymbol{R_4}\mathrm{d}\boldsymbol{R}\nonumber\\
&\times G_{3}(\boldsymbol{R'},\boldsymbol{R_{1}},\frac{\Delta t}{2})G_{2}(\boldsymbol{R_{1}},\boldsymbol{R_{2}},\frac{\Delta t}{2})\nonumber\\
&\times G_{1}(\boldsymbol{R_{2}},\boldsymbol{R_{3}},\Delta t)\nonumber\\
&\times G_{2}(\boldsymbol{R_{3}},\boldsymbol{R_{4}},\frac{\Delta t}{2})G_{3}(\boldsymbol{R_{4}},\boldsymbol{R},\frac{\Delta t}{2})\nonumber\\
&\times \sum_{\alpha}[e^{-\mathbb{V}(\boldsymbol{R})\Delta t}]_{\alpha'\alpha} f_\alpha(\boldsymbol{R},t)\,,
\end{align}
with $V_{\alpha'\alpha}(\boldsymbol{R})=\chi_{\alpha'}^\dagger V(\boldsymbol{R})\chi_{\alpha}$. The matrix $\mathbb{V}(\boldsymbol{R})$ is composed of elements $V_{\alpha'\alpha}(\boldsymbol{R})$. The local energy is changed to $E_{L}(\boldsymbol{R})=\psi_T(\boldsymbol{R})^{-1}\hat{H}_{0}\psi_T(\boldsymbol{R})$.

Following the method in Refs.~\cite{Sanchez2018, Gordillo:2020sgc}, we define the quantity
\begin{equation}
\mathcal{F}(\boldsymbol{R},t)\equiv\sum_{\alpha}f_{\alpha}(\boldsymbol{R},t)\,.
\end{equation}
It propagates as
\begin{align}\label{evolutionF}
\mathcal{F}(\boldsymbol{R}',t+\Delta t)=&\int \mathrm{d}\boldsymbol{R} \langle \boldsymbol{R}'|e^{-A\Delta t}|\boldsymbol{R}\rangle\nonumber\\
&\times \sum_{\alpha'\alpha}[e^{-\mathbb{V}(\boldsymbol{R})\Delta t}]_{\alpha'\alpha} f_\alpha(\boldsymbol{R},t)\,.
\end{align}
To implement Eq.(\ref{evolutionF}) by algorithm, a set of coefficients $c_\alpha^{(i)}$ is attached to each walker. The Birth-Death process in Eq.(\ref{birth_death}) becomes
\begin{equation}\label{n}
n_{r}=\text{Floor}[\mathcal{B}_1\cdot \mathcal{B}_2 +u]\cdot\frac{N_0}{N}\,,
\end{equation}
with
\begin{eqnarray}
\mathcal{B}_1 &=&e^{-\left(\frac{E_L(\boldsymbol{R}')+E_L(\boldsymbol{R})}{2}-E_R\right)\Delta t}\,,\\
\mathcal{B}_2&=&\frac{\sum_{\alpha}c^{(i)\prime}_{\alpha}}{\sum_{\alpha}c^{(i)}_{\alpha}}\,,\quad c^{(i)\prime}_{\alpha} \equiv\sum_{\beta}[e^{-\mathbb{V}(\boldsymbol{R})\delta t}]_{\alpha\beta}c^{(i)}_{\beta}\,,
\end{eqnarray}
where $c^{(i)}_{\alpha}$ represents the proportion of the channel $\alpha$ among all the channels for the walker $i$. The mixed energy for the coupling channel is
\begin{align}
\langle E\rangle_{mixed}&=\frac{\int\mathcal{F}(\boldsymbol{R})E_{L}(\boldsymbol{R})\mathrm{d}\boldsymbol{R}+\int\sum_{\alpha\beta}V_{\alpha\beta}(\boldsymbol{R})f_{\beta}(\boldsymbol{R})\mathrm{d}\boldsymbol{R}}{\int\mathcal{F}(\boldsymbol{R})\mathrm{d}\boldsymbol{R}}\,.
\end{align}

\section{Hamiltonian}~\label{sec:Hamiltonian}

The nonrelativistic Hamiltonian of a 3-quark system reads
\begin{align}
H=\sum_i^3\left(m_i+\frac{\boldsymbol{p}_i^2}{2m_i}\right)-T_{CM}+V\,,
\end{align}
where $m_i$ and $\boldsymbol{p}_i$ are the mass and momentum of quark $i$. $T_{CM}$ is the center-of-mass kinematic energy, which automatically vanishes in the evolution, because the system will tend to the lowest energy state.

In order to investigate the effects of different confinement scenarios, we modify a nonrelativistic potential proposed in Ref.~\cite{Silvestre-Brac:1996myf}. The potential is divided into two parts, $V=V_0+V_\text{conf}$. For different scenarios, we choose the same $V_0$ terms,
\begin{eqnarray}
    V_{0}=&-&\frac{3}{16}\sum_{i<j}\lambda_i \cdot \lambda_j\left(-\frac{\kappa}{r_{ij}}+\frac{8\pi\kappa'}{3m_{i}m_{j}}\frac{\exp(-r_{ij}^{2}/r_{0}^{2})}{\pi^{3/2}r_{0}^{3}}\boldsymbol{s}_{i}\cdot\boldsymbol{s}_{j}\right)\nonumber
    \\&+&\frac{3}{16}\sum_{i<j}\lambda_i \cdot \lambda_j\Lambda-\frac{C}{m_{1}m_{2}m_{3}},
\end{eqnarray}
with
\begin{align}
r_0=A\left(\frac{2m_im_j}{m_i+m_j}\right)^{-B}\,,
\end{align}
where $\lambda_i$ is the $\text{SU}(3)$-color Gell-Mann matrix, $\boldsymbol{s}_i$ is the spin operator of quark $i$, and $r_{ij}$ is the relative distance between quark $i$ and $j$. In the above interaction, the Coulomb term and hyperfine term come from the one-gluon exchange interaction. The $\Lambda$ term is a pairwise constant interaction to shift the overall mass spectrum. The $C$ term is a three-body phenomenological contribution to obtain a better agreement with the experimental baryon masses. In Ref.~\cite{Silvestre-Brac:1996myf}, the parameters of the potential have been determined through fitting to a large number of mesons, and here we use the $AL$1 model parameters, which are listed in TABLE~\ref{AL1}.

\begin{table}[htbp] 
\centering
\caption{\label{AL1} $AL$1 quark model parameters taken from Ref.~\cite{Silvestre-Brac:1996myf}}
\begin{tabular*}{\hsize}{@{}@{\extracolsep{\fill}}cccc@{}}
   
\hline  \hline 
$m_u=m_d$ &   0.315 GeV        & $\lambda$   & 0.1653 $\mathrm{GeV}^2$   \\ 
$m_s$         &   0.577 GeV       & $\Lambda$  & 0.8321 $\mathrm{GeV}$ \\ 
$m_c$         &   1.836 GeV       & $A$             & 1.6553 $\mathrm{GeV}^{B-1}$\\ 
$m_b$         &   5.227 GeV      & $B$             & 0.2204\\ 
 $\kappa$     & 0.5069          &   $C$             &0.00202 $\mathrm{GeV}^{4}$ \\
   $\kappa'$    & 1.8609    &                       &  \\
\hline  \hline  
\end{tabular*}
\end{table}

In this work, we investigate two different confinement scenarios, as shown in Fig.~\ref{fig:two_confine}, the pairwise ($\Delta$-type) confinement and the three-body flux-tube (Y-type) confinement. In the first one, the two-body linear confinement term is introduced as
\begin{align}\label{Vconf}
V_{\mathrm{conf}}^{\Delta}=&-\frac{3}{16}\lambda \sum_{i<j}\lambda_i \cdot \lambda_j  r_{ij}\equiv\sigma_{\Delta}\sum_{i<j}r_{ij}.
\end{align}
where $\lambda$ is the confinement coupling constant in the $AL$1 model of Ref.~\cite{Silvestre-Brac:1996myf}. The value of $\lambda$ is presented in TABLE~\ref{AL1}. For the baryons, we introduce an alternative coupling constant $\sigma_{\Delta}$ to replace $\lambda$. Apparently, \clabel[perimeter]{the confinement potential is proportional to the length of the perimeter of the triangle connecting three quarks. }

In the second scenario, the confinement interaction is proportional to the minimal total length of the color flux tubes linking three quarks,
\begin{align}
V_{\mathrm{conf}}^Y=&\sigma_{Y} L_{min}\,,
\end{align}
where $\sigma_{Y}$ is the string tension. $L_{min}$ denotes the minimal length by tuning the joint point. In this scenario, three quarks are confined by the three-body force. In Refs.~\cite{Takahashi:2000te,Takahashi:2002bw}, the Y-type confinement interaction was favored by the lattice simulations. Another point of view is that the confinement in baryon is roughly one-half $\Delta$ and one-half Y-string \cite{koma2017precise,leech2021hyperspherical}, while it is not explored in this work.

We will take two $\sigma_{Y}$ values and give their results respectively later in this work. According to the lattice QCD result in Ref.~\cite{Takahashi:2000te}, a universal feature of the string tension $\sigma_{Y}\simeq\sigma_{Q\bar{Q}}$ is found. Herein for the first $\sigma_{Y}$ value, we naively take $\sigma_{Y}=\sigma_{Q\bar{Q}}=2\sigma_{\Delta}=\lambda$. However, the above value is only a rough approximation. To get a more accurate $\sigma_{Y}$ value, we adjust it to make the $\Omega^-(sss)$ mass coincide exactly with the experimental mass $1.672$ GeV. Since we expect the confinement coupling constants for the light quark systems and heavy quark systems could be different, the system composed of strange quarks (too heavy to be light and too light to be heavy) could be a good compromise. The benchmark gives us $\sigma_{Y}=0.9204\sigma_{Q\bar{Q}}=0.9204\lambda$. This coefficient is consistent with the best fitting parameters in lattice QCD simulation~\cite{Takahashi:2000te} with $\sigma_{Y}/\sigma_{Q\bar{Q}}=0.1524/0.1629=0.9355$.

For the three quark system, the minimal value of the total length $L_{min}$  linking three quarks can be obtained analytically~\cite{Takahashi:2002bw}. In this work, we choose a more general numerical algorithm to determine $L_{min}$ in order to extend the framework to the multiquark systems~\cite{Bai:2016int} in the future. This operation is essentially a Euclidean Steiner Tree Problem (ESTP)~\cite{smith1992find}. The ESTP seeks a network of the minimal length spanning a set of points by allowing the insertion of new points (Steiner points). The Smiths algorithm was designed for ESTP, which is an iterative method to optimize the search of coordinates of Steiner points in $d$-dimensional space, given a topology and terminal positions. The code of this program can be found in Ref.~\cite{ESMPgithub}.

\section{NUMERICAL RESULTS}~\label{sec:results}

In the simulation, we use $1\times10^4$ walkers to sample the $f(\bm{R},t)$ or $\mathcal{F}(\bm{R},t)$. We let the ensemble evolve $1\times10^4$ steps for the single-channel system and $2\times10^4$ steps for the coupling-channel systems with the $\Delta t=0.01\mathrm{GeV}^{-1}$ for each step to ensure stability. The resulting energy is averaged over the last 5000 steps to reduce fluctuation. \clabel[uncertainty]{To estimate the statistical uncertainty, the correlations among adjacent steps should be considered. To do this, we divide the steps into blocks and calculate the block averages. When the block size is taken large enough, the averages become uncorrelated and the uncertainty becomes independent of the block size. We find the blocks have become uncorrelated when taking the size as 500 steps. So we divide the last 5000 steps into 10 blocks of size 500 steps. Then by using the Jackknife resampling method \cite{gattringer2009quantum}, the uncertainty turns out to be less than 1 MeV. As an example, the uncertainty of $\Omega^-(sss)$ mass is 0.3 MeV. More details are given in Appendix~\ref{app:statistical_uncertainty}.}

Once the system is stabilized, the wave function will not change with time. We calculate the radial distribution $r_{ij}^2\rho(r_{ij})$, mean-square radius and charge mean-square radius using forward walking algorithm introduced in Sec.~\ref{forwardwalking}. The $\rho(r_{12})$ is defined as
\begin{align}\label{rhoij}
\rho(r_{12})=\int \mathrm{d}\hat{\boldsymbol{r}}_{12}\mathrm{d}\boldsymbol{r}_{3}|\Psi(\boldsymbol{r}_{1},\boldsymbol{r}_{2},\boldsymbol{r}_{3})|^{2}\,.
\end{align}
The square of the wave function $|\Psi(\boldsymbol{r}_{1},\boldsymbol{r}_{2},\boldsymbol{r}_{3})|^{2}$ can be obtained from Eq. (\ref{Phi02}). The definitions of $\rho(r_{13})$ and $\rho(r_{23})$ are similar. The root mean-square radius is defined as
\begin{align}\label{r}
\sqrt{\langle r_{ij}^{2}\rangle}&\equiv\sqrt{\langle\Psi|(\boldsymbol{r}_{i}-\boldsymbol{r}_{j})^{2}|\Psi\rangle}\,,
\end{align}
where $\boldsymbol{r}_{i}$ indicates the position of the $i$th quark. For the point-like quark, the charge mean-square radius $R_{c}^{2}$ is defined as 
\begin{align}\label{Rc2}
\langle R_{c}^{2}\rangle\equiv&\langle\Psi|\sum_{i=1}^{3}e_{i}(\boldsymbol{r}_{i}-\boldsymbol{R}_{CM})^{2}|\Psi\rangle\,+\sum_i^3 \langle R_c^2\rangle_{q_i},
\end{align}
where $\boldsymbol{R}_{CM}$ refers to the center of mass, and $e_i$ is the charge of the $i$th quark. In the practical calculation, the first term can be obtained from Eq.~\eqref{Apure}. The second term represent the effect of the charge radii of the constituent quarks. It is unreasonable to naively regard the constituent quarks as point-like particles. Details are given in Appendix~\ref{app:Rc2_size_contribution}. 

In this work, the charge radii of the constituent quarks are extracted from experiments and LQCD calculations. The contributions from the $u$ and $d$ quark size are extracted from the $p$ and $n$ experimental $\langle R_c^2\rangle$ values~\cite{ParticleDataGroup:2022pth}. In principle, one can extract the charge radius of the constituent $s$ quark from the experimental $\langle R_c^2\rangle$ of $\Sigma^-$~\cite{SELEX:2001fbx}. However, the present experimental uncertainty is large. Hence we choose to extract the $s$ and $c$ charge radii from the $\langle R_c^2\rangle$ of $\Omega(sss)$ and $\Omega(ccc)$ 
through lattice QCD simulations in Refs.~\cite{Can:2013tna,Can:2021ehb}. These contributions are listed in TABLE~\ref{Rc2_quark_size_contribution}. We can see that the quark charge radii vary with flavors, which is different from the scenario in Refs.~\cite{Wagner:1998fi,Wagner:2000ii}. Meanwhile, their signs are consistent with their charges. The electric size of the constituent quark decreases with the quark mass. In view of the small contribution of the $c$ quark size, we neglect the contribution from the $b$ quark.

\begin{table}[htbp] 
\centering
\caption{\label{Rc2_quark_size_contribution} The contributions of different quarks to the baryon $R_{c}^{2}$ due to the electromagnetic size of the constituent quark (in unit of $e\cdot\mathrm{ fm}^2$).}
\begin{tabular*}{\hsize}{@{}@{\extracolsep{\fill}}ccccc@{}}  
\hline  \hline 
$\langle R_c^2\rangle_u$              & $\langle R_c^2\rangle_d$    & $\langle R_c^2\rangle_s$      &  $\langle R_c^2\rangle_c$      & $\langle R_c^2\rangle_b$     \\ 
 $0.348$   & $-0.232$     & $-0.042$   & $0.009$                         & $0$     \\             
\hline  \hline     
\end{tabular*}
\end{table}

The distribution plots related to the wave function are averaged over 500 steps to reduce the noise and make it smooth. And the expectation value for the observables are averaged over 2000 steps.

\begin{figure}
    \centering
    \includegraphics[width=0.25\textwidth]{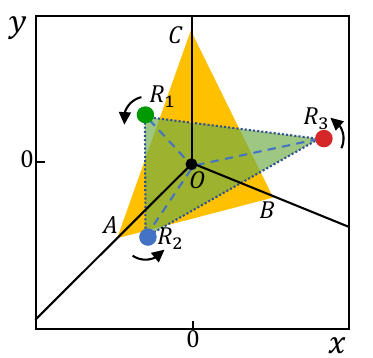}
    \caption{Operation to define the rotation-irrelevant distribution. For
each walker, the triangle $R_1R_2R_3$ connecting three
quarks is fixed into the $x-y$ plane with the center of mass at the origin $O$. Triangle $ABC$ is formed by three inner angle expectation values as a reference frame, with vertex $C$ on the $y$-axis. The rotating degree of freedom of triangle $R_1R_2R_3$ is fixed by minimizing the quantity $\angle R_1OA^2+\angle R_2OB^2+\angle R_3OC^2$. }
    \label{fig:rotation}
\end{figure}

In order to show the internal structure of the quarks more visually, we provide two additional plots, the angle distribution and rotation-irrelevant distribution following Ref.~\cite{Shen:2022bak}. For each walker, we introduce $\theta_1$, $\theta_2$ and $\theta_3=180^\circ-\theta_1-\theta_2$ to label the three inner angles of the triangle by linking three quarks. With the spatial probability distribution of walkers, we can easily get the angle distributions and the expectation values of the inner angles. To visualize the rotation-irrelevant structure, we define a way to eliminate the rotation degree of freedom. For each walker, we first put the triangle $R_1R_2R_3$ connecting three quarks into a 2D $x-y$ plane as shown in Fig.~\ref{fig:rotation}. We put its center of mass at the origin $O$. Now the only remaining degree of freedom for each walker is the rotation around the origin $O$. We then use the triangle $ABC$ formed by three inner angle expectation values to define a reference frame, where $A,B,C$ corresponds to $1,2,3$ quark respectively, and vertex $C$ is fixed on the y-axis. Finally we rotate the triangle $R_1R_2R_3$ to minimize the quantity $\angle R_1OA^2+\angle R_2OB^2+\angle R_3OC^2$. We can draw the new positions of quarks for all walkers in the plot and get a distribution reflecting the inner structure of the quarks. In principle, we can choose other angle-fixing strategies, for example minimizing $\angle R_1OA^4+\angle R_2OB^4+\angle R_3OC^4$. The different strategies will not change the qualitative properties. We will choose some typical baryons to show their angle distributions and rotation-irrelevant distributions. The complete distributions will be given in the Supplement material.

In the following sections, we will classify all of the ground state baryons according to the number of identical quarks inside them.  We treat the $u$ and $d$ quarks as identical particles under SU(2)-flavor symmetry. We will use $n$ to label them. The related spin matrix elements are presented in TABLE~\ref{spin_matrix_element}. For the the $AL$1 model, apart from the DMC, we use a variational method with $20^2$ Gaussian basis functions~\cite{Hiyama:2003cu} as a benchmark, which is denoted as GEM in the following. Our results are also compared with the Faddeev formalism results in Ref.~\cite{Silvestre-Brac:1996myf}. Furthermore, for the flux-tube confinement model, we use two sets of parameters as mentioned above. Flux-tube I refers to the universal tension parameter with $\sigma_{Y}=\lambda$, and Flux-tube II refers to $\sigma_{Y}=0.9204\lambda$ determined by the experimental $\Omega^-(sss)$ mass. Both of them are solved using DMC algorithm.

\begin{table}[htbp] 
\centering
\caption{\label{spin_matrix_element} The spin matrix element $\langle\chi_{\mathrm{spin}}|\boldsymbol{s}_i\cdot\boldsymbol{s}_j|\chi_{\mathrm{spin}}\rangle$ for the $(i,j)$ pair of quarks. }
\begin{tabular*}{\hsize}{@{}@{\extracolsep{\fill}}cccc@{}}
   
\hline  \hline                                                                                     
& $(1,2)$                & $(1,3)$                         & $(2,3)$             \\ 
\hline
$\langle[(12)_1 3]_{\frac{3}{2}}|\boldsymbol{s}_i\cdot\boldsymbol{s}_j|[(12)_1 3]_{\frac{3}{2}}\rangle$             & $\frac{1}{4}$        & $\frac{1}{4}$                 & $\frac{1}{4}$      \\ 
$\langle[(12)_1 3]_{\frac{1}{2}}|\boldsymbol{s}_i\cdot\boldsymbol{s}_j|[(12)_1 3]_{\frac{1}{2}}\rangle$              & $\frac{1}{4}$        & $-\frac{1}{2}$               & $-\frac{1}{2}$     \\ 
$\langle[(12)_0 3]_{\frac{1}{2}}|\boldsymbol{s}_i\cdot\boldsymbol{s}_j|[(12)_0 3]_{\frac{1}{2}}\rangle$             & $-\frac{3}{4}$      & $0$                             & $0$                   \\ 
$\langle[(12)_0 3]_{\frac{1}{2}}|\boldsymbol{s}_i\cdot\boldsymbol{s}_j|[(12)_1 3]_{\frac{1}{2}}\rangle$              &$0$                     & $-\frac{\sqrt{3}}{4}$      & $\frac{\sqrt{3}}{4}$     \\ 
\hline  \hline  
\end{tabular*}
\end{table}

\subsection{$J^P=\frac{3}{2}^+$ without identical quarks}~\label{sec:1q_3over2}

We label three quarks in mass ascending order with 1,2 and 3. Since there are no identical particles, there is no need to satisfy the Pauli principle. For the ground spin-$3\over 2$ system (assuming no orbital excitation), the spin wave function is symmetric and unique, whose spin matrix elements can be found in TABLE~\ref{spin_matrix_element}.

\begin{figure*}[htbp]
  \centering
  \includegraphics[width=0.9\textwidth]{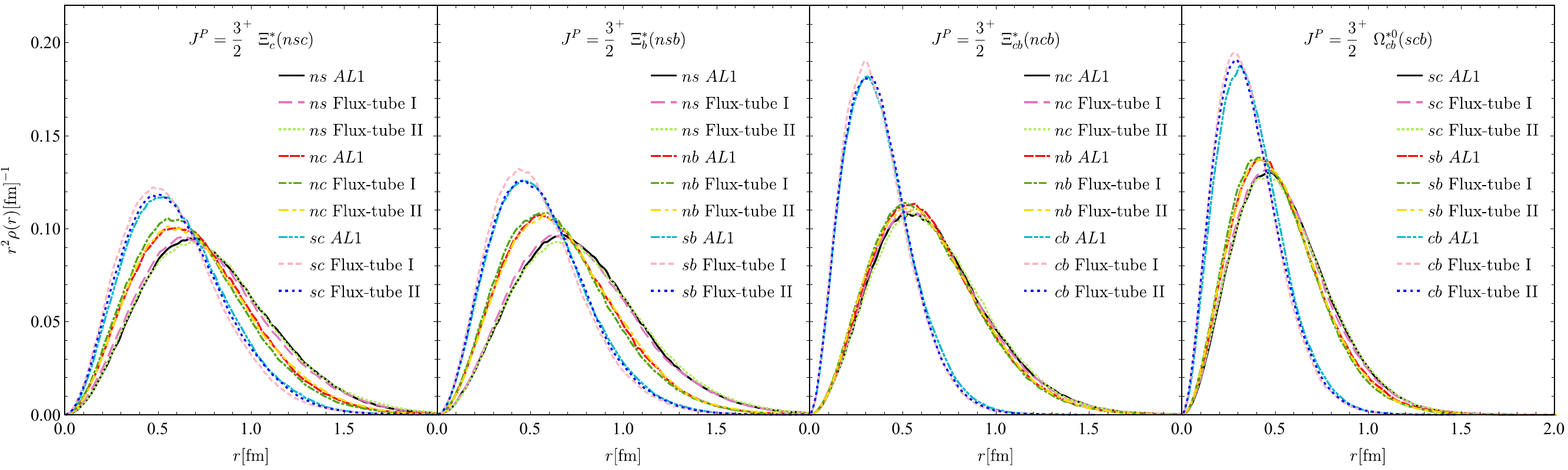} 
  \caption{\label{1q_3over2_r2rho} The $r^2\rho(r)$ distributions for the $J^P=\frac{3}{2}^+$ baryons without identical quarks. }
    \setlength{\belowdisplayskip}{1pt}
\end{figure*}

\begin{figure*}[htb] 
\centering
\includegraphics[width=1\textwidth]{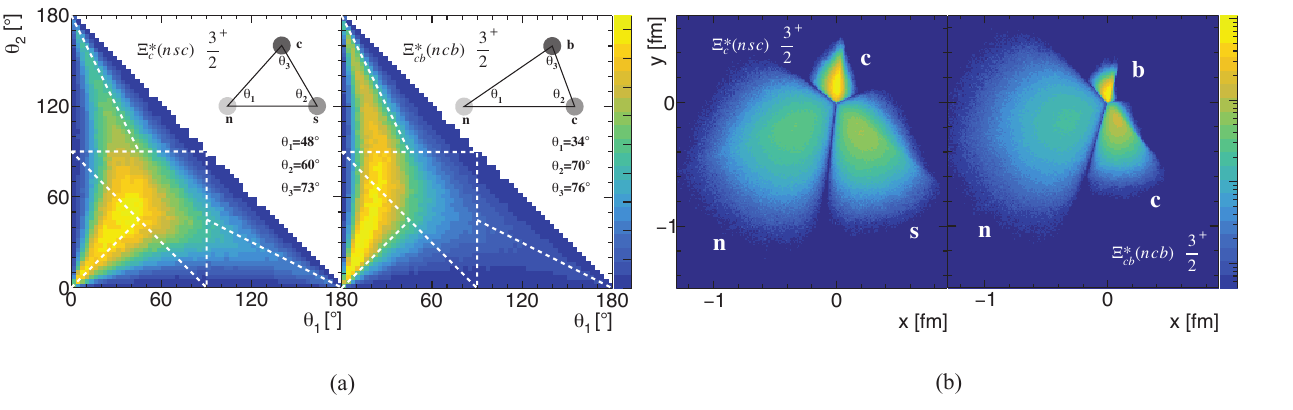} 
\caption{\label{1q_2D_nsc_ncb} (a) Internal angle distribution of quarks in the $\Xi{}_{c}^{*}(nsc)$ and $\Xi_{cb}^{*}(ncb)$ respectively in the $AL$1 model.  The white dashed triangle in (a) indicates three quarks form a right-angled triangle. Walkers appearing inside this triangle refer to acute triangles, and outside refer to obtuse triangles. The other three white dashed lines correspond to obtuse isosceles triangles. The black triangle drawn in the upper right corner indicates the calculated expectation values of the inner angles (averaged over identical quarks if it has). The yellow (blue) color signals a high (low) probability. (b) 2D wave function probability distribution of the $\Xi{}_{c}^{*}(nsc)$ and $\Xi_{cb}^{*}(ncb)$ baryon respectively in the $AL$1 model. The color bar represents the logarithmic coordinates in this subfigure.}
\end{figure*}

The masses and radii are listed in TABLES ~\ref{1q_3over2_mass} and~\ref{1q_3over2_expect}, and the $r^2_{ij}\rho(r_{ij})$ distributions are shown in Fig.~\ref{1q_3over2_r2rho}. In Fig.~\ref{1q_2D_nsc_ncb}, we use $\Xi{}_{c}^{*}(nsc)$ and $\Xi_{cb}^{*}(ncb)$ as two examples to illustrate the distribution of inner angles and the 2D wave function probability distribution. Here only the $AL$1 model results are shown because the three models have little difference. 

One can see that the results for the $AL$1 model from the DMC and GEM are consistent. Furthermore, for the flux-tube model, our results show the Flux-tube I with the naive universal tension $\sigma_{Y}=\sigma_{Q\bar{Q}}$ is not as good as the Flux-tube II. Compared with the experimental data, the Flux-tube II and $AL$1 results, the Flux-tube I overestimates the mass and underestimates the sizes. From the $r_{ij}^2\rho(r_{ij})$ distributions, one can see that the more massive quark pair tend to get closer. Meanwhile, the rotation-irrelevant distributions shows that the heavier quark will be closer to the center of mass.

\begin{table}[htbp] 
\centering
\caption{\label{1q_3over2_mass} Masses of the $J^P=\frac{3}{2}^+$ baryons in MeV. The interaction models include the $AL$1, Flux-tube I (FT I) and Flux-tube II (FT II). The DMC method and variational method (Gaussian expansion method, GEM) are used to solve the three-body problem. The Faddeev equation (FAD) results are presented if they were provided in Ref.~\cite{Silvestre-Brac:1996myf}. The experimental results (EXP) (if they exist) are averaged over the isospin multiples.}
\begin{tabular*}{\hsize}{@{}@{\extracolsep{\fill}}cccccccc@{}}
  
\hline  \hline 
\multirow{2}*{$J^{P}=\frac{3}{2}^{+}$}   &\multicolumn{3}{c}{$AL$1}                                                          & FT I      & FT II         & \multirow{2}*{EXP\cite{ParticleDataGroup:2022pth}}  \\
\cline{2-4}
  ~                                                     &  DMC          &  VAR    &  FAD~\cite{Silvestre-Brac:1996myf}     &DMC                &DMC                   &  ~                              \\
\hline 
$\Xi{}_{c}^{*}(nsc)$ & 2646 & 2645&   /  & 2725 & 2650 & 2646\\
$\Xi{}_{b}^{*}(nsb)$ & 5972 & 5971&   /  & 6046 & 5974 & 5954\\
$\Xi_{cb}^{*}(ncb)$ & 6990 & 6989&   /  & 7047 & 6986 & /\\
$\Omega_{cb}^{*0}(scb)$ & 7072 & 7070&   /  & 7121 & 7071 & /\\
\hline 
$\Sigma^{*}(nns)$   & 1404  & 1402   &   /                 & 1504              & 1412                    & 1385                  \\
$\Sigma_{c}^{*}(nnc)$  & 2535             & 2535  &   /                  & 2624             & 2540                    & 2518                  \\
$\Sigma_{b}^{*}(nnb)$                        & 5875             & 5874      &   /             & 5959              & 5878                   & 5833                  \\
\hline
$\Xi^{*}(ssn)$                                    & 1541             & 1540         &   /          & 1633              & 1549                    & 1533                  \\
$\Xi_{cc}^{*}(ccn)$   & 3701             & 3700   &   /        & 3765             & 3700	                   &   /                       \\
$\Xi_{bb}^{*}(bbn)$       & 10233            & 10232    &   /    & 10275            & 10222	                   &   /                       \\
\hline
$\Omega_{c}^{*0}(ssc)$                       & 2749            & 2749   &   /       & 2821               & 2755                   & 2766                   \\
$\Omega_{cc}^{*+}(ccs)$                     & 3791             & 3790   &   /        & 3849             & 3794	                   &   /                       \\
$\Omega_{b}^{*-}(ssb)$                       & 6064            & 6063   &   /       & 6129              & 6067                   &   /                      \\
$\Omega_{ccb}^{*+}	(ccb)$                   & 8046             & 8046  &   /         & 8087            & 8049	                   &   /                       \\
$\Omega_{bb}^{*-}(bbs)$                     & 10304            & 10303   &   /        & 10344           & 10299	                   &   /                       \\
$\Omega_{cbb}^{*0}	(bbc)$                   & 11247             & 11247  &   /      & 11280             & 11248	                   &   /                       \\
\hline 
$\Delta(nnn)$                                    & 1243             & 1242                 &   /                                          & 1353              & 1253                    & 1232                  \\
$\Omega^{-}(sss)$                             & 1664            & 1664                  & 1663                                       & 1749               & 1672                   & 1672                   \\
$\Omega_{ccc}^{++}(ccc)$                  & 4799            & 4798                  & 4799                                      & 4848              & 4803                   &   /                      \\
$\Omega_{bbb}^{-}(bbb)$                   & 14398           & 14398                & 14398                                     & 14424             & 14400	            &   /                       \\
\hline  \hline  
\end{tabular*}
\end{table}

\begin{table*}[htbp] 
\centering
\caption{\label{1q_3over2_expect} Root mean square radii and charge mean-square radii $R_{c}^{2}$ expectation values of the $J^P=\frac{3}{2}^+$ baryons without identical quarks. ``FT I" is short for ``Flux-tube I", and ``FT II" is short for ``Flux-tube II".}
\begin{tabular*}{\hsize}{@{}@{\extracolsep{\fill}}ccccccccccccc@{}}
  
\hline  \hline 
\multirow{2}*{$J^{P}=\frac{3}{2}^{+}$}  &\multicolumn{3}{c}{$\sqrt{\langle r_{12}^{2}\rangle} [\mathrm{fm}]$}    &\multicolumn{3}{c}{$\sqrt{\langle r_{13}^{2}\rangle} [\mathrm{fm}]$}      &\multicolumn{3}{c}{$\sqrt{\langle r_{23}^{2}\rangle} [\mathrm{fm}]$}  &\multicolumn{3}{c}{$\langle R_{c}^{2}\rangle[e\cdot\mathrm{ fm^{2}}]$}  \\
\cline{2-4}\cline{5-7}\cline{8-10}\cline{11-13}
  ~                                             &$AL$1                        & FT I                  & FT II                 &$AL$1            & FT I      & FT II      &$AL$1                        & FT I                  & FT II                &$AL$1                        & FT I                  & FT II \\
\hline 

$\Xi{}_{c}^{*0}(dsc)$ & \multirow{2}{*}{0.854} & \multirow{2}{*}{0.841} & \multirow{2}{*}{0.863} & \multirow{2}{*}{0.776} & \multirow{2}{*}{0.758} & \multirow{2}{*}{0.779} & \multirow{2}{*}{0.666} & \multirow{2}{*}{0.642} & \multirow{2}{*}{0.659} & $-0.471$ & $-0.462$ & $-0.471$\tabularnewline
$\Xi{}_{c}^{*+}(usc)$ &  &  &  &  &  &  &  &  &  & 0.541 & 0.533 & 0.549\tabularnewline
\cline{2-10}
$\Xi{}_{b}^{*-}(dsb)$ & \multirow{2}{*}{0.843} & \multirow{2}{*}{0.829} & \multirow{2}{*}{0.851} & \multirow{2}{*}{0.738} & \multirow{2}{*}{0.717} & \multirow{2}{*}{0.740} & \multirow{2}{*}{0.617} & \multirow{2}{*}{0.595} & \multirow{2}{*}{0.610} & $-0.535$ & $-0.519$ & $-0.534$\tabularnewline
$\Xi{}_{b}^{*0}(usb)$ &  &  &  &  &  &  &  &  &  & 0.521 & 0.511 & 0.524\tabularnewline
\cline{2-10}
$\Xi_{cb}^{*0}(dcb)$ & \multirow{2}{*}{0.727} & \multirow{2}{*}{0.720} & \multirow{2}{*}{0.737} & \multirow{2}{*}{0.702} & \multirow{2}{*}{0.693} & \multirow{2}{*}{0.709} & \multirow{2}{*}{0.420} & \multirow{2}{*}{0.404} & \multirow{2}{*}{0.411} & $-0.306$ & $-0.307$ & $-0.315$\tabularnewline
$\Xi_{cb}^{*+}(ucb)$ &  &  &  &  &  &  &  &  &  & 0.705 & 0.695 & 0.706\tabularnewline
\cline{2-10}
$\Omega_{cb}^{*0}(scb)$ & 0.601 & 0.592 & 0.603 & 0.569  & 0.558 & 0.566 & 0.408  & 0.394  & 0.401 & $-0.062$ & $-0.063$ & $-0.063$\tabularnewline

\hline  \hline  
\end{tabular*}
\end{table*}

\subsection{ $J^P=\frac{3}{2}^+$ with two identical quarks}~\label{sec:2q_3over2}

We label the two identical quarks as $1$ and $2$, and the remaining one as $3$. The spin wave function is still completely symmetric. The space wave function is symmetric since it is a ground state. As for the flavor part, the baryons with two identical $s,c,b$ quarks are apparently symmetric for exchanging $q_1$ and $q_2$. For the baryons with two identical $u,d$ quarks, it should be constructed symmetrically with $I=1$ to fulfill the Pauli principle.

\begin{figure*}[htbp]
  \centering
  \includegraphics[width=0.9\textwidth]{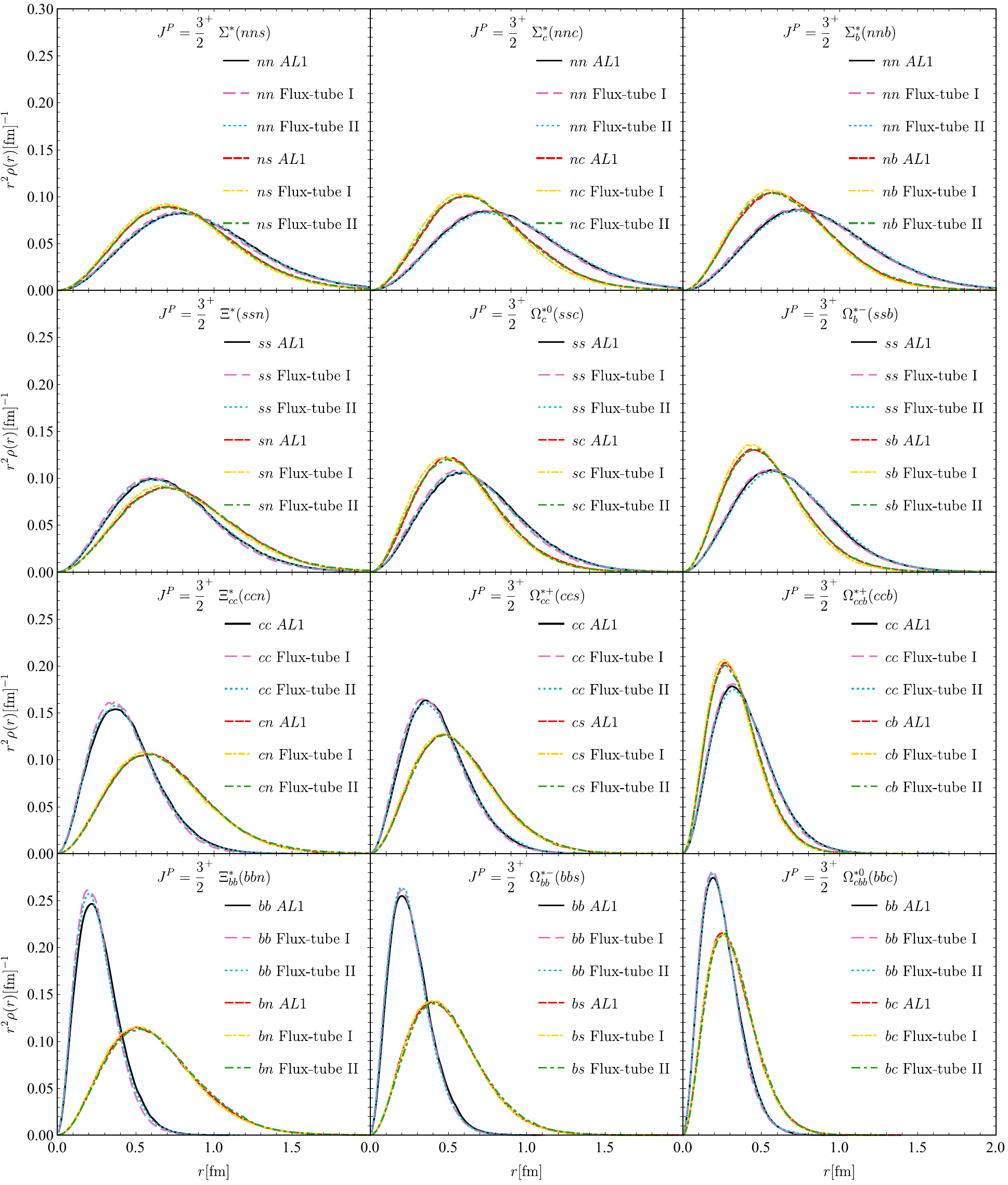} 

  \caption{\label{2q_3over2_r2rho} The $r^2\rho(r)$ distributions for the $J^P=\frac{3}{2}^+$ baryons with two identical quarks.}
    \setlength{\belowdisplayskip}{1pt}
\end{figure*}

The masses, root mean square radii and the charge mean-square radii of the $J^P=\frac{3}{2}^+$ baryons with two identical quarks are shown in TABLES~\ref{1q_3over2_mass} and~\ref{2q_3over2_expect}. The radial distributions are displayed in Fig.~\ref{2q_3over2_r2rho}. Again, the results from DMC and variational method for the $AL$1 model are consistent. The baryon masses from Flux-tube II are in better agreement with the experiments than those of Flux-tube I. For Flux-tube I, $\sqrt{\langle r^{2}\rangle}$ and $R_{c}^{2}$ are both smaller. For the radial distribution, there is still the general property that more massive quarks will get closer.

\begin{figure*}[htb] 
\centering
\includegraphics[width=0.75\textwidth]{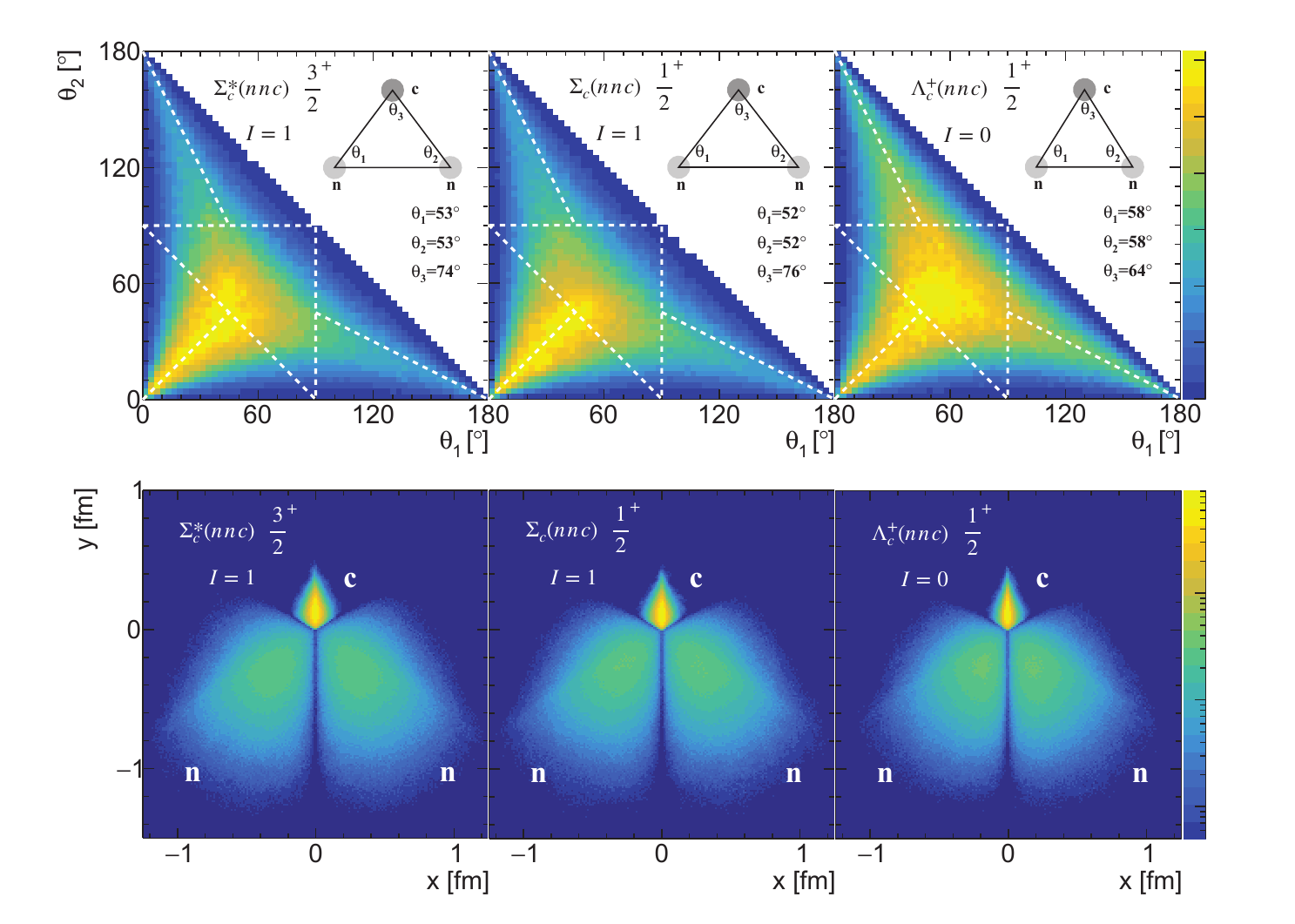} 

\caption{\label{2q_2D_nnc} Up: Internal angle distribution of the quarks in the $\Sigma_c^*(nnc)$, $\Sigma_c(nnc)$ and $\Lambda_c(nnc)$ baryon respectively in the $AL$1 model.  Low: 2D wave function probability distribution of the $\Sigma_c^*(nnc)$, $\Sigma_c(nnc)$ and $\Lambda_c(nnc)$  baryon respectively in the $AL$1 model. Other notations are the same as those in Fig.~\ref{1q_2D_nsc_ncb}.}
\end{figure*}

\begin{figure*}[htb] 
\centering
\includegraphics[width=1\textwidth]{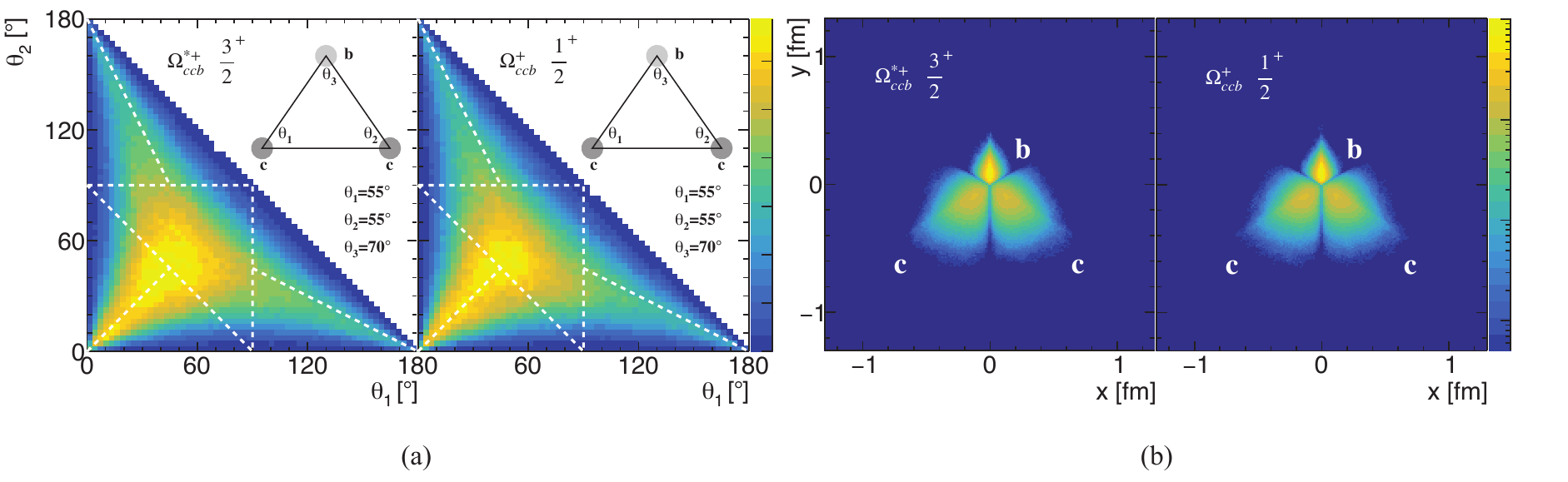} 
\caption{\label{2q_2D_ccb} (a) Internal angle distribution of quarks in the $\Omega_{ccb}^{*+}(ccb)$ and $\Omega_{ccb}^{+}(ccb)$ baryon respectively in the $AL$1 model.  (b) 2D wave function probability distribution of $\Omega_{ccb}^{*+}(ccb)$ and $\Omega_{ccb}^{+}(ccb)$ baryon respectively in $AL$1 model. Other notations are the same as those in Fig.~\ref{1q_2D_nsc_ncb}.}
\end{figure*}

As an example, the distribution of the inner angles and the 2D wave function probability distribution of the $\Sigma_c^*(nnc)$ are given in the left column of Fig.~\ref{2q_2D_nnc}. Likewise, only the $AL$1 model results are shown since the three models have little difference. For the baryons with two identical quarks, the shape of the triangle is isosceles triangle. Apparently, the heavier quark tends to get closer to the center of mass, thus, the angle with the heavier quark as the vertex is larger. Similarly, the distributions of the $\Omega_{ccb}^{*+}(ccb)$ are given in Fig.~\ref{2q_2D_ccb}.

\begin{table*}[htbp] 
\centering
\caption{\label{2q_3over2_expect} Root mean square radii and charge mean-square radii $R_{c}^{2}$ expectation values of the $J^P=\frac{3}{2}^+$ baryons with two identical quarks. The $\sqrt{\langle r_{13}^{2}\rangle}$ column is the average of $\sqrt{\langle r_{13}^{2}\rangle}$ and $\sqrt{\langle r_{23}^{2}\rangle}$, since $1,2$ are the identical quarks. }
\begin{tabular*}{\hsize}{@{}@{\extracolsep{\fill}}cccccccccc@{}}
  
\hline  \hline 
\multirow{2}*{$J^{P}=\frac{3}{2}^{+}$}   &\multicolumn{3}{c}{$\sqrt{\langle r_{12}^{2}\rangle} [\mathrm{fm}]$}    &\multicolumn{3}{c}{$\sqrt{\langle r_{13}^{2}\rangle} [\mathrm{fm}]$}    &\multicolumn{3}{c}{$\langle R_{c}^{2}\rangle[e\cdot\mathrm{ fm^{2}}]$}  \\
\cline{2-4}\cline{5-7}\cline{8-10}
  ~                                                     &$AL$1                        &FT I                  &FT II                 &$AL$1            &FT I      &FT II      &$AL$1                        &FT I                  &FT II \\
\midrule[0.5pt]

$\Sigma^{*-}(dds)$ &\multirow{3}*{0.986} &\multirow{3}*{0.962} &\multirow{3}*{0.986}   &\multirow{3}*{0.903}  &\multirow{3}*{0.875} &\multirow{3}*{0.899} &$-0.807$ &$-0.789$ &$-0.804$   \\
$\Sigma^{*0}(uds)$ & ~                               & ~                               &  ~                      & ~                               & ~                               &  ~                  &0.146  &0.144  & 0.146      \\
$\Sigma^{*+}(uus)$ & ~                               & ~                               &  ~                      & ~                               & ~                               &  ~                  &1.099  &1.079  & 1.102      \\
\cline{2-7}
$\Sigma_{c}^{*0}(ddc)$ &\multirow{3}*{0.954} &\multirow{3}*{0.940} &\multirow{3}*{0.965} &\multirow{3}*{0.795} &\multirow{3}*{0.771} &\multirow{3}*{0.791} &$-0.739$ &$-0.725$ &$-0.738$  \\
$\Sigma_{c}^{*+}(udc)$ & ~                            & ~                           &  ~                      & ~                               & ~                               &  ~                  &0.292  &0.285  & 0.293         \\
$\Sigma_{c}^{*++}(uuc)$ & ~                          & ~                           &  ~                      & ~                               & ~                               &  ~                  &1.324  &1.289  & 1.325        \\
\cline{2-7}
$\Sigma_{b}^{*-}(ddb)$ &\multirow{3}*{0.942} &\multirow{3}*{0.931} &\multirow{3}*{0.953} &\multirow{3}*{0.756} &\multirow{3}*{0.735} &\multirow{3}*{0.752} &$-0.800$ &$-0.782$ &$-0.797$  \\
$\Sigma_{b}^{*0}(udb)$ & ~                            & ~                           &  ~                      & ~                               & ~                               &  ~                  &0.278  &0.270  & 0.279        \\
$\Sigma_{b}^{*+}(uub)$ & ~                          & ~                           &  ~                      & ~                                & ~                                &  ~                  &1.363  &1.328  & 1.358       \\
\midrule[0.5pt]
$\Xi^{*-}(ssd)$ &\multirow{2}*{0.792} &\multirow{2}*{0.767} &\multirow{2}*{0.787} &\multirow{2}*{0.884} &\multirow{2}*{0.866} &\multirow{2}*{0.889} &$-0.567$ &$-0.555$ &$-0.569$  \\
$\Xi^{*0}(ssu)$ & ~                            & ~                           &  ~                      & ~                               & ~                            &  ~                        &0.398  &0.398  & 0.404        \\
\cline{2-7}
$\Xi_{cc}^{*+}(ccd)$          &\multirow{2}*{0.497} &\multirow{2}*{0.476} &\multirow{2}*{0.486} &\multirow{2}*{0.743} &\multirow{2}*{0.734} &\multirow{2}*{0.752} &$-0.266$ &$-0.270$ &$-0.275$ \\
$\Xi_{cc}^{*++}(ccu)$        & ~                            & ~                            & ~                            & ~                            & ~                            & ~        & 0.732 & 0.719 & 0.735\\
\cline{2-7}
$\Xi_{bb}^{*-}(bbd)$         &\multirow{2}*{0.304} &\multirow{2}*{0.289} &\multirow{2}*{0.295} &\multirow{2}*{0.680} &\multirow{2}*{0.675} &\multirow{2}*{0.692} &$-0.386$ &$-0.382$ &$-0.390$\\
$\Xi_{bb}^{*0}(bbu)$         & ~                            & ~                            & ~                            & ~                           &  ~                           & ~                         & 0.608 & 0.607 & 0.621\\
\midrule[0.5pt]
$\Omega_{c}^{*0}(ssc)$     & 0.747 & 0.734 & 0.755 & 0.648 & 0.631 & 0.651 & $-0.210$ & $-0.204$ & $-0.212$\\
$\Omega_{cc}^{*+}(ccs)$   & 0.483 & 0.465 & 0.481 & 0.618 & 0.609 & 0.624 & $-0.018$ & $-0.022$ & $-0.020$\\
$\Omega_{b}^{*-}(ssb)$     & 0.729 & 0.718 & 0.739 & 0.599 & 0.582 & 0.598 & $-0.275$ & $-0.265$ & $-0.275$\\
$\Omega_{ccb}^{*+}(ccb)$ & 0.435 & 0.427 & 0.438 & 0.379 & 0.371 & 0.379 & 0.121 & 0.117 & 0.122\\
$\Omega_{bb}^{*-}(bbs)$   & 0.295 & 0.287 & 0.287 & 0.543 & 0.538 & 0.549 & $-0.139$ & $-0.137$ & $-0.140$\\
$\Omega_{cbb}^{*0}(bbc)$  & 0.274 & 0.268 & 0.273 & 0.354 & 0.350 & 0.355 & 0.046 & 0.046 & 0.047\\

\hline  \hline  
\end{tabular*}
\end{table*}

\begin{figure*}[htb] 
\centering
\includegraphics[width=1\textwidth]{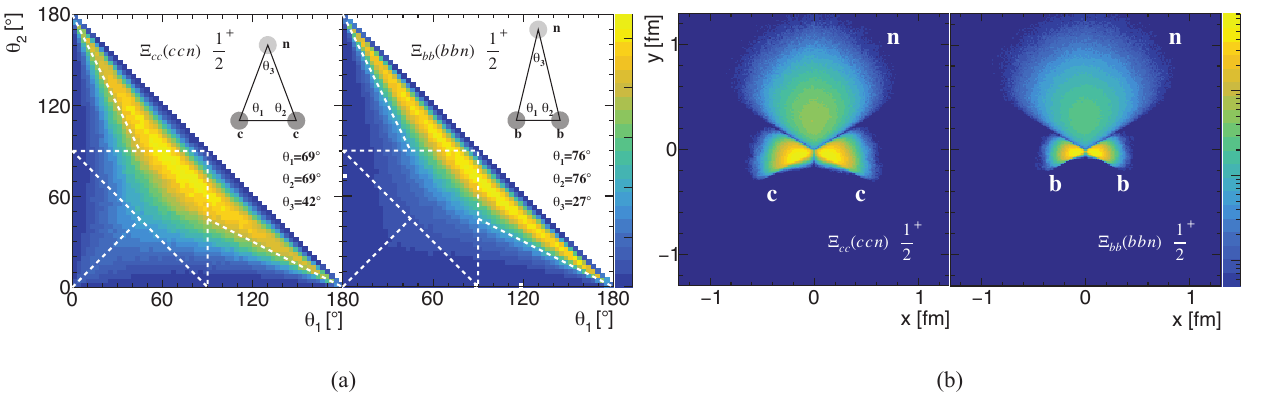} 
\caption{\label{2q_2D_ccn_bbn} (a) Internal angle distribution of quarks in the $\Xi_{cc}(ccn)$ and $\Xi_{bb}(bbn)$ baryon respectively in the $AL$1 model. (b) 2D wave function probability distribution of the $\Xi_{cc}(ccn)$ and $\Xi_{bb}(bbn)$ baryon respectively in the $AL$1 model. Other notations are the same as those in Fig.~\ref{1q_2D_nsc_ncb}.}
\end{figure*}

\subsection{$J^P=\frac{3}{2}^+$ with three identical quarks}~\label{sec:3q_3over2}

For the ground baryons composed of three identical quarks with $J^P=\frac{3}{2}^+$, the spin and spatial wave functions are both completely symmetric. The spin matrix elements $\langle\chi_{\mathrm{spin}}|\boldsymbol{s}_i\cdot\boldsymbol{s}_j|\chi_{\mathrm{spin}}\rangle$ for the $(ij)$ pair of quarks are shown in TABLE~\ref{spin_matrix_element}. As for the flavor part, $\Omega^{-}(sss)$, $\Omega_{ccc}^{++}(ccc)$, $\Omega_{bbb}^{-}(bbb)$, $\Delta^{-}(ddd)$ and $\Delta^{++}(uuu)$ are fully symmetric. Considering the SU(2)-flavor symmetry, the $u$ and $d$ are identical particles, thus the $\Delta^{0}(udd)$ and $\Delta^{+}(uud)$ should be constructed symmetrically to fulfill the Pauli principle.

\begin{figure*}[htbp]
  \centering
  \includegraphics[width=1\textwidth]{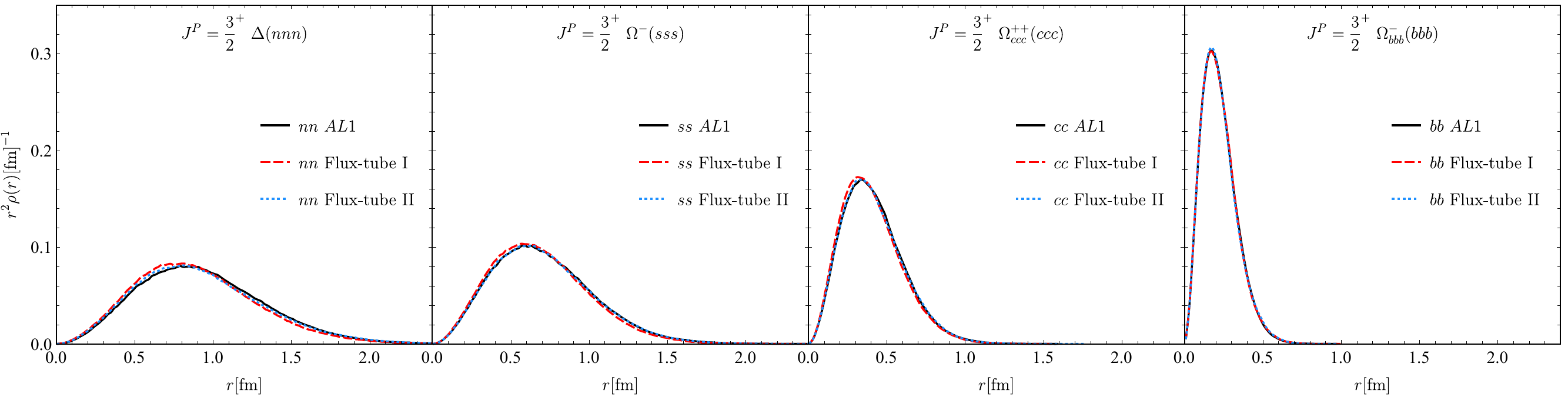} 
 
  \caption{\label{qqq_3over2_r2rho} The $r^2\rho(r)$ distributions for the $J^P=\frac{3}{2}^+$ baryons with three identical quarks.}
    \setlength{\belowdisplayskip}{1pt}
\end{figure*}

The masses of the $J^P={3\over 2}^+$ baryons with three identical quarks are shown in TABLE~\ref{1q_3over2_mass}. The root mean square radii and the charge mean-square radii are given in TABLE~\ref{qqq_3over2_expect}. The distributions for the $\frac{3}{2}^+$ baryons with three identical quarks are displayed in Fig.~\ref{qqq_3over2_r2rho}. We get consistent results for different few-body methods, DMC, GEM and Faddeev formalism. For the different interactions, the Flux-tube II works as good as the $AL$1 model and better than Flux-tube I.

\begin{figure}[htb] 
\centering
\includegraphics[width=0.5\textwidth]{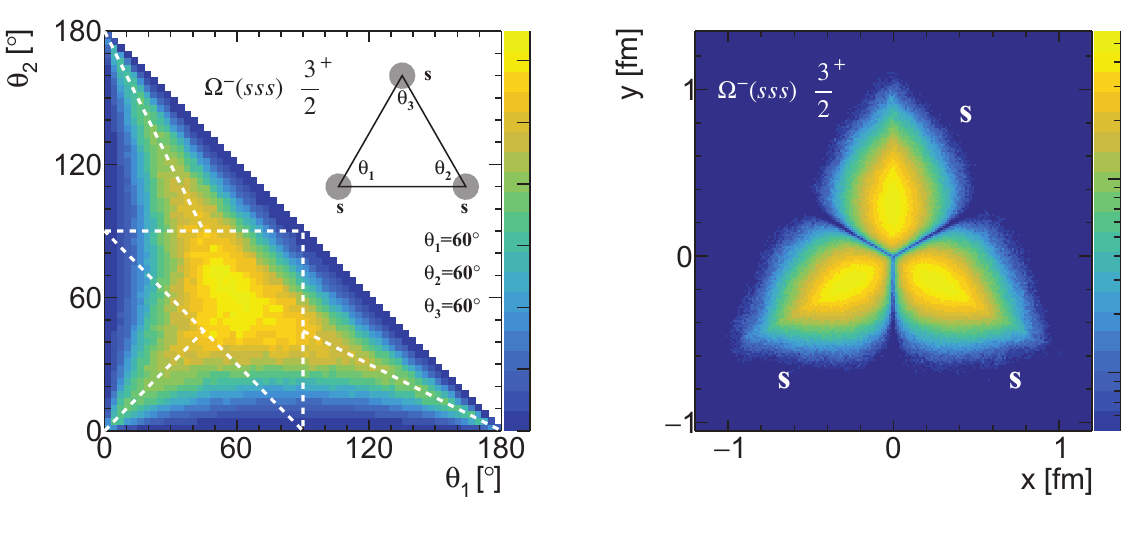} 
\caption{\label{qqq_3over2_2D_eg} Left: Internal angle distribution of quarks in the $\Omega^-(sss)$ baryon in the $AL$1 model. Other notations are the same as those in Fig.~\ref{1q_2D_nsc_ncb}.}
\end{figure}

In Fig.~\ref{qqq_3over2_2D_eg}, we use $\Omega^-(sss)$ as an example to illustrate the distribution of angle and the rotation-irrelevant distribution. Here only the $AL$1 model results are shown because the three models have little difference. In the left panel of Fig.~\ref{qqq_3over2_2D_eg}, the walkers are mainly distributed inside the white triangle and concentrated around $60^\circ$. This makes sense as they are three identical quarks. In the right panel of Fig.~\ref{qqq_3over2_2D_eg}, we can identify three identical regions stemming from three identical quarks. It is worthwhile mentioning that the almond shape depends on the specific angle-fixing strategy, which could be distorted if we chose different angle-fixing strategies.

\begin{table}[htbp] 
\centering
\caption{\label{qqq_3over2_expect} Root mean square radii and charge mean-square radii $R_{c}^{2}$ expectation values of the $J^P=\frac{3}{2}^+$ baryons with three identical quarks. The root mean square radii $\sqrt{\langle r^{2}\rangle}$ are averaged for the identical quarks. }
\begin{tabular*}{\hsize}{@{}@{\extracolsep{\fill}}ccccccc@{}}
  
\hline  \hline 
\multirow{2}*{$J^{P}=\frac{3}{2}^{+}$}   &\multicolumn{3}{c}{$\sqrt{\langle r^{2}\rangle} [\mathrm{fm}]$}     &\multicolumn{3}{c}{$\langle R_{c}^{2}\rangle[e\cdot\mathrm{ fm^{2}}]$}  \\
\cline{2-4}\cline{5-7}
  ~                                                     &$AL$1                        &FT I                  &FT II                 &$AL$1            &FT I      &FT II    \\
\hline 
$\Delta^-(ddd)$                                 & \multirow{4}*{1.003}    & \multirow{4}*{0.976}    & \multirow{4}*{1.001}      &$-1.034$          &$-1.013$             &$-1.031$            \\
$\Delta^0(udd)$                                 & ~                               & ~                               &  ~                               &$-0.116$             &$-0.118$             & $-0.117$             \\
$\Delta^+(uud)$                                 & ~                               & ~                               &  ~                               &0.796            &0.779                & 0.799            \\
$\Delta^{++}(uuu)$                             & ~                               & ~                               &  ~                               &1.715             &1.679               & 1.716             \\
$\Omega^{-}(sss)$                             & 0.775                         & 0.756                        & 0.777                          & $-0.325$           & $-0.316$            & $-0.326$            \\
$\Omega_{ccc}^{++}(ccc)$                  & 0.458                         & 0.447                        & 0.458                          & 0.168            & 0.161              &  0.168             \\
$\Omega_{bbb}^{-}(bbb)$                   & 0.249                         & 0.247                        & 0.250                          & $-0.021$          & $-0.020$            & $-0.021$            \\
\hline  \hline  
\end{tabular*}
\end{table}

\begin{figure*}[htbp]
  \centering
  \includegraphics[width=0.9\textwidth]{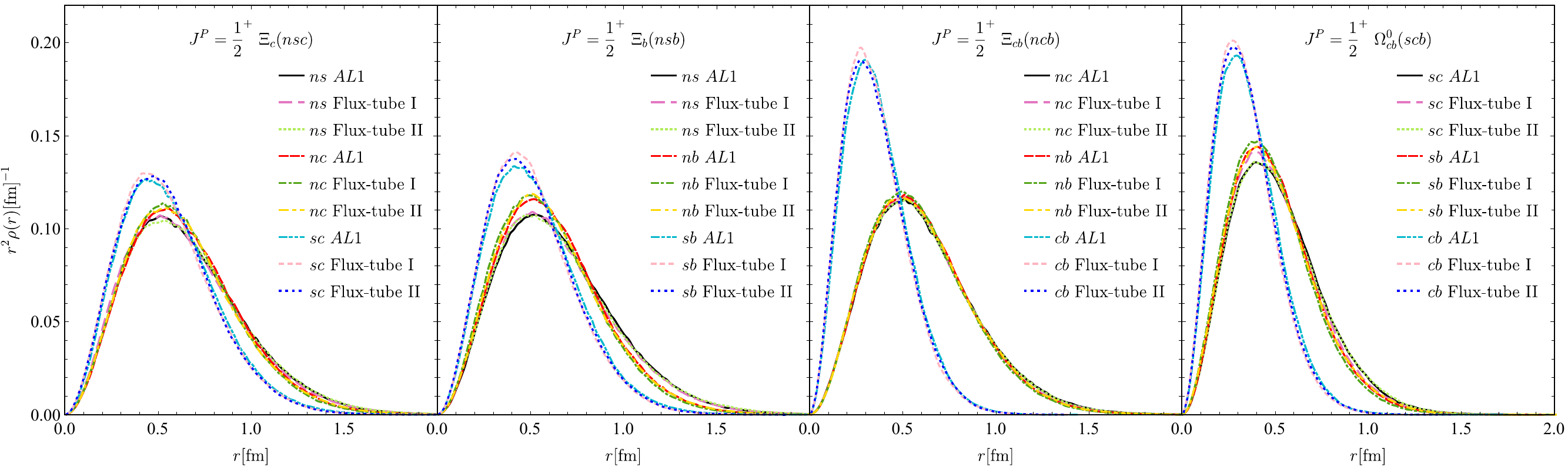} 
  \caption{\label{1q_1over2_r2rho} The $r^2\rho(r)$ distributions for the $J^P=\frac{1}{2}^+$ baryons without identical quarks.}
    \setlength{\belowdisplayskip}{1pt}
\end{figure*}

\subsection{$J^P={1\over 2}^+$ without identical quarks}~\label{sec:1q_1over2}

We use $1,2,3$ to label three quarks in the mass ascending order. For the ground spin-$1\over 2$ baryons (naively no orbital excitation), there are two spin channels,
\begin{equation}
   \chi_{s}^A(12;3)=\left[(12)_0 3\right]_{\frac{1}{2}}, \quad \chi_{s}^S(12;3)=\left[(12)_1 3\right]_{\frac{1}{2}}.  
\end{equation}
 The superscripts $A$ and $S$ indicate the symmetric and anti-symmetric wave functions, respectively. For the $J=\frac{1}{2}$ baryon system without identical quarks, the ground states should be the mixture of these two spin channels in principle. 

The masses calculated with the coupling-channel DMC are listed in TABLE~\ref{1q_1over2_mass}. In addition to the coupling-channels results, we also give the single channel masses for comparison. The single channel $\chi_{s}^A(12;3)$ masses are almost the same as the mixing ones, which indicates that the mixing state is almost entirely of the $\chi_{s}^A(12;3)$ component. In the coupling-channel DMC, we can only obtain the ground state. But we checked with the variational method and found that the first excited state in the coupled channel scheme is roughly the $\chi_{s}^S(12;3)$ state in either the energy sense or the wave function sense. Thus the following radii and distributions will all be presented using the single channel results. It is worthwhile to stress that one can choose $ \chi_{s}^{S,A}(23;1)$ or $ \chi_{s}^{S,A}(13;2)$ as spin channels alternatively. In these two schemes, the coupled-channel results will not change. However, the single-channel results will not be good approximations any more. In other words, for the ground spin-$1\over 2$ baryons with three different quarks, the first two low-lying states can be distinguished by the combined spin of the two lightest quarks approximately. 
Especially, the $\chi_{s}^A(12;3)$ state is the lighter one, which is the implication of the ``good" diquark introduced by Jaffe~\cite{Jaffe:2004ph}.

The radii expectation values and radial distributions are shown in TABLE~\ref{1q_1over2_expect} and Fig.\ref{1q_1over2_r2rho} respectively. As for the inner angle and 2D probability distributions, since the $\frac{1}{2}^+$ distributions are almost the same as the $\frac{3}{2}^+$ ones, they are not shown in the main text, and can be found in Supplement material.

\begin{table}[htbp] 
\centering
\caption{\label{1q_1over2_mass} Masses of the $J^P=\frac{1}{2}^+$ baryons without identical quarks in MeV. The notations are the same as those in Fig.~\ref{1q_1over2_mass}.}
\begin{tabular*}{\hsize}{@{}@{\extracolsep{\fill}}cccccccc@{}}
   
\hline  \hline 
\multirow{2}*{$J^{P}=\frac{1}{2}^{+}$}  &\multirow{2}*{channel} &\multicolumn{3}{c}{$AL$1}                                                          &FT I      &FT II         & \multirow{2}*{EXP\cite{ParticleDataGroup:2022pth}}  \\
\cline{3-5}
  ~                                               &      &  DMC          &  VAR    &  FAD~\cite{Silvestre-Brac:1996myf}     &DMC                &DMC                   &  ~                    \\
\hline 
\multirow{3}{*}{$\Xi_{c}(nsc)$} & $\chi_{s}^A(12;3)$ &2466 & 2465 & / & 2537 & 2470& 2469\tabularnewline
 & $\chi_{s}^S(12;3)$ & 2470& 2569 & / & 2643 &2573 & 2579\tabularnewline
 & mixing & 2465 & 2464 & 2467 & 2535 & 2469 & 2469\tabularnewline
 \hline 
\multirow{3}{*}{$\Xi_{b}(nsb)$} & $\chi_{s}^A(12;3)$ & 5803 & 5802& / & 5870& 5806 & 5794\tabularnewline
 & $\chi_{s}^S(12;3)$ & 5942 & 5941 & / & 6014& 5944 & 5935\tabularnewline
 & mixing & 5802 & 5802 & 5806 & 5870 & 5806 & 5794\tabularnewline
 \hline 
\multirow{3}{*}{$\Xi_{cb}(ncb)$} & $\chi_{s}^A(12;3)$ & 6915&6914 & / & 6968 & 6911 & /\tabularnewline
 & $\chi_{s}^S(12;3)$ & 6960 & 6959 & / & 7014 & 6955& /\tabularnewline
 & mixing & 6914 & 6914 & 6915 & 6967 & 6911 & /\tabularnewline
 \hline 
\multirow{3}{*}{$\Omega_{cb}^{0}(scb)$} & $\chi_{s}^A(12;3)$ & 7003 & 7002 & / & 7052 & 7004 & /\tabularnewline
 & $\chi_{s}^S(12;3)$ & 7041 & 7040 & / & 7091 & 7040 & /\tabularnewline
 & mixing & 7002 & 7002 & 7003 & 7052 & 7003 & /\tabularnewline

\hline  \hline  
\end{tabular*}
\end{table}

\begin{table*}[htbp] 
\centering
\caption{\label{1q_1over2_expect} Root mean square radii and charge mean-square radii $R_{c}^{2}$ expectation values of the $J^P=\frac{1}{2}^+$ baryons without identical quarks. ``FT I" is short for ``Flux-tube I", and ``FT II" is short for ``Flux-tube II".}
\begin{tabular*}{\hsize}{@{}@{\extracolsep{\fill}}ccccccccccccc@{}}
   
\hline  \hline 
\multirow{2}*{$J^{P}=\frac{1}{2}^{+}$}  &\multicolumn{3}{c}{$\sqrt{\langle r_{12}^{2}\rangle} [\mathrm{fm}]$}    &\multicolumn{3}{c}{$\sqrt{\langle r_{13}^{2}\rangle} [\mathrm{fm}]$}      &\multicolumn{3}{c}{$\sqrt{\langle r_{23}^{2}\rangle} [\mathrm{fm}]$}  &\multicolumn{3}{c}{$\langle R_{c}^{2}\rangle[e\cdot\mathrm{ fm^{2}}]$}  \\
\cline{2-4}\cline{5-7}\cline{8-10}\cline{11-13}
  ~                                             &$AL$1                        & FT I                  & FT II                 &$AL$1            & FT I      & FT II      &$AL$1                        & FT I                  & FT II                &$AL$1                        & FT I                  & FT II \\
\hline 
$\Xi_{c}^{0}(dsc)$ & \multirow{2}{*}{0.728} & \multirow{2}{*}{0.717} & \multirow{2}{*}{0.730} & \multirow{2}{*}{0.705} & \multirow{2}{*}{0.691} & \multirow{2}{*}{0.706} & \multirow{2}{*}{0.614} & \multirow{2}{*}{0.601} & \multirow{2}{*}{0.608} & $-0.428$ & $-0.421$ & $-0.427$\tabularnewline
$\Xi_{c}^{+}(usc)$ &  &  &  &  &  &  &  &  &  & 0.495 & 0.489 & 0.496\tabularnewline
\cline{2-10}
$\Xi_{b}^{-}(dsb)$ & \multirow{2}{*}{0.719} & \multirow{2}{*}{0.709} & \multirow{2}{*}{0.725} & \multirow{2}{*}{0.672} & \multirow{2}{*}{0.656} & \multirow{2}{*}{0.672} & \multirow{2}{*}{0.572} & \multirow{2}{*}{0.554} & \multirow{2}{*}{0.567} & $-0.488$ & $-0.478$ & $-0.489$ \tabularnewline
$\Xi_{b}^{0}(usb)$ &  &  &  &  &  &  &  &  &  & 0.478 & 0.471 & 0.480\tabularnewline
\cline{2-10}
$\Xi_{cb}^{0}(dcb)$ & \multirow{2}{*}{0.678} & \multirow{2}{*}{0.670} & \multirow{2}{*}{0.684} & \multirow{2}{*}{0.666} & \multirow{2}{*}{0.656} & \multirow{2}{*}{0.669} & \multirow{2}{*}{0.403} & \multirow{2}{*}{0.390} & \multirow{2}{*}{0.397} & $-0.295$ & $-0.296$ & $-0.300$ \tabularnewline
$\Xi_{cb}^{+}(ucb)$ &  &  &  &  &  &  &  &  &  & 0.667 & 0.656 & 0.668\tabularnewline
\cline{2-10}
$\Omega_{cb}^{0}(scb)$ & 0.563 & 0.552 & 0.565 & 0.541 & 0.531 & 0.543 & 0.393 & 0.380 & 0.386 & $-0.057$ & $-0.058$ & $-0.060$\tabularnewline
\hline  \hline  
\end{tabular*}
\end{table*}

\subsection{ $J^P=\frac{1}{2}^+$ with two identical quarks}~\label{sec:2q_1over2}

We label the two identical quarks as $1$ and $2$, and the remaining one as $3$. In general, we can construct the following symmetric spatial-spin-flavor wave functions by exchanging 1 and 2 quarks,
\begin{eqnarray}
|B1\rangle&=&\text{\ensuremath{\chi_{s}^{S}(12;3)}}\chi_{f}^{S}(12;3)\psi_{1}^{S}(12;3),\nonumber \\
|B2\rangle&=&\text{\ensuremath{\chi_{s}^{S}(12;3)}}\chi_{f}^{A}(12;3)\psi_{2}^{A}(12;3),\nonumber \\
|B3\rangle&=&\text{\ensuremath{\chi_{s}^{A}(12;3)}}\chi_{f}^{A}(12;3)\psi_{3}^{S}(12;3),\nonumber \\
|B4\rangle&=&\text{\ensuremath{\chi_{s}^{A}(12;3)}}\chi_{f}^{S}(12;3)\psi_{4}^{A}(12;3),\label{eq:Bi}
\end{eqnarray}
where we use semicolon to separate the exchanging (anti-)symmetric part with the remaining part. The superscripts $A$ and $S$ indicate the symmetric and anti-symmetric wave functions, respectively. 

At first, we only include the $|B1\rangle$ and $|B3\rangle$ channels with symmetric spatial wave functions. In TABLE~\ref{2q_1over2_mass}, the masses calculated with different models and methods are shown. The results from the $AL$1 and Flux-tube II models are basically consistent with the experimental values.  Our energies for the $\Sigma$, $\Lambda$ and some other particles are a bit higher than the results from the Faddeev equation. But we have checked that our results will become equal to or even smaller than the results from the Faddeev equation once we include the $|B2\rangle$ and $|B4\rangle$ into the coupled-channel calculations. The root mean square radii and the charge mean-square radii are shown in TABLE~\ref{2q_1over2_expect}. The calculated $\Sigma^-$ charge radius $-0.746$ $\mathrm{fm}^2$ in Flux-tube II model is consistent with the experimental value $-0.61\pm0.12\pm0.09$ $\mathrm{fm}^2$ \cite{SELEX:2001fbx} within errors.

\begin{figure*}[htbp]
  \centering
  \includegraphics[width=0.9\textwidth]{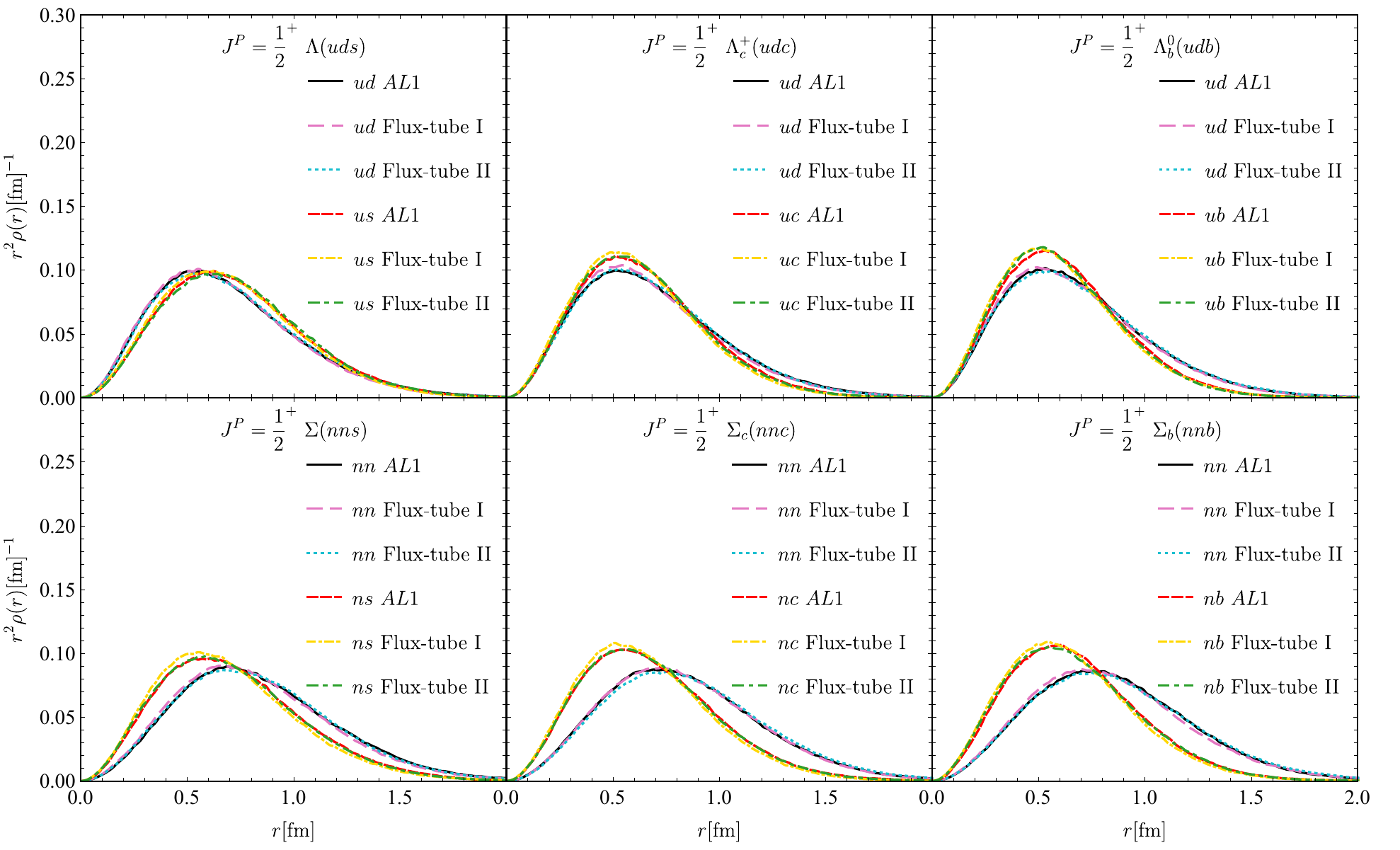} 

  \caption{\label{2q_1over2_nnq_r2rho} The $r^2\rho(r)$ distributions for the $J^P=\frac{1}{2}^+$ baryons with two identical $n$ quarks .}
    \setlength{\belowdisplayskip}{1pt}
\end{figure*}

\begin{figure*}[htbp]
  \centering
  \includegraphics[width=0.9\textwidth]{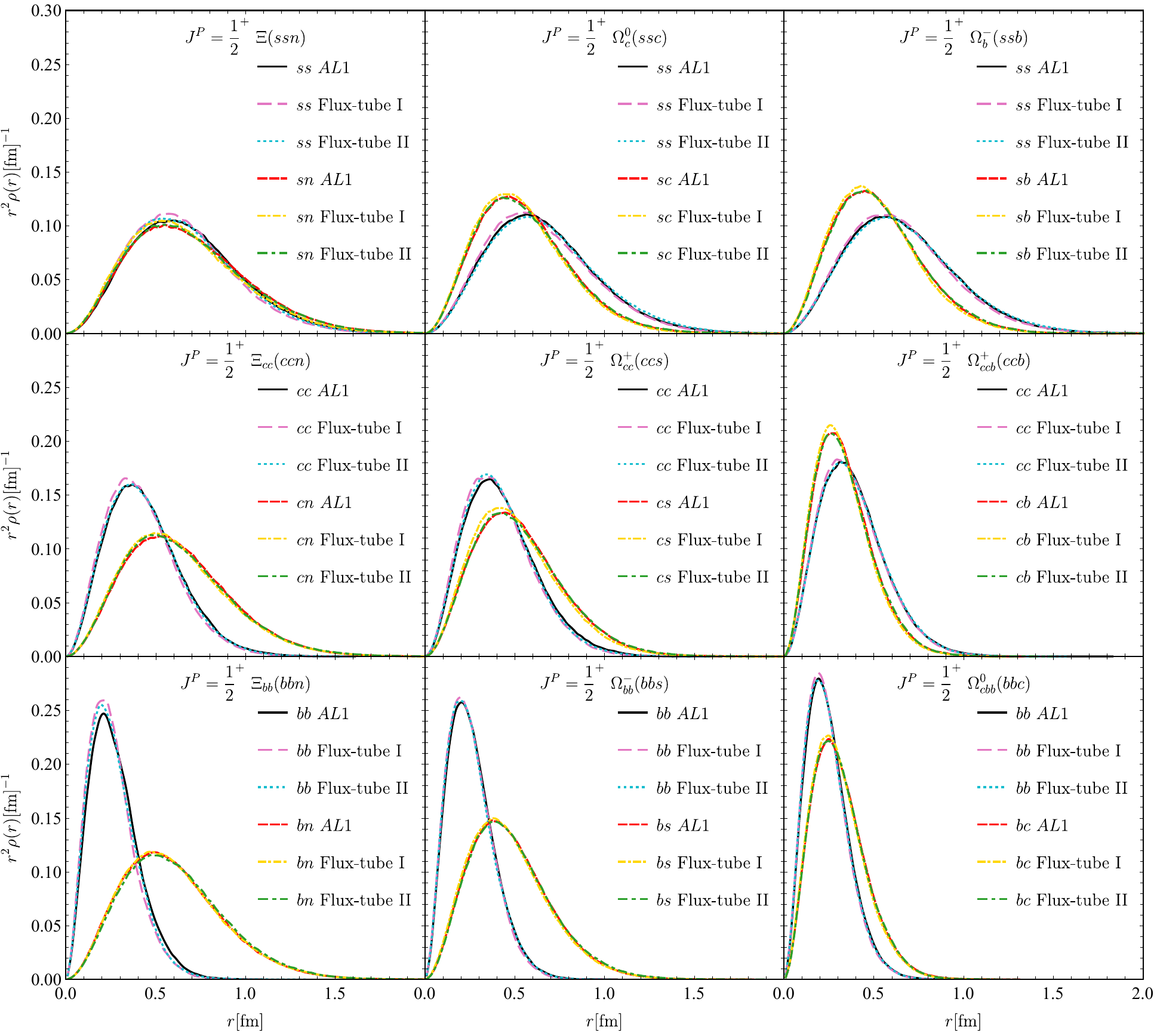} 

  \caption{\label{2q_1over2_S12_without_nnq_r2rho} The $r^2\rho(r)$ distributions for the $J^P=\frac{1}{2}^+$ baryons with two identical $s,c,b$ quarks.}
    \setlength{\belowdisplayskip}{1pt}
\end{figure*}

The radial distributions of the $\frac{1}{2}^+$ baryons with two identical $u,d$ quarks are displayed in Fig.~\ref{2q_1over2_nnq_r2rho}. It can be seen that the distance between the $nn$ pair is closer when they are in the spin $S = 0$ state than in the $S=1$ one, which means the ``good" diquark is a more compact object than the ``bad" diquark. In the $\Lambda(nns)$ baryon, the $r_{nn}$ is even smaller than $r_{ns}$. Similarly, the radial distributions of the $\frac{1}{2}^+$ baryons with two identical $s,c,b$ quarks are shown in Fig.~\ref{2q_1over2_S12_without_nnq_r2rho}.

The distribution of the inner angles and the 2D wave function probability distribution of the $\Sigma_c(nnc)$ and $\Lambda_{c}^{+}(nnc)$ are given in the middle and right columns of Fig.~\ref{2q_2D_nnc} respectively. There is hardly any obvious difference among the three $nnc$ states in this figure. Another $\frac{1}{2}^+$ baryon $\Omega_{ccb}^{+}(ccb)$ distribution is given in Fig.~\ref{2q_2D_ccb}. Again, there is no obvious difference between the $\frac{1}{2}^+$ and $\frac{3}{2}^+$ states. But actually, the $\frac{3}{2}^+$ state is a little bit more extended than the $\frac{1}{2}^+$ one, which can be seen by comparing their root mean square radii values in TABLE~\ref{2q_3over2_expect} and TABLE~\ref{2q_1over2_expect}. We also give the comparison between $\Xi_{cc}(ccn)$ and $\Xi_{bb}(bbn)$ in Fig.~\ref{2q_2D_ccn_bbn}, which shows that the angle with the  heavier quark as the vertex tends to be larger.

\begin{table}[htbp] 
\centering
\caption{\label{2q_1over2_mass} Masses of the $\frac{1}{2}^+$ baryons with two identical quarks in MeV. The $I$ column is the isospin. The notations are the same as those in Fig.~\ref{1q_1over2_mass}. }
\begin{tabular*}{\hsize}{@{}@{\extracolsep{\fill}}cccccccc@{}}
  
\hline  \hline 
\multirow{2}*{$J^{P}=\frac{1}{2}^{+}$}  &\multirow{2}*{$I$} &\multicolumn{3}{c}{$AL$1}                                                          &FT I      &FT II         & \multirow{2}*{EXP\cite{ParticleDataGroup:2022pth}}  \\
\cline{3-5}
  ~                                               &      &  DMC          &  VAR    &  FAD~\cite{Silvestre-Brac:1996myf}     &DMC                &DMC                   &  ~                    \\
\hline 
$\Lambda(nns)$ & \multirow{3}{*}{0} & 1125 & 1123 & 1119 & 1209 & 1131 & 1116\tabularnewline
$\Lambda_{c}^{+}(nnc)$ &  & 2281 & 2280 & 2285 & 2359 & 2288 & 2286\tabularnewline
$\Lambda_{b}^{0}(nnb)$ &  & 5633 & 5632 & 5638 & 5707 & 5639 & 5620\tabularnewline
\hline 
$\Sigma(nns)$ & \multirow{3}{*}{1} & 1206 & 1204 & 1196 & 1292 & 1211 & 1193\tabularnewline
$\Sigma_{c}(nnc)$ &  & 2457 & 2456 & 2455 & 2540 & 2460 & 2454\tabularnewline
$\Sigma_{b}(nnb)$ &  & 5845 & 5844& 5845 & 5927 & 5848 & 5813\tabularnewline
\hline 
$\Xi(ssn)$ & \multirow{3}{*}{$\frac{1}{2}$} & 1331 & 1329 & 1324& 1410 & 1337 & 1318\tabularnewline
$\Xi_{cc}(ccn)$ &  & 3607 &3607 & 3607 & 3668 & 3608& 3621\tabularnewline
$\Xi_{bb}(bbn)$ &  & 10194& 10193 & 10194 & 10236 &10185 & /\tabularnewline
\hline 
$\Omega_{c}^{0}(ssc)$ & \multirow{6}{*}{0} & 2676 & 2675 & 2675 & 2744 & 2680 & 2695\tabularnewline
$\Omega_{cc}^{+}(ccs)$ &  & 3709 & 3708 & 3710 & 3764 & 3712 & /\tabularnewline
$\Omega_{b}^{-}(ssb)$ &  & 6033 & 6033 & 6034 & 6097 & 6036 & 6046\tabularnewline
$\Omega_{ccb}^{+}(ccb)$ &  & 8018 & 8017 & 8019 & 8058 & 8019 & /\tabularnewline
$\Omega_{bb}^{-}(bbs)$ &  & 10267 & 10266 & 10267 & 10306 & 10262 & /\tabularnewline
$\Omega_{cbb}^{0}(bbc)$ &  & 11215 & 11215 & 11217 & 11247 & 11216 &/\tabularnewline

\hline  \hline  
\end{tabular*}
\end{table}

\begin{table*}[htbp] 
\centering
\caption{\label{2q_1over2_expect} Root mean square radii and charge mean-square radii $R_{c}^{2}$ expectation values of the $\frac{1}{2}^+$ baryons with two identical quarks. The $\sqrt{\langle r_{13}^{2}\rangle}$ column is the average of the $\sqrt{\langle r_{13}^{2}\rangle}$ and $\sqrt{\langle r_{23}^{2}\rangle}$, since $1,2$ are identical quarks. }
\begin{tabular*}{\hsize}{@{}@{\extracolsep{\fill}}ccccccccccc@{}}
 
\hline  \hline 
\multirow{2}*{$J^{P}=\frac{1}{2}^{+}$}  &\multirow{2}*{$I$} &\multicolumn{3}{c}{$\sqrt{\langle r_{12}^{2}\rangle} [\mathrm{fm}]$}    &\multicolumn{3}{c}{$\sqrt{\langle r_{13}^{2}\rangle} [\mathrm{fm}]$}    &\multicolumn{3}{c}{$\langle R_{c}^{2}\rangle[e\cdot\mathrm{ fm^{2}}]$}  \\
\cline{3-5}\cline{6-8}\cline{9-11}
  ~                                                &     &$AL$1                        &FT I                  &FT II                 &$AL$1            &FT I      &FT II      &$AL$1                        &FT I                  &FT II \\
\midrule[0.5pt]
$\Lambda(nns)$ & \multirow{3}{*}{0} & 0.785 & 0.764 & 0.783 & 0.802  & 0.784 & 0.802 & 0.118 & 0.115 & 0.120\tabularnewline
$\Lambda_{c}^{+}(nnc)$ &  & 0.764 & 0.752 & 0.769 & 0.709 & 0.693 & 0.711 & 0.255 & 0.250 & 0.257\tabularnewline
$\Lambda_{b}^{0}(nnb)$ &  & 0.761 & 0.750  & 0.767 & 0.679 & 0.663 & 0.678 & 0.247 & 0.242 & 0.245\tabularnewline
\midrule[0.5pt]
$\Sigma^{-}(dds)$ & \multirow{9}{*}{1} & \multirow{3}{*}{0.907} & \multirow{3}{*}{0.890} & \multirow{3}{*}{0.913} & \multirow{3}{*}{0.788} & \multirow{3}{*}{0.769} & \multirow{3}{*}{0.787} & $-0.744$ & $-0.733$ & $-0.745$ \tabularnewline
$\Sigma^{0}(uds)$ &  &  &  &  &  &  &  & 0.136 & 0.134 & 0.137\tabularnewline
$\Sigma^{+}(uus)$ &  &  &  &  &  &  &  & 1.018 & 1.004 & 1.022\tabularnewline
\cline{3-8}
$\Sigma_{c}^{0}(ddc)$ &  & \multirow{3}{*}{0.922} & \multirow{3}{*}{0.907} & \multirow{3}{*}{0.933} & \multirow{3}{*}{0.751} & \multirow{3}{*}{0.729} & \multirow{3}{*}{0.750} & $-0.712$ & $-0.697$ & $-0.712$\tabularnewline
$\Sigma_{c}^{+}(udc)$ &  &  &  &  &  &  &  & 0.277 & 0.270 & 0.274\tabularnewline
$\Sigma_{c}^{++}(uuc)$ &  &  &  &  &  &  &  & 1.263 & 1.236 & 1.267\tabularnewline
\cline{3-8}
$\Sigma_{b}^{-}(ddb)$ &  & \multirow{3}{*}{0.929} & \multirow{3}{*}{0.916} & \multirow{3}{*}{0.940} & \multirow{3}{*}{0.739} & \multirow{3}{*}{0.719} & \multirow{3}{*}{0.735} & $-0.784$ & $-0.768$ & $-0.783$\tabularnewline
$\Sigma_{b}^{0}(udb)$ &  &  &  &  &  &  &  & 0.273 & 0.265 & 0.278\tabularnewline
$\Sigma_{b}^{+}(uub)$ &  &  &  &  &  &  &  & 1.335 & 1.299 & 1.329 \tabularnewline
\midrule[0.5pt]
$\Xi^{-}(ssd)$ & \multirow{6}{*}{$\frac{1}{2}$} & \multirow{2}{*}{0.734} & \multirow{2}{*}{0.710} & \multirow{2}{*}{0.729} & \multirow{2}{*}{0.763} & \multirow{2}{*}{0.741} & \multirow{2}{*}{0.762} & $-0.511$ & $-0.500$ & $-0.510$\tabularnewline
$\Xi^{0}(ssu)$ &  &  &  &  &  &  &  & 0.347 & 0.341 & 0.346\tabularnewline
\cline{3-8}
$\Xi_{cc}^{+}(ccd)$ &  & \multirow{2}{*}{0.482} & \multirow{2}{*}{0.465} & \multirow{2}{*}{0.476} & \multirow{2}{*}{0.689} & \multirow{2}{*}{0.679} & \multirow{2}{*}{0.694} & $-0.250$ & $-0.253$ & $-0.255$\tabularnewline
$\Xi_{cc}^{++}(ccu)$ &  &  &  &  &  &  &  & 0.685 & 0.672 & 0.687\tabularnewline
\cline{3-8}
$\Xi_{bb}^{-}(bbd)$ &  & \multirow{2}{*}{0.305} & \multirow{2}{*}{0.290} & \multirow{2}{*}{0.294} & \multirow{2}{*}{0.659} & \multirow{2}{*}{0.653} & \multirow{2}{*}{0.670} & $-0.377$ & $-0.373$ & $-0.381$\tabularnewline
$\Xi_{bb}^{0}(bbu)$ &  &  &  &  &  &  &  & 0.589 & 0.589 & 0.601\tabularnewline
\midrule[0.5pt]
$\Omega_{c}^{0}(ssc)$ & \multirow{6}{*}{0} & 0.721 & 0.708 & 0.730 & 0.611 & 0.596 & 0.610 & $-0.198$ & $-0.192$ & $-0.199$\tabularnewline
$\Omega_{b}^{-}(ssb)$ &  & 0.721 & 0.711 & 0.731 & 0.583 & 0.568 & 0.584 & $-0.266$ & $-0.257$ & $-0.267$\tabularnewline
$\Omega_{cc}^{+}(ccs)$ &  & 0.469 & 0.455 & 0.463 & 0.576 & 0.564 & 0.579 & $-0.012$ & $-0.014$ & $-0.015$\tabularnewline
$\Omega_{ccb}^{+}(ccb)$ &  & 0.427 & 0.421 & 0.429 & 0.370 & 0.362 & 0.370 & 0.117 & 0.113 & 0.117\tabularnewline
$\Omega_{bb}^{-}(bbs)$ &  & 0.291 & 0.283 & 0.286 & 0.525 & 0.519 & 0.530 & $-0.133$ & $-0.131$ & $-0.134$\tabularnewline
$\Omega_{cbb}^{0}(bbc)$ &  & 0.271 & 0.266 & 0.271 & 0.344 & 0.338 & 0.345 & 0.043 & 0.042 & 0.044\tabularnewline

\hline  \hline  
\end{tabular*}
\end{table*}

\subsection{ $J^P=\frac{1}{2}^+$ with three identical quarks}~\label{sec:3q_1over2}

\begin{table*}[htbp] 
\centering
\caption{\label{qqq_1over2_fac} Masses, root mean square radii and charge mean-square radii $R_{c}^{2}$ expectation values of the $\frac{1}{2}^+$ baryons with three identical quarks calculated using the factorization formalism with the DMC method. The root mean square radii $\sqrt{\langle r^{2}\rangle}$ are averaged for the identical quarks. }
\begin{tabular*}{\hsize}{@{}@{\extracolsep{\fill}}cccccccccc@{}}
 
\hline  \hline 
\multirow{2}*{$J^{P}=\frac{1}{2}^{+}$}  &\multicolumn{3}{c}{Mass [MeV]}    &\multicolumn{3}{c}{$\sqrt{\langle r^{2}\rangle} [\mathrm{fm}]$}     &\multicolumn{3}{c}{$\langle R_{c}^{2}\rangle[e\cdot\mathrm{ fm^{2}}]$}  \\
\cline{2-4}\cline{5-7}\cline{8-10}
  ~                &$AL$1                  & FT I                & FT II         &$AL$1               & FT I         & FT II                 &$AL$1     & FT I    & FT II \\
\hline 
$n(udd)$       & \multirow{2}*{968}&\multirow{2}*{1059}&\multirow{2}*{975}&\multirow{2}*{0.855}&\multirow{2}*{0.837}&\multirow{2}*{0.856}  &$-0.116$           &$-0.116$           & $-0.116$  \\
$p(uud)$       & ~                         & ~                            &  ~                         & ~                            & ~                             &  ~                 &0.707          &0.698          & 0.707  \\
\hline  \hline  
\end{tabular*}
\end{table*}

\begin{figure}[htbp]
  \centering
  \includegraphics[width=0.4\textwidth]{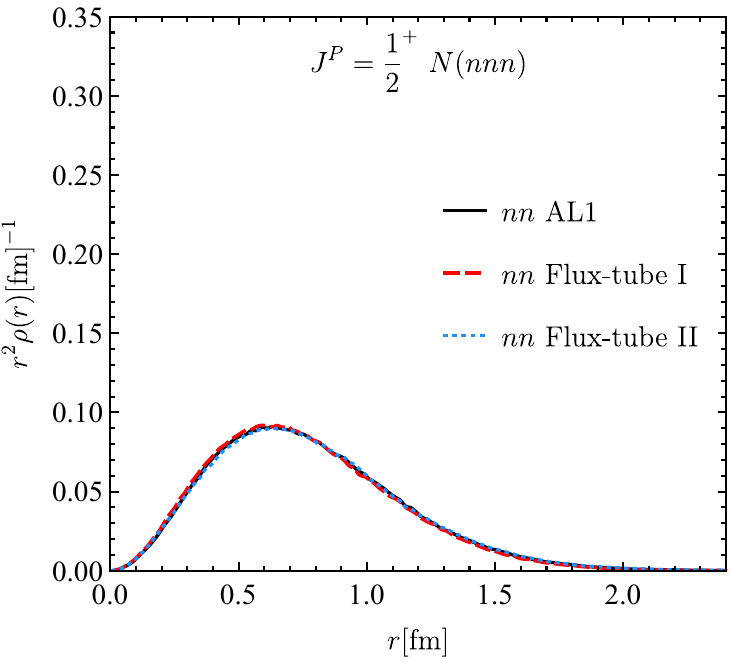} 
 
  \caption{\label{qqq_1over2_r2rho} The $r^2\rho(r)$ distributions for the $J^P=\frac{1}{2}^+$ baryons with three identical quarks.}
    \setlength{\belowdisplayskip}{1pt}
\end{figure}

The spin-$1\over 2$ baryons composed of three identical quarks are the nucleons, which are the lightest baryons.  In the following discussion, we omit the trivial color wave function. Naively, one can construct the spatial-spin-flavor wave function of the nucleon with a factorization formalism,
\begin{equation}
    |N\rangle_{\text{frac}}=\chi_{sf}^{S}(123)\psi^{S}(123),\label{eq:nucleon_fac}
\end{equation}
where the spatial wave function $\psi^{S}(123)$ and spin-flavor function are symmetrized separately. The masses, root mean square radii and charge mean-square radii calculated using the factorization formalism with the DMC method are given in TABLE~\ref{qqq_1over2_fac}. And the distribution is displayed in Fig.~\ref{qqq_1over2_r2rho}. However, the nucleon masses (e.g. 968 MeV in the $AL$1 model) obtained with the factorized wave function are larger than the results (e.g. 933 MeV in $AL$1 model) from the Faddeev equation using the same interaction by over 30 MeV. To obtain the lower mass results, we need to go beyond the above factorization wave function.

There is no reason to prevent the existence of the non-factorization spatial-flavor-spin wave functions. For the nucleon, one can obtain the totally symmetric spatial-spin-flavor wave function by permuting the quarks in $|Bi\rangle$ in Eqs.~\eqref{eq:Bi},
\begin{eqnarray}
    |Ci\rangle=|Bi\rangle+\text{even perm. (1,2,3)}
\end{eqnarray}
where ``\text{even perm. (1,2,3)}"  means summation over all the even permutation of (1,2,3) quarks. In general, the ground state of the nucleon could be obtained using all four $|Ci\rangle $ channels. In practice, some of them are less relevant. In TABLE~\ref{tab:proton_diff_method}, we present the results from the variational method, specifically GEM, with the $\{|C1\rangle\}$, $\{|C3\rangle\}$, $\{|C1\rangle,|C3\rangle\}$, and $\{|C1\rangle,|C2\rangle,|C3\rangle,|C4\rangle\}$ assignments respectively. We can see that the  $\{|C3\rangle\}$ is the most relevant assignment. Adding other channels only reduces the energy by 1 MeV. One can identify the $|C3\rangle$ is the symmetrized ``good" diquark configuration~\cite{Jaffe:2004ph}. The nucleon is more like a symmetrized $\Lambda$ baryon.  Apparently, the naive factorization assignment of the wave function is the special case of $|C3\rangle$ with $\psi_3^S(12;3)=\psi^S(123)$. This extra constraint prevents the naive factorization wave function from the lowest energy solution.  

In our present DMC algorithm, we have to introduce either the single channel or a series of orthogonal spin-flavor-color channels to perform simulations. For the naive factorization assignment of the wave function (single channel), we get consistent results with the variational method, which is much larger than the lowest solution. Although we have not presumed the clustering behavior in the coordinate space in the DMC method, the pre-assignment of the channels could still prevent us from the lowest mass solutions. This is the lesson about the DMC method from the baryon calculation. In Section~\ref{sec:disc}, we will see the pre-assignment of the channels could prevent us from the lowest solution in the tetraquark systems. In the assignments of $|Ci\rangle$, the $|Bi\rangle$ and its permutations are non-orthogonal channels. In our present coupled-channel formalism of DMC in Sec.~\ref{sec:couple-channel}, we are unable to deal with them directly.

Alternatively, we could take an indirect strategy. For example, we could start from the $\Lambda(nns)$ and $\Sigma(nns)$ baryons. If we decrease the strange quark mass to $m_{u,d}$ in the SU(3)-flavor limit, the masses of the $\Lambda(nns)$ and $\Sigma(nns)$ will approach the nucleon mass. In this process, the baryon mass will change continuously. Since we treat $m_s-m_{u,d}$ as a perturbation, the first order correction to the eigenenergy is proportional to the perturbation. In practice, we could use the wave functions $|B_i\rangle$ in Eqs.~\eqref{eq:Bi} to perform the calculation. More generally, we could expect the mass of the spin-$1\over 2$ $\Xi_c(usc)$ to reduce to the nucleon mass in the SU(4)-flavor limit by taking $m_c=m_s=m_{u,d}$. Thus we could use the wave function without any exchanging symmetry,
\begin{eqnarray}
|A1\rangle&=&\text{\ensuremath{\chi_{s}^{S}(12;3)}}\chi_{f}^{S}(12;3)\psi_{1}(1;2;3),\nonumber\\
|A2\rangle&=&\text{\ensuremath{\chi_{s}^{S}(12;3)}}\chi_{f}^{A}(12;3)\psi_{2}(1;2;3),\nonumber\\
|A3\rangle&=&\text{\ensuremath{\chi_{s}^{A}(12;3)}}\chi_{f}^{A}(12;3)\psi_{3}(1;2;3),\nonumber\\
|A4\rangle&=&\text{\ensuremath{\chi_{s}^{A}(12;3)}}\chi_{f}^{S}(12;3)\psi_{4}(1;2;3).~\label{eq:Ai}
\end{eqnarray}
In TABLE~\ref{tab:proton_diff_method}, we use both the DMC algorithm and variational method to calculate energies using wave functions in Eqs.~\eqref{eq:Bi} and \eqref{eq:Ai}. One can see the results for the DMC and variational method agree well with each other and obtain the same lowest solution as the $|Ci\rangle$ coupled-channel result. 

The indirect strategy could not obtain the correct wave function directly. For example, solving the $\{|A1\rangle,|A2\rangle,|A3\rangle,|A4\rangle\}$ coupled-channel problem can not ensure 
the wave function $|A^{math}_{ground}\rangle$ has the correct exchanging symmetry. We call  $|A_{ground}^{math}\rangle$ together with its corresponding eigenvalue $E^{math}_0$ as the mathematical ground state. Since the Hamiltonian has the exchange symmetry, i.e. $\left[\hat{H},\hat{P}_{ij}\right]=0$, the permutation of the obtained wave function is still the solution of this Hamiltonian at the same energy level, i.e. 
\begin{equation}
    \hat{H}\hat{P}_{ij}|A^{math}_{ground}\rangle=\hat{P}_{ij}\hat{H}|A^{math}_{ground}\rangle=E^{math}_{0}\hat{P}_{ij}|A^{math}_{ground}\rangle
\end{equation}
In principle, the mathematical ground states could be degenerate. We could try to combine $|A^{math}_{ground}\rangle$ and its degenerate states  $\hat{P}_{ij}|A_{ground}\rangle$ to construct the physical ground state with correct symmetry. There are two fates of the mathematical ground states. If the $E^{math}_{0}$ is exactly the physical ground energy, one could obtain a non-vanishing wave function after symmetrizing  the mathematical ground states. If $E^{math}_{0}$ is not the physical ground 
energy, one should obtain a null wave function after symmetrization. Fortunately, for the nucleon system, we obtain the physical wave function by symmetrizing $|A^{math}_{ground}\rangle$. 

The final wave functions do have small difference from the results of the naive factorization scheme in Eq.~\eqref{eq:nucleon_fac}. We do not show the distributions and expectation values because there is no qualitative distinction. However, the mass differences from different wave function assignments should be attached enough importance. For the tetraquark system, the wave function differences could be more significant because of more clustering possibilities.

It is worth mentioning that if the replication number $n_{r}$ in Eq.(\ref{n}) becomes negative for a walker, it will be discarded to ensure the stability of the energy. This will result in a reduction of the wave function space, but not much, because such walkers are rare.

\begin{table}[htp]
    \centering
        \caption{ Masses of the $\frac{1}{2}^+$ baryons with three identical quarks in MeV using different wave functions.}
\begin{tabular*}{\hsize}{@{}@{\extracolsep{\fill}}lcccccc@{}}
\hline \hline 
 & \multicolumn{3}{c}{AL1} & \multirow{2}{*}{FT I}& \multirow{2}{*}{FT II}& \multirow{2}{*}{Exp}\tabularnewline
\cline{2-4} \cline{3-4} \cline{4-4} 
 & DMC & VAR & Faddeev & \tabularnewline
\hline 
$|N(123)\rangle_{\text{fac}}$ & 968 & 966 & \multirow{7}{*}{933} &1059 &975 & \multirow{7}{*}{939}\tabularnewline
$|C1\rangle$ & / & 944 &  & / & / & \tabularnewline
$|C3\rangle$ & / & 931 &  & / & / & \tabularnewline
$|C1\rangle$, $|C3\rangle$ & / & 930 &  & / & / & \tabularnewline
$|C1\rangle$, $|C2\rangle$, $|C3\rangle$, $|C4\rangle$ & / & 930 &  & / & / & \tabularnewline
\cline{1-3} \cline{2-3} \cline{3-3}  \cline{5-6}\cline{5-6}\cline{5-6}\cline{6-6}
$|B1\rangle$, $|B2\rangle$, $|B3\rangle$, $|B4\rangle$ & / & 930 &  & / & / & \tabularnewline
$|A1\rangle$, $|A2\rangle$, $|A3\rangle$, $|A4\rangle$ & 930 & 930 &  &1019 &936 & \tabularnewline
\hline  \hline 
\end{tabular*}
    \label{tab:proton_diff_method}
\end{table}

\subsection{Comparison of the charge radius $\langle R_c^2\rangle$ with experiment data and lattice QCD simulations}~\label{sec:Rc2_comparison}

\begin{figure}[htb] 
\flushleft
\includegraphics[width=0.5\textwidth]{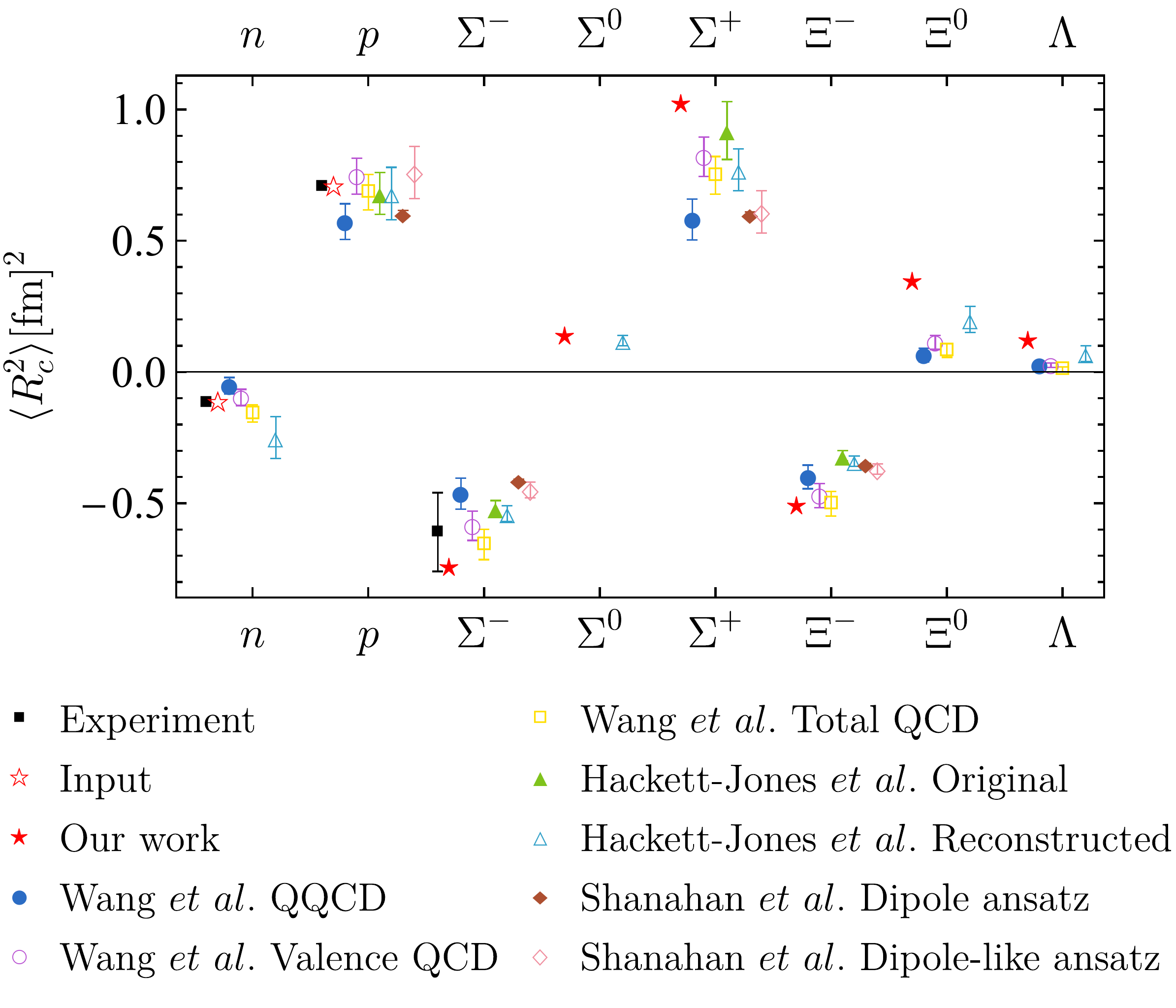} 
\caption{\label{Rc2_compare_baryon_octet} $\langle R_c^2\rangle$ for the $\frac{1}{2}^+$ baryon octet in this work compared with experimental data and lattice QCD results~\cite{ParticleDataGroup:2022pth,SELEX:2001fbx,Wang:2008vb,Hackett-Jones:2000qxh,Shanahan:2014cga}.}
\end{figure}

In Fig.~\ref{Rc2_compare_baryon_octet}, the $\frac{1}{2}^+$ baryon octet charge radii calculated in this work are compared with the experimental data and lattice QCD results in Refs.~\cite{ParticleDataGroup:2022pth,SELEX:2001fbx,Wang:2008vb,Hackett-Jones:2000qxh,Shanahan:2014cga}. The black square symbols are experimental values. The red stars are the Flux-tube II model values calculated with the DMC technique in this work, where the neutron and proton charge radii are inputs taken from the experimental values \cite{ParticleDataGroup:2022pth} and shown with the hollow red stars. The blue solid circle, purple hollow circle and yellow box symbols are from Ref.~\cite{Wang:2008vb}, representing the quenched QCD, valence sector and full-QCD extrapolation results respectively. The green solid triangle and light blue hollow triangle symbols are from Ref.~\cite{Hackett-Jones:2000qxh}, representing the charge radii from the original extrapolation and the charge radii reconstructed from the sum of separate quark sector extrapolations. The brown solid rhombus and pink hollow rhombus are from Ref.~\cite{Shanahan:2014cga}, representing charge radii based on a dipole or dipole-like fit to the extrapolated lattice simulation results respectively. It can be seen from the figure that our results are consistent with experimental data and lattice QCD results.

\begin{figure}[htb] 
\centering
\includegraphics[width=0.47\textwidth]{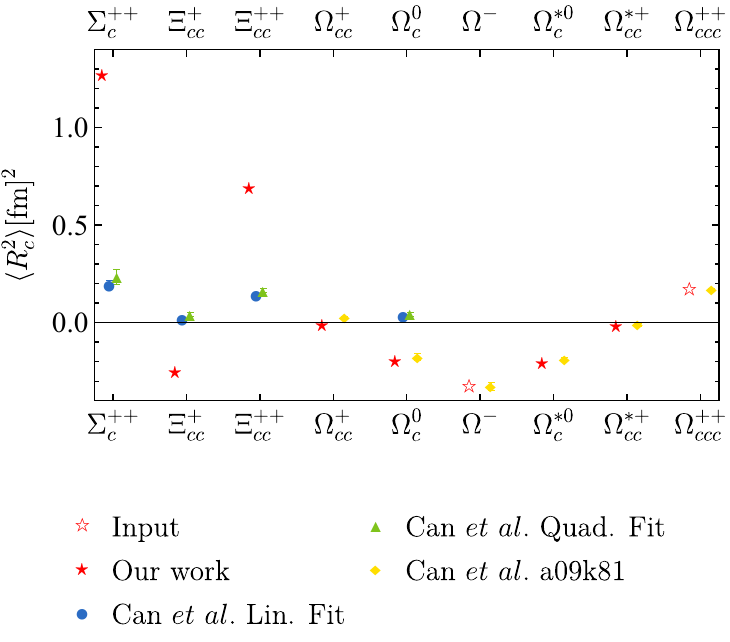} 
\caption{\label{Rc2_compare_baryon_charmed_strange} $\langle R_c^2\rangle$ for the charmed-strange baryons in this work compared with the lattice QCD results in Ref.~\cite{Can:2021ehb}.}
\end{figure}

In Fig.~\ref{Rc2_compare_baryon_charmed_strange}, we compare the charmed-strange baryon charge radii calculated in this work with the lattice QCD results in Ref.~\cite{Can:2021ehb}. Two inputs taken from Ref.~\cite{Can:2021ehb} are shown as the hollow red stars. The red solid stars are the Flux-tube II model values calculated with the DMC technique in this work. The blue circle, green triangle and yellow rhombus symbols are from Ref.~\cite{Can:2021ehb}, representing extrapolations linear and quadratic in the pion mass-squared, and the near-physical-point ensemble a09k81 results. When the mass difference among the constituent quarks is not large, our result is consistent with theirs. But for the large mass difference baryons, the difference in results becomes obvious.

\section{Discussion}~\label{sec:disc}

In the variational method, one has to introduce the trial wave functions (basis functions). If they are improper, one could obtain misleading solutions. Unlike the variational method, no specific basis of wave functions or presumed clustering behavior of the spatial wave function is introduced in the DMC method. However, the wave functions of the DMC are still not the most general one due to the coupled-channel scheme and sign problem. 

In the present DMC method in Sec.~\ref{sec:couple-channel}, one has to assign several orthogonal channels of the discrete quantum numbers before simulation. One lesson from the baryon calculation (especially the nucleon mass) with DMC is that, the pre-assignment of the channels could prevent us from getting the real ground state. In fact, this flaw could appear in the calculation of the multiquark states, such as the tetraquark system in Ref.~\cite{Gordillo:2020sgc} and the hexaquark system in Ref.~\cite{Alcaraz-Pelegrina:2022fsi}.

\begin{figure}[htb] 
\centering
\includegraphics[width=0.49\textwidth]{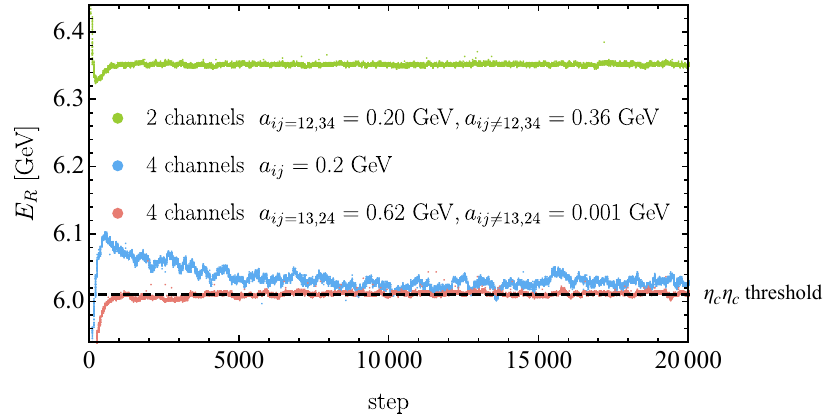} 
\caption{\label{cccc_energy} Energies of the $cc\bar{c}\bar{c}$ system with $J^{PC}=0^{++}$ with different spin-color configurations and $a_{ij}$ values. The green line shows the change of energy $E_R$ with increasing steps using two antisymmetric spin-color configuration channels. The pink line is the result using four complete spin-color configuration channels with optimal $a_{ij=13,24}=0.62$  $\mathrm{GeV}$ and $a_{ij\neq13,24}=0.001$ $\mathrm{GeV}$ in $\psi_T$. The blue line shows the result using four complete spin-color configuration channels with all $a_{ij}=0.2$ $\mathrm{GeV}$.}
\end{figure}

To make this clearer, we take the $cc\bar{c}\bar{c}$ system with $J^{PC}=0^{++}$ as an example.  The mass of this state in Ref.~\cite{Gordillo:2020sgc} using the DMC method is 6351 MeV, which is much higher than the $\eta_c\eta_c$ threshold 6010 MeV. In this calculation, the authors assumed the symmetric spatial wave function and introduced two discrete channels
\begin{eqnarray}
   |T1\rangle &=&  \left[(12)_{\bar{3}_{c}}^{0_{s}}(34)_{3_{c}}^{0_{s}}\right]_{1_{c}}^{0_{s}}\psi_1^{SS}(12;34),\\
     |T2\rangle &=&\left[(12)_{6_{c}}^{1_{s}}(34)_{\bar{6}_{c}}^{1_{s}}\right]_{1_{c}}^{0_{s}}\psi_2^{SS}(12;34)
\end{eqnarray}
where Labels 1,2 are for $c$ and 3,4 for $\bar{c}$. We repeated this calculation in the same channels and got the same result. In Fig.~\ref{cccc_energy}, the green line shows the change of the energy $E_R$ with the increasing steps, which stabilizes at 6351 MeV. This result is close to that obtained using the variational method in Ref.~\cite{Wang:2019rdo}. However, the variational method calculation has been improved in Ref.~\cite{Wang:2022yes}, where the extra di-meson channels are included,
\begin{eqnarray}
    |T_a\rangle &=\left[(13)_{1_{c}}^{0_{s}}(24)_{1_{c}}^{0_{s}}\right]_{1_{c}}^{0_{s}}\psi^{SS}(13;24),\\
    |T_b\rangle&=\left[(14)_{1_{c}}^{0_{s}}(23)_{1_{c}}^{0_{s}}\right]_{1_{c}}^{0_{s}}\psi^{SS}(14;23),
\end{eqnarray}
The updated results in the variational method show that the energy levels above the di-meson threshold in Ref.~\cite{Wang:2019rdo} will become either the continuum spectrum (scattering states) or resonances (obtained from complex-scaling method). Therefore, we can conjecture that the $|T1\rangle$ and $|T2\rangle$ wave functions are not general enough to get the $\eta_c \eta_c$ state threshold (which is actually the ground state) even if we adopt DMC.

Apparently, it is very easy to reproduce the $\eta_c\eta_c$ threshold either in the DMC method or variational method if we only include the $|T_a\rangle$ or $|T_b\rangle$ channel. After the single channel calculation, we can mix the degenerate $|T_a\rangle$ and $|T_b\rangle$ states to satisfy the Pauli principle and only one mixing state survives. In other words, we can get the real ground state ($\eta_c\eta_c$ threshold) in the DMC method by choosing the channels properly.

If we insist on choosing the diquark basis in the DMC method, we should include the $|T3\rangle$ and $|T4\rangle$ channels in addition to the $|T1\rangle$ and $|T2\rangle$ channels,
\begin{eqnarray}
       |T3\rangle &=&  \left[(12)_{\bar{3}_{c}}^{1_{s}}(34)_{3_{c}}^{1_{s}}\right]_{1_{c}}^{0_{s}}\psi_3^{AA}(12;34),\\
     |T4\rangle &=&\left[(12)_{6_{c}}^{0_{s}}(34)_{\bar{6}_{c}}^{0_{s}}\right]_{1_{c}}^{0_{s}}\psi_4^{AA}(12;34) 
\end{eqnarray}
In our practical DMC simulation, we do not constrain the exchange symmetry of the spatial wave function. Instead, we will symmetrize the math ground state after simulation to satisfy the Pauli principle. The pink line in Fig.~\ref{cccc_energy} is the result after adjusting $a_{ij}$ in Eq. (\ref{psi_T}) to minimize fluctuations. The energy $E_R$ reaches threshold around 1000 steps. With these optimal $a_{ij}$ values, i.e., $a_{ij=13,24}=0.62$ $\mathrm{GeV}$ and $a_{ij\neq13,24}=0.001$ $\mathrm{GeV}$, the importance function Eq. (\ref{psi_T}) becomes a di-meson form. To avoid the possible bias of the importance function on the result, another set of no-clustering $a_{ij}$ values, i.e., all $a_{ij}=0.2$ $\mathrm{GeV}$, are set. Its result is shown with the blue line in Fig.~\ref{cccc_energy}. Since it is not the optimal set of parameters, the energy stabilizes more slowly and fluctuates more obviously, but is still very close to the threshold and well below the green line. That is to say, starting from four di-quark spin-color basis, one will automatically obtain the di-meson configuration with the lowest threshold energy, which is hard to achieve in the variational method in the same basis~\cite{Barnea:2006sd}.

Apart from the pre-assignment channels, the sign problem~\cite{toulouse2016introduction} could also prevent the DMC method from getting the general wave functions. It occurs when the wave function of the Fermionic ground state has nodes. Naively, the walkers can only sample the positive-definite functions. In this case, the DMC methods will lead to a lower energy Bosonic state rather than the expected Fermionic state. When there are multiple channels, this problem becomes more complex. The scheme used in Ref. \cite{Sanchez2018} and \cite{Gordillo:2020sgc} that sums over all the spatial wave functions of each channel alleviates this problem but does not solve it perfectly. Because the sum $\mathcal{F}(\boldsymbol{R},t)$ may still be negative in some regions, yet these negative walkers are discarded. Due to the existence of the sign problem, the wave function space of the DMC method is limited. Fortunately, in the simulation of the $cc\bar{c}\bar{c}$ system with $J^{PC}=0^{++}$, after we put all of the four spin-color channels into the calculation and set the importance function parameters properly, the quantity $\mathcal{F}(\boldsymbol{R},t)$ behaves positive-definite, thus the threshold value 6010 MeV is reliable. 

One should be cautious about the above two flaws of DMC in the simulation. In fact, some strategies have been developed to better solve the above problems in the field of nuclear physics, such as the fixed-node approximation~\cite{anderson1976quantum} for the sign problem and the auxiliary field diffusion Monte Carlo~\cite{Gandolfi:2007ed} to avoid preassignment of the channels by sampling the spin and isospin. In hadronic physics, the DMC scheme in use is still a relatively simple one at the present stage. But considering its advantages, especially when the number of particles increases, it is still a promising method and worth further development. 

\section{Summary}~\label{sec:sum}

We have made a systematical diffusion Monte Carlo calculation for all the ground state baryons in a nonrelativistic quark model which contains the Coulomb term and hyperfine
term from the one-gluon exchange interaction. We have considered two different confinement scenarios, i.e., the pairwise ($\Delta$-type) confinement and the three-body flux-tube (Y-type) confinement respectively. The mass, mean-square radius and charge mean-square radius are calculated for each baryon. \clabel[JackknifeSummary]{The statistical uncertainty of the mass reaches less than 1 MeV given by the Jackknife resampling method.}  And the radial distribution, angle distribution and rotation-irrelevant distribution are also shown. We illustrate a feasible procedure to calculate the few-quark spectrum including the few-body confinement force which is favored by the lattice QCD simulations. The procedure could be easily extended to calculate the multi-quark states. We get a important cautionary experience when we investigate the nucleon mass. The DMC method with the pre-assignment of the coupled-channels gives the real ground states only when the channels are chosen properly or completely. The DMC method presumes the wave functions of discrete quantum numbers, although the spatial wave function is unconstrained. With this lesson, we illustrate how to obtain the real ground state (the $\eta_c\eta_c$ threshold rather than an above-threshold energy in Ref.~\cite{Gordillo:2020sgc}) of the $\bar{c}\bar{c}cc$ with $J^{PC}=0^{++}$. 

We use the $AL$1 model in the $\Delta$-type confinement scenario and two flux-tube models in Y-type scenario. Our results show that the differences between the pairwise confinement and three-body confinement could only be neglected when the tension strength of the flux-tube is finely determined. In the flux-tube I we choose a universal tension strength for the mesons and baryons $\sigma_Y=\sigma_{Q\bar{Q}}$. In the flux-tube II we fix the $\sigma_Y= 0.9204 \sigma_{Q\bar{Q}}$ from the $\Omega_{sss}$ experimental mass. Compared with experimental results, the Flux-tube II and AL1 model are in good agreement, while the naive parameterized Flux-tube I model overestimates the mass and underestimates the sizes. We prefer the tension parameter in the flux-tube II, which also coincides with the best fit of the lattice QCD result $\sigma_Y= 0.9355 \sigma_{Q\bar{Q}}$, in Ref.~\cite{Takahashi:2000te}.

We compare the charge radii calculated in this work with the experimental data and lattice QCD results. When the mass difference among the constituent quarks is not
large, our results are consistent with theirs.

The baryon system  is a relatively simple three body system, in which both the variational method and DMC method achieve a similar accuracy. But when the number of particles increases, the advantages of DMC will appear. The DMC method avoids the exponentially increasing number of the basis and the complicated integral related to the few-body force in the variational method. Meanwhile, the DMC method is easily parallelized to carry out high-performance simulations~\cite{Kosztin:1996fh,krogel2012population}. Considering the above advantages, it is worth further developing the application of the DMC method in the hadronic systems. We shall investigate the tetraquark bound states and resonances in the flux-tube confinement potentials via DMC in the near future.

\begin{appendix}

\section{Convection–diffusion equation}~\label{app:conv-diff-eq}
The Eq.~\eqref{schrodingerForf} is the analog of the convection–diffusion equation,
\begin{equation}
    \frac{\partial c(\bm{r},t)}{\partial t}=D\nabla^{2}c-\nabla\cdot(\bm{v}c)+R(\bm{r},t),
\end{equation}
where, for example, $c$ is the concentration field of the salt in a river. $\bm{v} $ is the velocity field of the river water. $D$ is the diffusion coefficient of salt in the water. $R(\bm{r},t)$ describes sources or sinks, where the salt can be generated or absorbed, respectively. Obviously, the three terms of the right-hand side correspond to the the random diffusion of salt, driven effect by the moving water and  source or sink effect, respectively.

\section{Statistical uncertainty analysis}~\label{app:statistical_uncertainty}
\clabel[app_uncertainty]{There exist correlations among results from adjacent steps. We divide a set of data ${X_i}$ with number $MR$ into $R$ blocks with block size $M$. We denote the block mean values as $X_{i}^{(1)}$. If $M$ is large enough, these new variables $X_{i}^{(1)}$ become uncorrelated. One way to test whether $M$ is large enough is by making a further blocking operation with block size $K$ and get the block mean values
\begin{align}\label{X2}
X_{i}^{(2)}&=\frac{1}{K}\sum_{j=1}^K X_{(i-1)K+j}^{(1)}.
\end{align}
When $X_{i}^{(1)}$ is uncorrelated, the variance of $X_{i}^{(2)}$ reads
\begin{align}\label{VarK}
Var[X_{i}^{(2)}]&=Var\left[\frac{1}{K}\sum_{j=1}^{K}X_{(i-1)K+j}^{(1)}\right]=\frac{1}{K}Var\left[X_{i}^{(1)}\right]\,,
\end{align}
where the variance of $X_{i}^{(2)}$ can be estimated by
\begin{align}\label{Var}
Var[X_{i}^{(2)}]&=\frac{1}{K-1}\sum_{i=1}^{K}(X_{i}^{(2)}-\overline{X^{(2)}})^2\,,
\end{align}
with $\overline{X^{(2)}}=\frac{1}{K}\sum_{j=1}^KX_j^{(2)}$.
The value $\delta\equiv\sqrt{\frac{Var[X_{i}^{(2)}]}{R/K}}$ should be approximately a constant, because the $Var[X_i^{(1)}]$ is fixed.  }

\change{To determine the block size making $X_i^{(1)}$ uncorrelated, we take $\Omega^-(sss)$ as an example and get $150000$ stable steps. When we set $M=500$, the change of value $\delta$ with $K$ is shown in Fig.~\ref{blocking_sss}.
\begin{figure}[htb] 
\centering
\includegraphics[width=0.49\textwidth]{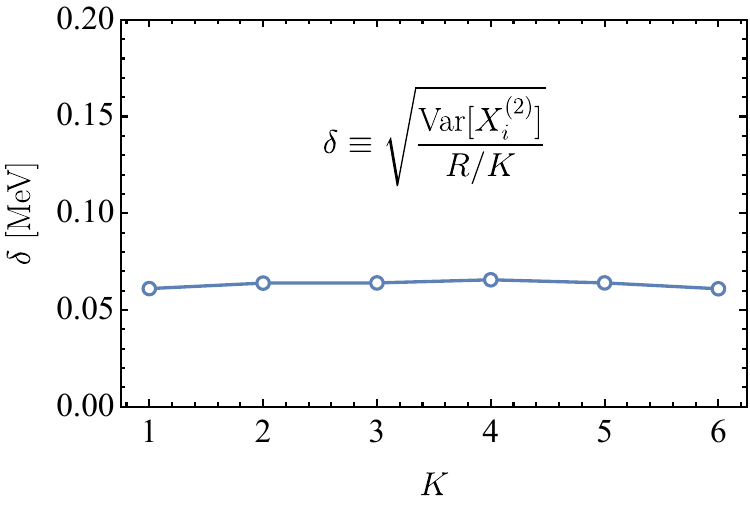} 
\caption{\label{blocking_sss} $\delta\equiv\sqrt{\frac{Var[X_{i}^{(2)}]}{R/K}}$ value changes with $K$ when $M=500$.}
\end{figure}
This nearly flat behavior means that the $X_{i}^{(1)}$ have become uncorrelated when taking the size as $500$ steps.}

\change{The simulations in the main text include 5000 stable steps.  We divide them into $R=10$ blocks and calculate the block averages $X_{i}^{(1)}$. The uncertainty of the total mean value $\bar{X}=\overline{X_i^{(1)}}=\frac{1}{n}\sum_i^n X_i$ will be
\begin{align}\label{error}
\sigma[\bar{X}]&=\sqrt{\frac{1}{R(R-1)}\sum_i^R(X_{i}^{(1)}-\bar{X})^2}\\
&=\sqrt{\frac{R-1}{R}\sum_i^R(\bar{X}_{(i)\rm{jack}}-\bar{X}_{\rm{jack}})^2}\,.\label{jack}
\end{align}
The Eq.(\ref{jack}) is from the Jackknife resampling method, where $\bar{X}_{(i)\rm{jack}}=\frac{1}{R-1}\sum_{j\neq i}^R X_j^{(1)}$, $\bar{X}_{\rm{jack}}=\bar{X}$. The calculated uncertainty of the $\Omega^-(sss)$ mass is 0.3 MeV. For other systems, the uncertainties are all within 1 MeV.}

\section{Size contribution to the charge radius}~\label{app:Rc2_size_contribution}

The charge form factor of the baryon is~\cite{Wagner:2000ii}
\begin{align}\label{charge_form_factor}
F(\boldsymbol{q}^2)&=\sqrt{4\pi}\langle \Psi|\frac{1}{4\pi}\int \mathrm{d}\Omega_{\boldsymbol{q}}\cdot Y_{00}(\hat{\boldsymbol{q}})\rho(\boldsymbol{q})|\Psi\rangle\,,
\end{align}
where $\boldsymbol{q}$ indicates the momentum transfer, $|\Psi\rangle$ represents the total baryon wavefunction, $\rho(\boldsymbol{q})$ is the charge density. The baryon charge radius can be obtained from this charge form factor with
\begin{align}\label{charge_radius_slope}
R_c^2&=-6\frac{\partial}{\partial \boldsymbol{q}^2}F(\boldsymbol{q}^2)|_{\boldsymbol{q}^2=0}\,.
\end{align}
Here we only consider the contribution from the one-body operator. Thus the total result can be obtained by summing over the single quark contribution,
\begin{align}
\rho(\boldsymbol{q})&=\sum_{i=1}^3\rho_{i}(\boldsymbol{q})\,.
\end{align}

When regarding the quarks as the point-like particles, the charge density $\rho_{0i}(\boldsymbol{q})$ will be
\begin{align}
\rho_{0i}(\boldsymbol{q})=e_{i}\mathrm{e}^{i\boldsymbol{q}\cdot\boldsymbol{r}_{i}}\,,
\end{align}
where $e_i$ and $\boldsymbol{r}_{i}$ indicate the charge and coordinate of the $i$th quark. 
In this way Eq. (\ref{charge_radius_slope}) becomes
\begin{align}\label{charge_radius_slope_0_sumi}
R_{c0}^2&=-6\sum_i^3\frac{\partial}{\partial \boldsymbol{q}^2}F_{0i}(\boldsymbol{q}^2)|_{\boldsymbol{q}^2=0}\,,
\end{align}
with
\begin{align}\label{charge_form_factor_i}
F_{0i}(\boldsymbol{q}^2)&=\sqrt{4\pi}\langle \Psi|\frac{1}{4\pi}\int \mathrm{d}\Omega_{\boldsymbol{q}}\cdot Y_{00}(\hat{\boldsymbol{q}})\rho_{0i}(\boldsymbol{q})|\Psi\rangle\,.
\end{align}
In the coordinate space, the contribution reads,
\begin{equation}
 R_{c0}^2=   \langle\Psi|\sum_{i=1}^{3}e_{i}(\boldsymbol{r}_{i}-\boldsymbol{R}_{CM})^{2}|\Psi\rangle
\end{equation}

However, if the electromagnetic size of the quark is introduced, the $q_iq_i\gamma^*$ interaction vertex will change from $ie_i\gamma_\mu$ to $ie_if_i(\boldsymbol{q})\gamma_\mu$ with $f_i(\boldsymbol{0})=1$, where $e_if_i(\bm{q})$ is the electric form factor of the constituent quark. Thus the charge density becomes
\begin{align}
\rho_{i}(\boldsymbol{q})=e_{i}f_i(\boldsymbol{q})\mathrm{e}^{i\boldsymbol{q}\cdot\boldsymbol{r}_{i}}\,,
\end{align}
and the charge radius turns into
\begin{align}\label{charge_radius_slope_sumi}
R_{c}^2&=-6\sum_i^3\frac{\partial F_{0i}(\boldsymbol{q}^2)}{\partial \boldsymbol{q}^2}|_{\boldsymbol{q}^2=0}f_i(\boldsymbol{0})-6\sum_i^3F_{0i}(0)\frac{\partial{f_i(\boldsymbol{q})}}{\partial \boldsymbol{q}^2}|_{\boldsymbol{q}^2=0}\nonumber\\
&=-6\sum_i^3\frac{\partial F_{0i}(\boldsymbol{q}^2)}{\partial \boldsymbol{q}^2}|_{\boldsymbol{q}^2=0}-6\sum_i^3e_i\frac{\partial{f_i(\boldsymbol{q})}}{\partial \boldsymbol{q}^2}|_{\boldsymbol{q}^2=0}\, \nonumber\\
&=R_{c0}^2+\sum_i^3 R_c^{q_i}.
\end{align}
The first term corresponds to Eq. (\ref{Rc2}), and the second term contribution is extracted from experiments and lattice QCD calculations as described in Sec.~\ref{sec:results}.

\end{appendix}

\begin{acknowledgements}
We are grateful to  Xin-Zhen Weng, Guang-Juan Wang and Zi-Yang Lin for the helpful discussions. L. M. is grateful to Jorge Segovia for the helpful communications. Y. M. would like to express sincere gratitude to Prof. Ya-Jun Mao for his invaluable advice and continuous support in all aspects. This project was supported by the National
Natural Science Foundation of China (11975033 and 12070131001). This
project was also funded by the Deutsche Forschungsgemeinschaft (DFG,
German Research Foundation, Project ID 196253076-TRR 110). 
\end{acknowledgements}

\bibliographystyle{apsrev4-2}
\bibliography{Ref}

\begin{thebibliography}{79}%
\makeatletter
\providecommand \@ifxundefined [1]{%
 \@ifx{#1\undefined}
}%
\providecommand \@ifnum [1]{%
 \ifnum #1\expandafter \@firstoftwo
 \else \expandafter \@secondoftwo
 \fi
}%
\providecommand \@ifx [1]{%
 \ifx #1\expandafter \@firstoftwo
 \else \expandafter \@secondoftwo
 \fi
}%
\providecommand \natexlab [1]{#1}%
\providecommand \enquote  [1]{``#1''}%
\providecommand \bibnamefont  [1]{#1}%
\providecommand \bibfnamefont [1]{#1}%
\providecommand \citenamefont [1]{#1}%
\providecommand \href@noop [0]{\@secondoftwo}%
\providecommand \href [0]{\begingroup \@sanitize@url \@href}%
\providecommand \@href[1]{\@@startlink{#1}\@@href}%
\providecommand \@@href[1]{\endgroup#1\@@endlink}%
\providecommand \@sanitize@url [0]{\catcode `\\12\catcode `\$12\catcode
  `\&12\catcode `\#12\catcode `\^12\catcode `\_12\catcode `\%12\relax}%
\providecommand \@@startlink[1]{}%
\providecommand \@@endlink[0]{}%
\providecommand \url  [0]{\begingroup\@sanitize@url \@url }%
\providecommand \@url [1]{\endgroup\@href {#1}{\urlprefix }}%
\providecommand \urlprefix  [0]{URL }%
\providecommand \Eprint [0]{\href }%
\providecommand \doibase [0]{https://doi.org/}%
\providecommand \selectlanguage [0]{\@gobble}%
\providecommand \bibinfo  [0]{\@secondoftwo}%
\providecommand \bibfield  [0]{\@secondoftwo}%
\providecommand \translation [1]{[#1]}%
\providecommand \BibitemOpen [0]{}%
\providecommand \bibitemStop [0]{}%
\providecommand \bibitemNoStop [0]{.\EOS\space}%
\providecommand \EOS [0]{\spacefactor3000\relax}%
\providecommand \BibitemShut  [1]{\csname bibitem#1\endcsname}%
\let\auto@bib@innerbib\@empty
\bibitem [{\citenamefont {Richard}(2012)}]{Richard:2012xw}%
  \BibitemOpen
  \bibfield  {author} {\bibinfo {author} {\bibfnamefont {J.-M.}\ \bibnamefont
  {Richard}},\ }in\ \href@noop {} {\emph {\bibinfo {booktitle} {{Ferrara
  International School Niccol\`o Cabeo 2012: Hadronic spectroscopy}}}}\
  (\bibinfo {year} {2012})\ \Eprint {https://arxiv.org/abs/1205.4326}
  {arXiv:1205.4326 [hep-ph]} \BibitemShut {NoStop}%
\bibitem [{\citenamefont {Richard}(2016)}]{Richard:2016eis}%
  \BibitemOpen
  \bibfield  {author} {\bibinfo {author} {\bibfnamefont {J.-M.}\ \bibnamefont
  {Richard}},\ }\href {https://doi.org/10.1007/s00601-016-1159-0} {\bibfield
  {journal} {\bibinfo  {journal} {Few Body Syst.}\ }\textbf {\bibinfo {volume}
  {57}},\ \bibinfo {pages} {1185} (\bibinfo {year} {2016})},\ \Eprint
  {https://arxiv.org/abs/1606.08593} {arXiv:1606.08593 [hep-ph]} \BibitemShut
  {NoStop}%
\bibitem [{\citenamefont {Meng}\ \emph {et~al.}(2022)\citenamefont {Meng},
  \citenamefont {Wang}, \citenamefont {Wang},\ and\ \citenamefont
  {Zhu}}]{Meng:2022ozq}%
  \BibitemOpen
  \bibfield  {author} {\bibinfo {author} {\bibfnamefont {L.}~\bibnamefont
  {Meng}}, \bibinfo {author} {\bibfnamefont {B.}~\bibnamefont {Wang}}, \bibinfo
  {author} {\bibfnamefont {G.-J.}\ \bibnamefont {Wang}},\ and\ \bibinfo
  {author} {\bibfnamefont {S.-L.}\ \bibnamefont {Zhu}},\ }\href@noop {} {\
  (\bibinfo {year} {2022})},\ \Eprint {https://arxiv.org/abs/2204.08716}
  {arXiv:2204.08716 [hep-ph]} \BibitemShut {NoStop}%
\bibitem [{\citenamefont {Chen}\ \emph {et~al.}(2022)\citenamefont {Chen},
  \citenamefont {Chen}, \citenamefont {Liu}, \citenamefont {Liu},\ and\
  \citenamefont {Zhu}}]{Chen:2022asf}%
  \BibitemOpen
  \bibfield  {author} {\bibinfo {author} {\bibfnamefont {H.-X.}\ \bibnamefont
  {Chen}}, \bibinfo {author} {\bibfnamefont {W.}~\bibnamefont {Chen}}, \bibinfo
  {author} {\bibfnamefont {X.}~\bibnamefont {Liu}}, \bibinfo {author}
  {\bibfnamefont {Y.-R.}\ \bibnamefont {Liu}},\ and\ \bibinfo {author}
  {\bibfnamefont {S.-L.}\ \bibnamefont {Zhu}},\ }\href@noop {} {\  (\bibinfo
  {year} {2022})},\ \Eprint {https://arxiv.org/abs/2204.02649}
  {arXiv:2204.02649 [hep-ph]} \BibitemShut {NoStop}%
\bibitem [{\citenamefont {Liu}\ \emph {et~al.}(2019)\citenamefont {Liu},
  \citenamefont {Chen}, \citenamefont {Chen}, \citenamefont {Liu},\ and\
  \citenamefont {Zhu}}]{Liu:2019zoy}%
  \BibitemOpen
  \bibfield  {author} {\bibinfo {author} {\bibfnamefont {Y.-R.}\ \bibnamefont
  {Liu}}, \bibinfo {author} {\bibfnamefont {H.-X.}\ \bibnamefont {Chen}},
  \bibinfo {author} {\bibfnamefont {W.}~\bibnamefont {Chen}}, \bibinfo {author}
  {\bibfnamefont {X.}~\bibnamefont {Liu}},\ and\ \bibinfo {author}
  {\bibfnamefont {S.-L.}\ \bibnamefont {Zhu}},\ }\href
  {https://doi.org/10.1016/j.ppnp.2019.04.003} {\bibfield  {journal} {\bibinfo
  {journal} {Prog. Part. Nucl. Phys.}\ }\textbf {\bibinfo {volume} {107}},\
  \bibinfo {pages} {237} (\bibinfo {year} {2019})},\ \Eprint
  {https://arxiv.org/abs/1903.11976} {arXiv:1903.11976 [hep-ph]} \BibitemShut
  {NoStop}%
\bibitem [{\citenamefont {Chen}\ \emph {et~al.}(2017)\citenamefont {Chen},
  \citenamefont {Chen}, \citenamefont {Liu}, \citenamefont {Liu},\ and\
  \citenamefont {Zhu}}]{Chen:2016spr}%
  \BibitemOpen
  \bibfield  {author} {\bibinfo {author} {\bibfnamefont {H.-X.}\ \bibnamefont
  {Chen}}, \bibinfo {author} {\bibfnamefont {W.}~\bibnamefont {Chen}}, \bibinfo
  {author} {\bibfnamefont {X.}~\bibnamefont {Liu}}, \bibinfo {author}
  {\bibfnamefont {Y.-R.}\ \bibnamefont {Liu}},\ and\ \bibinfo {author}
  {\bibfnamefont {S.-L.}\ \bibnamefont {Zhu}},\ }\href
  {https://doi.org/10.1088/1361-6633/aa6420} {\bibfield  {journal} {\bibinfo
  {journal} {Rept. Prog. Phys.}\ }\textbf {\bibinfo {volume} {80}},\ \bibinfo
  {pages} {076201} (\bibinfo {year} {2017})},\ \Eprint
  {https://arxiv.org/abs/1609.08928} {arXiv:1609.08928 [hep-ph]} \BibitemShut
  {NoStop}%
\bibitem [{\citenamefont {Chen}\ \emph {et~al.}(2016)\citenamefont {Chen},
  \citenamefont {Chen}, \citenamefont {Liu},\ and\ \citenamefont
  {Zhu}}]{Chen:2016qju}%
  \BibitemOpen
  \bibfield  {author} {\bibinfo {author} {\bibfnamefont {H.-X.}\ \bibnamefont
  {Chen}}, \bibinfo {author} {\bibfnamefont {W.}~\bibnamefont {Chen}}, \bibinfo
  {author} {\bibfnamefont {X.}~\bibnamefont {Liu}},\ and\ \bibinfo {author}
  {\bibfnamefont {S.-L.}\ \bibnamefont {Zhu}},\ }\href
  {https://doi.org/10.1016/j.physrep.2016.05.004} {\bibfield  {journal}
  {\bibinfo  {journal} {Phys. Rept.}\ }\textbf {\bibinfo {volume} {639}},\
  \bibinfo {pages} {1} (\bibinfo {year} {2016})},\ \Eprint
  {https://arxiv.org/abs/1601.02092} {arXiv:1601.02092 [hep-ph]} \BibitemShut
  {NoStop}%
\bibitem [{\citenamefont {Godfrey}\ and\ \citenamefont
  {Isgur}(1985)}]{Godfrey:1985xj}%
  \BibitemOpen
  \bibfield  {author} {\bibinfo {author} {\bibfnamefont {S.}~\bibnamefont
  {Godfrey}}\ and\ \bibinfo {author} {\bibfnamefont {N.}~\bibnamefont
  {Isgur}},\ }\href {https://doi.org/10.1103/PhysRevD.32.189} {\bibfield
  {journal} {\bibinfo  {journal} {Phys. Rev. D}\ }\textbf {\bibinfo {volume}
  {32}},\ \bibinfo {pages} {189} (\bibinfo {year} {1985})}\BibitemShut
  {NoStop}%
\bibitem [{\citenamefont {Eichten}\ \emph {et~al.}(1975)\citenamefont
  {Eichten}, \citenamefont {Gottfried}, \citenamefont {Kinoshita},
  \citenamefont {Kogut}, \citenamefont {Lane},\ and\ \citenamefont
  {Yan}}]{Eichten:1974af}%
  \BibitemOpen
  \bibfield  {author} {\bibinfo {author} {\bibfnamefont {E.}~\bibnamefont
  {Eichten}}, \bibinfo {author} {\bibfnamefont {K.}~\bibnamefont {Gottfried}},
  \bibinfo {author} {\bibfnamefont {T.}~\bibnamefont {Kinoshita}}, \bibinfo
  {author} {\bibfnamefont {J.~B.}\ \bibnamefont {Kogut}}, \bibinfo {author}
  {\bibfnamefont {K.~D.}\ \bibnamefont {Lane}},\ and\ \bibinfo {author}
  {\bibfnamefont {T.-M.}\ \bibnamefont {Yan}},\ }\href
  {https://doi.org/10.1103/PhysRevLett.34.369} {\bibfield  {journal} {\bibinfo
  {journal} {Phys. Rev. Lett.}\ }\textbf {\bibinfo {volume} {34}},\ \bibinfo
  {pages} {369} (\bibinfo {year} {1975})},\ \bibinfo {note} {[Erratum:
  Phys.Rev.Lett. 36, 1276 (1976)]}\BibitemShut {NoStop}%
\bibitem [{\citenamefont {Eichten}\ \emph {et~al.}(1978)\citenamefont
  {Eichten}, \citenamefont {Gottfried}, \citenamefont {Kinoshita},
  \citenamefont {Lane},\ and\ \citenamefont {Yan}}]{Eichten:1978tg}%
  \BibitemOpen
  \bibfield  {author} {\bibinfo {author} {\bibfnamefont {E.}~\bibnamefont
  {Eichten}}, \bibinfo {author} {\bibfnamefont {K.}~\bibnamefont {Gottfried}},
  \bibinfo {author} {\bibfnamefont {T.}~\bibnamefont {Kinoshita}}, \bibinfo
  {author} {\bibfnamefont {K.~D.}\ \bibnamefont {Lane}},\ and\ \bibinfo
  {author} {\bibfnamefont {T.-M.}\ \bibnamefont {Yan}},\ }\href
  {https://doi.org/10.1103/PhysRevD.17.3090} {\bibfield  {journal} {\bibinfo
  {journal} {Phys. Rev. D}\ }\textbf {\bibinfo {volume} {17}},\ \bibinfo
  {pages} {3090} (\bibinfo {year} {1978})},\ \bibinfo {note} {[Erratum:
  Phys.Rev.D 21, 313 (1980)]}\BibitemShut {NoStop}%
\bibitem [{\citenamefont {Eichten}\ \emph {et~al.}(1980)\citenamefont
  {Eichten}, \citenamefont {Gottfried}, \citenamefont {Kinoshita},
  \citenamefont {Lane},\ and\ \citenamefont {Yan}}]{Eichten:1979ms}%
  \BibitemOpen
  \bibfield  {author} {\bibinfo {author} {\bibfnamefont {E.}~\bibnamefont
  {Eichten}}, \bibinfo {author} {\bibfnamefont {K.}~\bibnamefont {Gottfried}},
  \bibinfo {author} {\bibfnamefont {T.}~\bibnamefont {Kinoshita}}, \bibinfo
  {author} {\bibfnamefont {K.~D.}\ \bibnamefont {Lane}},\ and\ \bibinfo
  {author} {\bibfnamefont {T.-M.}\ \bibnamefont {Yan}},\ }\href
  {https://doi.org/10.1103/PhysRevD.21.203} {\bibfield  {journal} {\bibinfo
  {journal} {Phys. Rev. D}\ }\textbf {\bibinfo {volume} {21}},\ \bibinfo
  {pages} {203} (\bibinfo {year} {1980})}\BibitemShut {NoStop}%
\bibitem [{\citenamefont {Capstick}\ and\ \citenamefont
  {Isgur}(1986)}]{Capstick:1986ter}%
  \BibitemOpen
  \bibfield  {author} {\bibinfo {author} {\bibfnamefont {S.}~\bibnamefont
  {Capstick}}\ and\ \bibinfo {author} {\bibfnamefont {N.}~\bibnamefont
  {Isgur}},\ }\href {https://doi.org/10.1103/physrevd.34.2809} {\bibfield
  {journal} {\bibinfo  {journal} {Phys. Rev. D}\ }\textbf {\bibinfo {volume}
  {34}},\ \bibinfo {pages} {2809} (\bibinfo {year} {1986})}\BibitemShut
  {NoStop}%
\bibitem [{\citenamefont {Semay}\ and\ \citenamefont
  {Silvestre-Brac}(1992)}]{Semay:1992xq}%
  \BibitemOpen
  \bibfield  {author} {\bibinfo {author} {\bibfnamefont {C.}~\bibnamefont
  {Semay}}\ and\ \bibinfo {author} {\bibfnamefont {B.}~\bibnamefont
  {Silvestre-Brac}},\ }\href {https://doi.org/10.1103/PhysRevD.46.5177}
  {\bibfield  {journal} {\bibinfo  {journal} {Phys. Rev. D}\ }\textbf {\bibinfo
  {volume} {46}},\ \bibinfo {pages} {5177} (\bibinfo {year}
  {1992})}\BibitemShut {NoStop}%
\bibitem [{\citenamefont {Semay}\ and\ \citenamefont
  {Silvestre-Brac}(1994)}]{Semay:1994ht}%
  \BibitemOpen
  \bibfield  {author} {\bibinfo {author} {\bibfnamefont {C.}~\bibnamefont
  {Semay}}\ and\ \bibinfo {author} {\bibfnamefont {B.}~\bibnamefont
  {Silvestre-Brac}},\ }\href {https://doi.org/10.1007/BF01413104} {\bibfield
  {journal} {\bibinfo  {journal} {Z. Phys. C}\ }\textbf {\bibinfo {volume}
  {61}},\ \bibinfo {pages} {271} (\bibinfo {year} {1994})}\BibitemShut
  {NoStop}%
\bibitem [{\citenamefont {Silvestre-Brac}(1996)}]{Silvestre-Brac:1996myf}%
  \BibitemOpen
  \bibfield  {author} {\bibinfo {author} {\bibfnamefont {B.}~\bibnamefont
  {Silvestre-Brac}},\ }\href {https://doi.org/10.1007/s006010050028} {\bibfield
   {journal} {\bibinfo  {journal} {Few Body Syst.}\ }\textbf {\bibinfo {volume}
  {20}},\ \bibinfo {pages} {1} (\bibinfo {year} {1996})}\BibitemShut {NoStop}%
\bibitem [{\citenamefont {Li}\ and\ \citenamefont {Chao}(2009)}]{Li:2009zu}%
  \BibitemOpen
  \bibfield  {author} {\bibinfo {author} {\bibfnamefont {B.-Q.}\ \bibnamefont
  {Li}}\ and\ \bibinfo {author} {\bibfnamefont {K.-T.}\ \bibnamefont {Chao}},\
  }\href {https://doi.org/10.1103/PhysRevD.79.094004} {\bibfield  {journal}
  {\bibinfo  {journal} {Phys. Rev. D}\ }\textbf {\bibinfo {volume} {79}},\
  \bibinfo {pages} {094004} (\bibinfo {year} {2009})},\ \Eprint
  {https://arxiv.org/abs/0903.5506} {arXiv:0903.5506 [hep-ph]} \BibitemShut
  {NoStop}%
\bibitem [{\citenamefont {Bali}(2001)}]{Bali:2000gf}%
  \BibitemOpen
  \bibfield  {author} {\bibinfo {author} {\bibfnamefont {G.~S.}\ \bibnamefont
  {Bali}},\ }\href {https://doi.org/10.1016/S0370-1573(00)00079-X} {\bibfield
  {journal} {\bibinfo  {journal} {Phys. Rept.}\ }\textbf {\bibinfo {volume}
  {343}},\ \bibinfo {pages} {1} (\bibinfo {year} {2001})},\ \Eprint
  {https://arxiv.org/abs/hep-ph/0001312} {arXiv:hep-ph/0001312} \BibitemShut
  {NoStop}%
\bibitem [{\citenamefont {Artru}(1975)}]{Artru:1974zn}%
  \BibitemOpen
  \bibfield  {author} {\bibinfo {author} {\bibfnamefont {X.}~\bibnamefont
  {Artru}},\ }\href {https://doi.org/10.1016/0550-3213(75)90019-X} {\bibfield
  {journal} {\bibinfo  {journal} {Nucl. Phys. B}\ }\textbf {\bibinfo {volume}
  {85}},\ \bibinfo {pages} {442} (\bibinfo {year} {1975})}\BibitemShut
  {NoStop}%
\bibitem [{\citenamefont {Hasenfratz}\ \emph {et~al.}(1980)\citenamefont
  {Hasenfratz}, \citenamefont {Horgan}, \citenamefont {Kuti},\ and\
  \citenamefont {Richard}}]{Hasenfratz:1980ka}%
  \BibitemOpen
  \bibfield  {author} {\bibinfo {author} {\bibfnamefont {P.}~\bibnamefont
  {Hasenfratz}}, \bibinfo {author} {\bibfnamefont {R.~R.}\ \bibnamefont
  {Horgan}}, \bibinfo {author} {\bibfnamefont {J.}~\bibnamefont {Kuti}},\ and\
  \bibinfo {author} {\bibfnamefont {J.~M.}\ \bibnamefont {Richard}},\ }\href
  {https://doi.org/10.1016/0370-2693(80)90906-5} {\bibfield  {journal}
  {\bibinfo  {journal} {Phys. Lett. B}\ }\textbf {\bibinfo {volume} {94}},\
  \bibinfo {pages} {401} (\bibinfo {year} {1980})}\BibitemShut {NoStop}%
\bibitem [{\citenamefont {Richard}\ and\ \citenamefont
  {Taxil}(1983)}]{Richard:1983mu}%
  \BibitemOpen
  \bibfield  {author} {\bibinfo {author} {\bibfnamefont {J.~M.}\ \bibnamefont
  {Richard}}\ and\ \bibinfo {author} {\bibfnamefont {P.}~\bibnamefont
  {Taxil}},\ }\href {https://doi.org/10.1016/0003-4916(83)90009-X} {\bibfield
  {journal} {\bibinfo  {journal} {Annals Phys.}\ }\textbf {\bibinfo {volume}
  {150}},\ \bibinfo {pages} {267} (\bibinfo {year} {1983})}\BibitemShut
  {NoStop}%
\bibitem [{\citenamefont {Blask}\ \emph {et~al.}(1990)\citenamefont {Blask},
  \citenamefont {Bohn}, \citenamefont {Huber}, \citenamefont {Metsch},\ and\
  \citenamefont {Petry}}]{Blask:1990ez}%
  \BibitemOpen
  \bibfield  {author} {\bibinfo {author} {\bibfnamefont {W.~H.}\ \bibnamefont
  {Blask}}, \bibinfo {author} {\bibfnamefont {U.}~\bibnamefont {Bohn}},
  \bibinfo {author} {\bibfnamefont {M.~G.}\ \bibnamefont {Huber}}, \bibinfo
  {author} {\bibfnamefont {B.~C.}\ \bibnamefont {Metsch}},\ and\ \bibinfo
  {author} {\bibfnamefont {H.~R.}\ \bibnamefont {Petry}},\ }\href
  {https://doi.org/10.1007/BF01289701} {\bibfield  {journal} {\bibinfo
  {journal} {Z. Phys. A}\ }\textbf {\bibinfo {volume} {337}},\ \bibinfo {pages}
  {327} (\bibinfo {year} {1990})}\BibitemShut {NoStop}%
\bibitem [{\citenamefont {Carlson}\ \emph {et~al.}(1983)\citenamefont
  {Carlson}, \citenamefont {Kogut},\ and\ \citenamefont
  {Pandharipande}}]{Carlson:1982xi}%
  \BibitemOpen
  \bibfield  {author} {\bibinfo {author} {\bibfnamefont {J.}~\bibnamefont
  {Carlson}}, \bibinfo {author} {\bibfnamefont {J.~B.}\ \bibnamefont {Kogut}},\
  and\ \bibinfo {author} {\bibfnamefont {V.~R.}\ \bibnamefont
  {Pandharipande}},\ }\href {https://doi.org/10.1103/PhysRevD.27.233}
  {\bibfield  {journal} {\bibinfo  {journal} {Phys. Rev. D}\ }\textbf {\bibinfo
  {volume} {27}},\ \bibinfo {pages} {233} (\bibinfo {year} {1983})}\BibitemShut
  {NoStop}%
\bibitem [{\citenamefont {Sartor}\ and\ \citenamefont
  {Stancu}(1985)}]{Sartor:1985ss}%
  \BibitemOpen
  \bibfield  {author} {\bibinfo {author} {\bibfnamefont {R.}~\bibnamefont
  {Sartor}}\ and\ \bibinfo {author} {\bibfnamefont {F.}~\bibnamefont
  {Stancu}},\ }\href {https://doi.org/10.1103/PhysRevD.31.128} {\bibfield
  {journal} {\bibinfo  {journal} {Phys. Rev. D}\ }\textbf {\bibinfo {volume}
  {31}},\ \bibinfo {pages} {128} (\bibinfo {year} {1985})}\BibitemShut
  {NoStop}%
\bibitem [{\citenamefont {Stancu}\ and\ \citenamefont
  {Stassart}(1988)}]{Stancu:1988gb}%
  \BibitemOpen
  \bibfield  {author} {\bibinfo {author} {\bibfnamefont {F.}~\bibnamefont
  {Stancu}}\ and\ \bibinfo {author} {\bibfnamefont {P.}~\bibnamefont
  {Stassart}},\ }\href {https://doi.org/10.1103/PhysRevD.38.233} {\bibfield
  {journal} {\bibinfo  {journal} {Phys. Rev. D}\ }\textbf {\bibinfo {volume}
  {38}},\ \bibinfo {pages} {233} (\bibinfo {year} {1988})}\BibitemShut
  {NoStop}%
\bibitem [{\citenamefont {Stancu}\ and\ \citenamefont
  {Stassart}(1989)}]{Stancu:1989iu}%
  \BibitemOpen
  \bibfield  {author} {\bibinfo {author} {\bibfnamefont {F.}~\bibnamefont
  {Stancu}}\ and\ \bibinfo {author} {\bibfnamefont {P.}~\bibnamefont
  {Stassart}},\ }\href {https://doi.org/10.1103/PhysRevD.39.343} {\bibfield
  {journal} {\bibinfo  {journal} {Phys. Rev. D}\ }\textbf {\bibinfo {volume}
  {39}},\ \bibinfo {pages} {343} (\bibinfo {year} {1989})}\BibitemShut
  {NoStop}%
\bibitem [{\citenamefont {Takahashi}\ \emph {et~al.}(2001)\citenamefont
  {Takahashi}, \citenamefont {Matsufuru}, \citenamefont {Nemoto},\ and\
  \citenamefont {Suganuma}}]{Takahashi:2000te}%
  \BibitemOpen
  \bibfield  {author} {\bibinfo {author} {\bibfnamefont {T.~T.}\ \bibnamefont
  {Takahashi}}, \bibinfo {author} {\bibfnamefont {H.}~\bibnamefont
  {Matsufuru}}, \bibinfo {author} {\bibfnamefont {Y.}~\bibnamefont {Nemoto}},\
  and\ \bibinfo {author} {\bibfnamefont {H.}~\bibnamefont {Suganuma}},\ }\href
  {https://doi.org/10.1103/PhysRevLett.86.18} {\bibfield  {journal} {\bibinfo
  {journal} {Phys. Rev. Lett.}\ }\textbf {\bibinfo {volume} {86}},\ \bibinfo
  {pages} {18} (\bibinfo {year} {2001})},\ \Eprint
  {https://arxiv.org/abs/hep-lat/0006005} {arXiv:hep-lat/0006005} \BibitemShut
  {NoStop}%
\bibitem [{\citenamefont {Takahashi}\ \emph {et~al.}(2002)\citenamefont
  {Takahashi}, \citenamefont {Suganuma}, \citenamefont {Nemoto},\ and\
  \citenamefont {Matsufuru}}]{Takahashi:2002bw}%
  \BibitemOpen
  \bibfield  {author} {\bibinfo {author} {\bibfnamefont {T.~T.}\ \bibnamefont
  {Takahashi}}, \bibinfo {author} {\bibfnamefont {H.}~\bibnamefont {Suganuma}},
  \bibinfo {author} {\bibfnamefont {Y.}~\bibnamefont {Nemoto}},\ and\ \bibinfo
  {author} {\bibfnamefont {H.}~\bibnamefont {Matsufuru}},\ }\href
  {https://doi.org/10.1103/PhysRevD.65.114509} {\bibfield  {journal} {\bibinfo
  {journal} {Phys. Rev. D}\ }\textbf {\bibinfo {volume} {65}},\ \bibinfo
  {pages} {114509} (\bibinfo {year} {2002})},\ \Eprint
  {https://arxiv.org/abs/hep-lat/0204011} {arXiv:hep-lat/0204011} \BibitemShut
  {NoStop}%
\bibitem [{\citenamefont {Dmitrasinovic}\ \emph {et~al.}(2009)\citenamefont
  {Dmitrasinovic}, \citenamefont {Sato},\ and\ \citenamefont
  {Suvakov}}]{Dmitrasinovic:2009dy}%
  \BibitemOpen
  \bibfield  {author} {\bibinfo {author} {\bibfnamefont {V.}~\bibnamefont
  {Dmitrasinovic}}, \bibinfo {author} {\bibfnamefont {T.}~\bibnamefont
  {Sato}},\ and\ \bibinfo {author} {\bibfnamefont {M.}~\bibnamefont
  {Suvakov}},\ }\href {https://doi.org/10.1140/epjc/s10052-009-1050-y}
  {\bibfield  {journal} {\bibinfo  {journal} {Eur. Phys. J. C}\ }\textbf
  {\bibinfo {volume} {62}},\ \bibinfo {pages} {383} (\bibinfo {year} {2009})},\
  \Eprint {https://arxiv.org/abs/0906.2327} {arXiv:0906.2327 [hep-ph]}
  \BibitemShut {NoStop}%
\bibitem [{\citenamefont {Bicudo}\ and\ \citenamefont
  {Cardoso}(2016)}]{Bicudo:2015bra}%
  \BibitemOpen
  \bibfield  {author} {\bibinfo {author} {\bibfnamefont {P.}~\bibnamefont
  {Bicudo}}\ and\ \bibinfo {author} {\bibfnamefont {M.}~\bibnamefont
  {Cardoso}},\ }\href {https://doi.org/10.1103/PhysRevD.94.094032} {\bibfield
  {journal} {\bibinfo  {journal} {Phys. Rev. D}\ }\textbf {\bibinfo {volume}
  {94}},\ \bibinfo {pages} {094032} (\bibinfo {year} {2016})},\ \Eprint
  {https://arxiv.org/abs/1509.04943} {arXiv:1509.04943 [hep-ph]} \BibitemShut
  {NoStop}%
\bibitem [{\citenamefont {Vary}\ \emph {et~al.}(2009)\citenamefont {Vary},
  \citenamefont {Maris}, \citenamefont {Ng}, \citenamefont {Yang},\ and\
  \citenamefont {Sosonkina}}]{Vary:2009qp}%
  \BibitemOpen
  \bibfield  {author} {\bibinfo {author} {\bibfnamefont {J.~P.}\ \bibnamefont
  {Vary}}, \bibinfo {author} {\bibfnamefont {P.}~\bibnamefont {Maris}},
  \bibinfo {author} {\bibfnamefont {E.}~\bibnamefont {Ng}}, \bibinfo {author}
  {\bibfnamefont {C.}~\bibnamefont {Yang}},\ and\ \bibinfo {author}
  {\bibfnamefont {M.}~\bibnamefont {Sosonkina}},\ }\href
  {https://doi.org/10.1088/1742-6596/180/1/012083} {\bibfield  {journal}
  {\bibinfo  {journal} {J. Phys. Conf. Ser.}\ }\textbf {\bibinfo {volume}
  {180}},\ \bibinfo {pages} {012083} (\bibinfo {year} {2009})},\ \Eprint
  {https://arxiv.org/abs/0907.0209} {arXiv:0907.0209 [nucl-th]} \BibitemShut
  {NoStop}%
\bibitem [{\citenamefont {Kosztin}\ \emph {et~al.}(1996)\citenamefont
  {Kosztin}, \citenamefont {Faber},\ and\ \citenamefont
  {Schulten}}]{Kosztin:1996fh}%
  \BibitemOpen
  \bibfield  {author} {\bibinfo {author} {\bibfnamefont {I.}~\bibnamefont
  {Kosztin}}, \bibinfo {author} {\bibfnamefont {B.}~\bibnamefont {Faber}},\
  and\ \bibinfo {author} {\bibfnamefont {K.}~\bibnamefont {Schulten}},\ }\href
  {https://doi.org/10.1119/1.18168} {\bibfield  {journal} {\bibinfo  {journal}
  {Am. J. Phys.}\ }\textbf {\bibinfo {volume} {64}},\ \bibinfo {pages} {633}
  (\bibinfo {year} {1996})},\ \Eprint {https://arxiv.org/abs/physics/9702023}
  {arXiv:physics/9702023} \BibitemShut {NoStop}%
\bibitem [{\citenamefont {Krogel}\ and\ \citenamefont
  {Ceperley}(2012)}]{krogel2012population}%
  \BibitemOpen
  \bibfield  {author} {\bibinfo {author} {\bibfnamefont {J.~T.}\ \bibnamefont
  {Krogel}}\ and\ \bibinfo {author} {\bibfnamefont {D.~M.}\ \bibnamefont
  {Ceperley}},\ }in\ \href
  {https://pubs.acs.org/doi/abs/10.1021/bk-2012-1094.ch002} {\emph {\bibinfo
  {booktitle} {Advances in Quantum Monte Carlo}}}\ (\bibinfo  {publisher} {ACS
  Publications},\ \bibinfo {year} {2012})\ pp.\ \bibinfo {pages}
  {13--26}\BibitemShut {NoStop}%
\bibitem [{\citenamefont {Suhm}\ and\ \citenamefont
  {Watts}(1991)}]{suhm1991quantum}%
  \BibitemOpen
  \bibfield  {author} {\bibinfo {author} {\bibfnamefont {M.~A.}\ \bibnamefont
  {Suhm}}\ and\ \bibinfo {author} {\bibfnamefont {R.~O.}\ \bibnamefont
  {Watts}},\ }\href@noop {} {\bibfield  {journal} {\bibinfo  {journal} {Physics
  Reports}\ }\textbf {\bibinfo {volume} {204}},\ \bibinfo {pages} {293}
  (\bibinfo {year} {1991})}\BibitemShut {NoStop}%
\bibitem [{\citenamefont {Foulkes}\ \emph {et~al.}(2001)\citenamefont
  {Foulkes}, \citenamefont {Mitas}, \citenamefont {Needs},\ and\ \citenamefont
  {Rajagopal}}]{Foulkes:2001zz}%
  \BibitemOpen
  \bibfield  {author} {\bibinfo {author} {\bibfnamefont {W.~M.~C.}\
  \bibnamefont {Foulkes}}, \bibinfo {author} {\bibfnamefont {L.}~\bibnamefont
  {Mitas}}, \bibinfo {author} {\bibfnamefont {R.~J.}\ \bibnamefont {Needs}},\
  and\ \bibinfo {author} {\bibfnamefont {G.}~\bibnamefont {Rajagopal}},\ }\href
  {https://doi.org/10.1103/RevModPhys.73.33} {\bibfield  {journal} {\bibinfo
  {journal} {Rev. Mod. Phys.}\ }\textbf {\bibinfo {volume} {73}},\ \bibinfo
  {pages} {33} (\bibinfo {year} {2001})}\BibitemShut {NoStop}%
\bibitem [{\citenamefont {Carlson}\ \emph {et~al.}(2015)\citenamefont
  {Carlson}, \citenamefont {Gandolfi}, \citenamefont {Pederiva}, \citenamefont
  {Pieper}, \citenamefont {Schiavilla}, \citenamefont {Schmidt},\ and\
  \citenamefont {Wiringa}}]{Carlson:2014vla}%
  \BibitemOpen
  \bibfield  {author} {\bibinfo {author} {\bibfnamefont {J.}~\bibnamefont
  {Carlson}}, \bibinfo {author} {\bibfnamefont {S.}~\bibnamefont {Gandolfi}},
  \bibinfo {author} {\bibfnamefont {F.}~\bibnamefont {Pederiva}}, \bibinfo
  {author} {\bibfnamefont {S.~C.}\ \bibnamefont {Pieper}}, \bibinfo {author}
  {\bibfnamefont {R.}~\bibnamefont {Schiavilla}}, \bibinfo {author}
  {\bibfnamefont {K.~E.}\ \bibnamefont {Schmidt}},\ and\ \bibinfo {author}
  {\bibfnamefont {R.~B.}\ \bibnamefont {Wiringa}},\ }\href
  {https://doi.org/10.1103/RevModPhys.87.1067} {\bibfield  {journal} {\bibinfo
  {journal} {Rev. Mod. Phys.}\ }\textbf {\bibinfo {volume} {87}},\ \bibinfo
  {pages} {1067} (\bibinfo {year} {2015})},\ \Eprint
  {https://arxiv.org/abs/1412.3081} {arXiv:1412.3081 [nucl-th]} \BibitemShut
  {NoStop}%
\bibitem [{\citenamefont {Bai}\ \emph {et~al.}(2019)\citenamefont {Bai},
  \citenamefont {Lu},\ and\ \citenamefont {Osborne}}]{Bai:2016int}%
  \BibitemOpen
  \bibfield  {author} {\bibinfo {author} {\bibfnamefont {Y.}~\bibnamefont
  {Bai}}, \bibinfo {author} {\bibfnamefont {S.}~\bibnamefont {Lu}},\ and\
  \bibinfo {author} {\bibfnamefont {J.}~\bibnamefont {Osborne}},\ }\href
  {https://doi.org/10.1016/j.physletb.2019.134930} {\bibfield  {journal}
  {\bibinfo  {journal} {Phys. Lett. B}\ }\textbf {\bibinfo {volume} {798}},\
  \bibinfo {pages} {134930} (\bibinfo {year} {2019})},\ \Eprint
  {https://arxiv.org/abs/1612.00012} {arXiv:1612.00012 [hep-ph]} \BibitemShut
  {NoStop}%
\bibitem [{\citenamefont {Gordillo}\ \emph {et~al.}(2020)\citenamefont
  {Gordillo}, \citenamefont {De~Soto},\ and\ \citenamefont
  {Segovia}}]{Gordillo:2020sgc}%
  \BibitemOpen
  \bibfield  {author} {\bibinfo {author} {\bibfnamefont {M.~C.}\ \bibnamefont
  {Gordillo}}, \bibinfo {author} {\bibfnamefont {F.}~\bibnamefont {De~Soto}},\
  and\ \bibinfo {author} {\bibfnamefont {J.}~\bibnamefont {Segovia}},\ }\href
  {https://doi.org/10.1103/PhysRevD.102.114007} {\bibfield  {journal} {\bibinfo
   {journal} {Phys. Rev. D}\ }\textbf {\bibinfo {volume} {102}},\ \bibinfo
  {pages} {114007} (\bibinfo {year} {2020})},\ \Eprint
  {https://arxiv.org/abs/2009.11889} {arXiv:2009.11889 [hep-ph]} \BibitemShut
  {NoStop}%
\bibitem [{\citenamefont {Wang}\ \emph {et~al.}(2019)\citenamefont {Wang},
  \citenamefont {Meng},\ and\ \citenamefont {Zhu}}]{Wang:2019rdo}%
  \BibitemOpen
  \bibfield  {author} {\bibinfo {author} {\bibfnamefont {G.-J.}\ \bibnamefont
  {Wang}}, \bibinfo {author} {\bibfnamefont {L.}~\bibnamefont {Meng}},\ and\
  \bibinfo {author} {\bibfnamefont {S.-L.}\ \bibnamefont {Zhu}},\ }\href
  {https://doi.org/10.1103/PhysRevD.100.096013} {\bibfield  {journal} {\bibinfo
   {journal} {Phys. Rev. D}\ }\textbf {\bibinfo {volume} {100}},\ \bibinfo
  {pages} {096013} (\bibinfo {year} {2019})},\ \Eprint
  {https://arxiv.org/abs/1907.05177} {arXiv:1907.05177 [hep-ph]} \BibitemShut
  {NoStop}%
\bibitem [{\citenamefont {Gordillo}\ \emph {et~al.}(2021)\citenamefont
  {Gordillo}, \citenamefont {De~Soto},\ and\ \citenamefont
  {Segovia}}]{Gordillo:2021bra}%
  \BibitemOpen
  \bibfield  {author} {\bibinfo {author} {\bibfnamefont {M.~C.}\ \bibnamefont
  {Gordillo}}, \bibinfo {author} {\bibfnamefont {F.}~\bibnamefont {De~Soto}},\
  and\ \bibinfo {author} {\bibfnamefont {J.}~\bibnamefont {Segovia}},\ }\href
  {https://doi.org/10.1103/PhysRevD.104.054036} {\bibfield  {journal} {\bibinfo
   {journal} {Phys. Rev. D}\ }\textbf {\bibinfo {volume} {104}},\ \bibinfo
  {pages} {054036} (\bibinfo {year} {2021})},\ \Eprint
  {https://arxiv.org/abs/2105.11976} {arXiv:2105.11976 [hep-ph]} \BibitemShut
  {NoStop}%
\bibitem [{\citenamefont {Alcaraz-Pelegrina}\ and\ \citenamefont
  {Gordillo}(2022)}]{Alcaraz-Pelegrina:2022fsi}%
  \BibitemOpen
  \bibfield  {author} {\bibinfo {author} {\bibfnamefont {J.~M.}\ \bibnamefont
  {Alcaraz-Pelegrina}}\ and\ \bibinfo {author} {\bibfnamefont {M.~C.}\
  \bibnamefont {Gordillo}},\ }\href@noop {} {\  (\bibinfo {year} {2022})},\
  \Eprint {https://arxiv.org/abs/2205.13886} {arXiv:2205.13886 [hep-ph]}
  \BibitemShut {NoStop}%
\bibitem [{\citenamefont {Gordillo}\ \emph {et~al.}(2022)\citenamefont
  {Gordillo}, \citenamefont {De~Soto},\ and\ \citenamefont
  {Segovia}}]{Gordillo:2022nnj}%
  \BibitemOpen
  \bibfield  {author} {\bibinfo {author} {\bibfnamefont {M.~C.}\ \bibnamefont
  {Gordillo}}, \bibinfo {author} {\bibfnamefont {F.}~\bibnamefont {De~Soto}},\
  and\ \bibinfo {author} {\bibfnamefont {J.}~\bibnamefont {Segovia}},\ }\href
  {https://doi.org/10.1103/PhysRevD.106.094004} {\bibfield  {journal} {\bibinfo
   {journal} {Phys. Rev. D}\ }\textbf {\bibinfo {volume} {106}},\ \bibinfo
  {pages} {094004} (\bibinfo {year} {2022})},\ \Eprint
  {https://arxiv.org/abs/2209.04221} {arXiv:2209.04221 [hep-ph]} \BibitemShut
  {NoStop}%
\bibitem [{\citenamefont {Durgut
  (CMS~Collaboration)}(2018)}]{durgut2018search}%
  \BibitemOpen
  \bibfield  {author} {\bibinfo {author} {\bibfnamefont {S.}~\bibnamefont
  {Durgut (CMS~Collaboration)}},\ }in\ \href
  {https://meetings.aps.org/Meeting/APR18/Session/U09.6} {\emph {\bibinfo
  {booktitle} {APS April Meeting Abstracts}}}\ (\bibinfo {year} {2018})\ pp.\
  \bibinfo {pages} {U09--006}\BibitemShut {NoStop}%
\bibitem [{\citenamefont {Durgut}(2018)}]{durgut2018evidence}%
  \BibitemOpen
  \bibfield  {author} {\bibinfo {author} {\bibfnamefont {S.}~\bibnamefont
  {Durgut}},\ }\emph {\bibinfo {title} {Evidence of a Narrow Structure in
  $\Upsilon$ (1S) l+ l-Mass Spectrum and CMS Phase I and II Silicon Detector
  Upgrade Studies}},\ \href@noop {} {Ph.D. thesis},\ \bibinfo  {school} {The
  University of Iowa} (\bibinfo {year} {2018})\BibitemShut {NoStop}%
\bibitem [{\citenamefont {Aaij}\ \emph {et~al.}(2018)\citenamefont {Aaij} \emph
  {et~al.}}]{LHCb:2018uwm}%
  \BibitemOpen
  \bibfield  {author} {\bibinfo {author} {\bibfnamefont {R.}~\bibnamefont
  {Aaij}} \emph {et~al.} (\bibinfo {collaboration} {LHCb}),\ }\href
  {https://doi.org/10.1007/JHEP10(2018)086} {\bibfield  {journal} {\bibinfo
  {journal} {JHEP}\ }\textbf {\bibinfo {volume} {10}},\ \bibinfo {pages}
  {086}},\ \Eprint {https://arxiv.org/abs/1806.09707} {arXiv:1806.09707
  [hep-ex]} \BibitemShut {NoStop}%
\bibitem [{\citenamefont {Aaij}\ \emph {et~al.}(2020)\citenamefont {Aaij} \emph
  {et~al.}}]{LHCb:2020bwg}%
  \BibitemOpen
  \bibfield  {author} {\bibinfo {author} {\bibfnamefont {R.}~\bibnamefont
  {Aaij}} \emph {et~al.} (\bibinfo {collaboration} {LHCb}),\ }\href
  {https://doi.org/10.1016/j.scib.2020.08.032} {\bibfield  {journal} {\bibinfo
  {journal} {Sci. Bull.}\ }\textbf {\bibinfo {volume} {65}},\ \bibinfo {pages}
  {1983} (\bibinfo {year} {2020})},\ \Eprint {https://arxiv.org/abs/2006.16957}
  {arXiv:2006.16957 [hep-ex]} \BibitemShut {NoStop}%
\bibitem [{\citenamefont {Yi}()}]{yiCMSResults}%
  \BibitemOpen
  \bibfield  {author} {\bibinfo {author} {\bibfnamefont {K.}~\bibnamefont
  {Yi}},\ }\href {https://agenda.infn.it/event/28874/contributions/170300/}
  {\bibinfo {title} {Recent {CMS} results on exotic resonances}}\BibitemShut
  {NoStop}%
\bibitem [{\citenamefont {Bouhova-Thacker}()}]{bouhovaATLAS}%
  \BibitemOpen
  \bibfield  {author} {\bibinfo {author} {\bibfnamefont {E.}~\bibnamefont
  {Bouhova-Thacker}},\ }\href
  {https://agenda.infn.it/event/28874/contributions/170298/} {\bibinfo {title}
  {{ATLAS} results on exotic hadronic resonances}}\BibitemShut {NoStop}%
\bibitem [{\citenamefont {Wang}\ \emph {et~al.}(2022)\citenamefont {Wang},
  \citenamefont {Meng},\ and\ \citenamefont {Oka}}]{Wang:2022yes}%
  \BibitemOpen
  \bibfield  {author} {\bibinfo {author} {\bibfnamefont {G.-J.}\ \bibnamefont
  {Wang}}, \bibinfo {author} {\bibfnamefont {Q.}~\bibnamefont {Meng}},\ and\
  \bibinfo {author} {\bibfnamefont {M.}~\bibnamefont {Oka}},\ }\href
  {https://doi.org/10.1103/PhysRevD.106.096005} {\bibfield  {journal} {\bibinfo
   {journal} {Phys. Rev. D}\ }\textbf {\bibinfo {volume} {106}},\ \bibinfo
  {pages} {096005} (\bibinfo {year} {2022})},\ \Eprint
  {https://arxiv.org/abs/2208.07292} {arXiv:2208.07292 [hep-ph]} \BibitemShut
  {NoStop}%
\bibitem [{\citenamefont {Klos}\ \emph {et~al.}(2016)\citenamefont {Klos},
  \citenamefont {Lynn}, \citenamefont {Tews}, \citenamefont {Gandolfi},
  \citenamefont {Gezerlis}, \citenamefont {Hammer}, \citenamefont
  {Hoferichter},\ and\ \citenamefont {Schwenk}}]{Klos:2016fdb}%
  \BibitemOpen
  \bibfield  {author} {\bibinfo {author} {\bibfnamefont {P.}~\bibnamefont
  {Klos}}, \bibinfo {author} {\bibfnamefont {J.~E.}\ \bibnamefont {Lynn}},
  \bibinfo {author} {\bibfnamefont {I.}~\bibnamefont {Tews}}, \bibinfo {author}
  {\bibfnamefont {S.}~\bibnamefont {Gandolfi}}, \bibinfo {author}
  {\bibfnamefont {A.}~\bibnamefont {Gezerlis}}, \bibinfo {author}
  {\bibfnamefont {H.~W.}\ \bibnamefont {Hammer}}, \bibinfo {author}
  {\bibfnamefont {M.}~\bibnamefont {Hoferichter}},\ and\ \bibinfo {author}
  {\bibfnamefont {A.}~\bibnamefont {Schwenk}},\ }\href
  {https://doi.org/10.1103/PhysRevC.94.054005} {\bibfield  {journal} {\bibinfo
  {journal} {Phys. Rev. C}\ }\textbf {\bibinfo {volume} {94}},\ \bibinfo
  {pages} {054005} (\bibinfo {year} {2016})},\ \Eprint
  {https://arxiv.org/abs/1604.01387} {arXiv:1604.01387 [nucl-th]} \BibitemShut
  {NoStop}%
\bibitem [{\citenamefont {Gandolfi}\ \emph {et~al.}(2017)\citenamefont
  {Gandolfi}, \citenamefont {Hammer}, \citenamefont {Klos}, \citenamefont
  {Lynn},\ and\ \citenamefont {Schwenk}}]{Gandolfi:2016bth}%
  \BibitemOpen
  \bibfield  {author} {\bibinfo {author} {\bibfnamefont {S.}~\bibnamefont
  {Gandolfi}}, \bibinfo {author} {\bibfnamefont {H.~W.}\ \bibnamefont
  {Hammer}}, \bibinfo {author} {\bibfnamefont {P.}~\bibnamefont {Klos}},
  \bibinfo {author} {\bibfnamefont {J.~E.}\ \bibnamefont {Lynn}},\ and\
  \bibinfo {author} {\bibfnamefont {A.}~\bibnamefont {Schwenk}},\ }\href
  {https://doi.org/10.1103/PhysRevLett.118.232501} {\bibfield  {journal}
  {\bibinfo  {journal} {Phys. Rev. Lett.}\ }\textbf {\bibinfo {volume} {118}},\
  \bibinfo {pages} {232501} (\bibinfo {year} {2017})},\ \Eprint
  {https://arxiv.org/abs/1612.01502} {arXiv:1612.01502 [nucl-th]} \BibitemShut
  {NoStop}%
\bibitem [{\citenamefont {Toulouse}\ \emph {et~al.}(2016)\citenamefont
  {Toulouse}, \citenamefont {Assaraf},\ and\ \citenamefont
  {Umrigar}}]{toulouse2016introduction}%
  \BibitemOpen
  \bibfield  {author} {\bibinfo {author} {\bibfnamefont {J.}~\bibnamefont
  {Toulouse}}, \bibinfo {author} {\bibfnamefont {R.}~\bibnamefont {Assaraf}},\
  and\ \bibinfo {author} {\bibfnamefont {C.~J.}\ \bibnamefont {Umrigar}},\ }in\
  \href {https://www.sciencedirect.com/science/article/pii/S0065327615000386}
  {\emph {\bibinfo {booktitle} {Advances in Quantum Chemistry}}},\
  Vol.~\bibinfo {volume} {73}\ (\bibinfo  {publisher} {Elsevier},\ \bibinfo
  {year} {2016})\ pp.\ \bibinfo {pages} {285--314}\BibitemShut {NoStop}%
\bibitem [{\citenamefont {Boronat}\ and\ \citenamefont
  {Casulleras}(1994)}]{Boronat1994}%
  \BibitemOpen
  \bibfield  {author} {\bibinfo {author} {\bibfnamefont {J.}~\bibnamefont
  {Boronat}}\ and\ \bibinfo {author} {\bibfnamefont {J.}~\bibnamefont
  {Casulleras}},\ }\href {https://doi.org/10.1103/PhysRevB.49.8920} {\bibfield
  {journal} {\bibinfo  {journal} {Phys. Rev. B}\ }\textbf {\bibinfo {volume}
  {49}},\ \bibinfo {pages} {8920} (\bibinfo {year} {1994})}\BibitemShut
  {NoStop}%
\bibitem [{\citenamefont {Kalos}\ \emph {et~al.}(1974)\citenamefont {Kalos},
  \citenamefont {Levesque},\ and\ \citenamefont {Verlet}}]{Kalos1974}%
  \BibitemOpen
  \bibfield  {author} {\bibinfo {author} {\bibfnamefont {M.~H.}\ \bibnamefont
  {Kalos}}, \bibinfo {author} {\bibfnamefont {D.}~\bibnamefont {Levesque}},\
  and\ \bibinfo {author} {\bibfnamefont {L.}~\bibnamefont {Verlet}},\ }\href
  {https://doi.org/10.1103/PhysRevA.9.2178} {\bibfield  {journal} {\bibinfo
  {journal} {Phys. Rev. A}\ }\textbf {\bibinfo {volume} {9}},\ \bibinfo {pages}
  {2178} (\bibinfo {year} {1974})}\BibitemShut {NoStop}%
\bibitem [{\citenamefont {Hjorth-Jensen}\ \emph {et~al.}(2017)\citenamefont
  {Hjorth-Jensen}, \citenamefont {Lombardo},\ and\ \citenamefont {van
  Kolck}}]{Hjorth-Jensen:2017gss}%
  \BibitemOpen
  \bibinfo {editor} {\bibfnamefont {M.}~\bibnamefont {Hjorth-Jensen}}, \bibinfo
  {editor} {\bibfnamefont {M.~P.}\ \bibnamefont {Lombardo}},\ and\ \bibinfo
  {editor} {\bibfnamefont {U.}~\bibnamefont {van Kolck}},\ eds.,\ \href
  {https://doi.org/10.1007/978-3-319-53336-0} {\emph {\bibinfo {title} {{An
  Advanced Course in Computational Nuclear Physics}}}},\ Vol.\ \bibinfo
  {volume} {936}\ (\bibinfo  {publisher} {Springer},\ \bibinfo {year}
  {2017})\BibitemShut {NoStop}%
\bibitem [{\citenamefont {Hammond}\ \emph {et~al.}(1994)\citenamefont
  {Hammond}, \citenamefont {Lester},\ and\ \citenamefont
  {Reynolds}}]{hammond1994monte}%
  \BibitemOpen
  \bibfield  {author} {\bibinfo {author} {\bibfnamefont {B.~L.}\ \bibnamefont
  {Hammond}}, \bibinfo {author} {\bibfnamefont {W.~A.}\ \bibnamefont
  {Lester}},\ and\ \bibinfo {author} {\bibfnamefont {P.~J.}\ \bibnamefont
  {Reynolds}},\ }\href@noop {} {\emph {\bibinfo {title} {Monte Carlo methods in
  ab initio quantum chemistry}}},\ Vol.~\bibinfo {volume} {1}\ (\bibinfo
  {publisher} {World Scientific},\ \bibinfo {year} {1994})\BibitemShut
  {NoStop}%
\bibitem [{\citenamefont {Barnett}\ \emph {et~al.}(1991)\citenamefont
  {Barnett}, \citenamefont {Reynolds},\ and\ \citenamefont
  {Lester}}]{BARNETT1991}%
  \BibitemOpen
  \bibfield  {author} {\bibinfo {author} {\bibfnamefont {R.}~\bibnamefont
  {Barnett}}, \bibinfo {author} {\bibfnamefont {P.}~\bibnamefont {Reynolds}},\
  and\ \bibinfo {author} {\bibfnamefont {W.}~\bibnamefont {Lester}},\ }\href
  {https://doi.org/https://doi.org/10.1016/0021-9991(91)90236-E} {\bibfield
  {journal} {\bibinfo  {journal} {Journal of Computational Physics}\ }\textbf
  {\bibinfo {volume} {96}},\ \bibinfo {pages} {258} (\bibinfo {year}
  {1991})}\BibitemShut {NoStop}%
\bibitem [{\citenamefont {Liu}\ \emph {et~al.}(1974)\citenamefont {Liu},
  \citenamefont {Kalos},\ and\ \citenamefont {Chester}}]{Liu1974}%
  \BibitemOpen
  \bibfield  {author} {\bibinfo {author} {\bibfnamefont {K.~S.}\ \bibnamefont
  {Liu}}, \bibinfo {author} {\bibfnamefont {M.~H.}\ \bibnamefont {Kalos}},\
  and\ \bibinfo {author} {\bibfnamefont {G.~V.}\ \bibnamefont {Chester}},\
  }\href {https://doi.org/10.1103/PhysRevA.10.303} {\bibfield  {journal}
  {\bibinfo  {journal} {Phys. Rev. A}\ }\textbf {\bibinfo {volume} {10}},\
  \bibinfo {pages} {303} (\bibinfo {year} {1974})}\BibitemShut {NoStop}%
\bibitem [{\citenamefont {Casulleras}\ and\ \citenamefont
  {Boronat}(1995)}]{Casulleras1995}%
  \BibitemOpen
  \bibfield  {author} {\bibinfo {author} {\bibfnamefont {J.}~\bibnamefont
  {Casulleras}}\ and\ \bibinfo {author} {\bibfnamefont {J.}~\bibnamefont
  {Boronat}},\ }\href {https://doi.org/10.1103/PhysRevB.52.3654} {\bibfield
  {journal} {\bibinfo  {journal} {Phys. Rev. B}\ }\textbf {\bibinfo {volume}
  {52}},\ \bibinfo {pages} {3654} (\bibinfo {year} {1995})}\BibitemShut
  {NoStop}%
\bibitem [{\citenamefont {S\'anchez-Baena}\ \emph {et~al.}(2018)\citenamefont
  {S\'anchez-Baena}, \citenamefont {Boronat},\ and\ \citenamefont
  {Mazzanti}}]{Sanchez2018}%
  \BibitemOpen
  \bibfield  {author} {\bibinfo {author} {\bibfnamefont {J.}~\bibnamefont
  {S\'anchez-Baena}}, \bibinfo {author} {\bibfnamefont {J.}~\bibnamefont
  {Boronat}},\ and\ \bibinfo {author} {\bibfnamefont {F.}~\bibnamefont
  {Mazzanti}},\ }\href {https://doi.org/10.1103/PhysRevA.98.053632} {\bibfield
  {journal} {\bibinfo  {journal} {Phys. Rev. A}\ }\textbf {\bibinfo {volume}
  {98}},\ \bibinfo {pages} {053632} (\bibinfo {year} {2018})}\BibitemShut
  {NoStop}%
\bibitem [{\citenamefont {Koma}\ and\ \citenamefont
  {Koma}(2017)}]{koma2017precise}%
  \BibitemOpen
  \bibfield  {author} {\bibinfo {author} {\bibfnamefont {Y.}~\bibnamefont
  {Koma}}\ and\ \bibinfo {author} {\bibfnamefont {M.}~\bibnamefont {Koma}},\
  }\href@noop {} {\bibfield  {journal} {\bibinfo  {journal} {Physical Review
  D}\ }\textbf {\bibinfo {volume} {95}},\ \bibinfo {pages} {094513} (\bibinfo
  {year} {2017})}\BibitemShut {NoStop}%
\bibitem [{\citenamefont {Leech}\ \emph {et~al.}(2021)\citenamefont {Leech},
  \citenamefont {{\v{S}}uvakov},\ and\ \citenamefont
  {Dmitra{\v{s}}inovi{\'c}}}]{leech2021hyperspherical}%
  \BibitemOpen
  \bibfield  {author} {\bibinfo {author} {\bibfnamefont {J.}~\bibnamefont
  {Leech}}, \bibinfo {author} {\bibfnamefont {M.}~\bibnamefont
  {{\v{S}}uvakov}},\ and\ \bibinfo {author} {\bibfnamefont {V.}~\bibnamefont
  {Dmitra{\v{s}}inovi{\'c}}},\ }\href@noop {} {\bibfield  {journal} {\bibinfo
  {journal} {The European Physical Journal C}\ }\textbf {\bibinfo {volume}
  {81}},\ \bibinfo {pages} {75} (\bibinfo {year} {2021})}\BibitemShut {NoStop}%
\bibitem [{\citenamefont {Smith}(1992)}]{smith1992find}%
  \BibitemOpen
  \bibfield  {author} {\bibinfo {author} {\bibfnamefont {W.~D.}\ \bibnamefont
  {Smith}},\ }\href@noop {} {\bibfield  {journal} {\bibinfo  {journal}
  {Algorithmica}\ }\textbf {\bibinfo {volume} {7}},\ \bibinfo {pages} {137}
  (\bibinfo {year} {1992})}\BibitemShut {NoStop}%
\bibitem [{ESM()}]{ESMPgithub}%
  \BibitemOpen
  \href {https://github.com/RasmusFonseca/ESMT-Smith} {\bibinfo {title}
  {https://github.com/rasmusfonseca/esmt-smith}},\ \bibinfo {note} {finding
  Steiner minimal trees in euclidean d-space.}\BibitemShut {Stop}%
\bibitem [{\citenamefont {Gattringer}\ and\ \citenamefont
  {Lang}(2009)}]{gattringer2009quantum}%
  \BibitemOpen
  \bibfield  {author} {\bibinfo {author} {\bibfnamefont {C.}~\bibnamefont
  {Gattringer}}\ and\ \bibinfo {author} {\bibfnamefont {C.}~\bibnamefont
  {Lang}},\ }\href@noop {} {\emph {\bibinfo {title} {Quantum chromodynamics on
  the lattice: an introductory presentation}}},\ Vol.\ \bibinfo {volume} {788}\
  (\bibinfo  {publisher} {Springer Science \& Business Media},\ \bibinfo {year}
  {2009})\BibitemShut {NoStop}%
\bibitem [{\citenamefont {Workman}\ \emph {et~al.}(2022)\citenamefont {Workman}
  \emph {et~al.}}]{ParticleDataGroup:2022pth}%
  \BibitemOpen
  \bibfield  {author} {\bibinfo {author} {\bibfnamefont {R.~L.}\ \bibnamefont
  {Workman}} \emph {et~al.} (\bibinfo {collaboration} {Particle Data Group}),\
  }\href {https://doi.org/10.1093/ptep/ptac097} {\bibfield  {journal} {\bibinfo
   {journal} {PTEP}\ }\textbf {\bibinfo {volume} {2022}},\ \bibinfo {pages}
  {083C01} (\bibinfo {year} {2022})}\BibitemShut {NoStop}%
\bibitem [{\citenamefont {Gough~Eschrich}\ \emph {et~al.}(2001)\citenamefont
  {Gough~Eschrich} \emph {et~al.}}]{SELEX:2001fbx}%
  \BibitemOpen
  \bibfield  {author} {\bibinfo {author} {\bibfnamefont {I.~M.}\ \bibnamefont
  {Gough~Eschrich}} \emph {et~al.} (\bibinfo {collaboration} {SELEX}),\ }\href
  {https://doi.org/10.1016/S0370-2693(01)01285-0} {\bibfield  {journal}
  {\bibinfo  {journal} {Phys. Lett. B}\ }\textbf {\bibinfo {volume} {522}},\
  \bibinfo {pages} {233} (\bibinfo {year} {2001})},\ \Eprint
  {https://arxiv.org/abs/hep-ex/0106053} {arXiv:hep-ex/0106053} \BibitemShut
  {NoStop}%
\bibitem [{\citenamefont {Can}\ \emph {et~al.}(2014)\citenamefont {Can},
  \citenamefont {Erkol}, \citenamefont {Isildak}, \citenamefont {Oka},\ and\
  \citenamefont {Takahashi}}]{Can:2013tna}%
  \BibitemOpen
  \bibfield  {author} {\bibinfo {author} {\bibfnamefont {K.~U.}\ \bibnamefont
  {Can}}, \bibinfo {author} {\bibfnamefont {G.}~\bibnamefont {Erkol}}, \bibinfo
  {author} {\bibfnamefont {B.}~\bibnamefont {Isildak}}, \bibinfo {author}
  {\bibfnamefont {M.}~\bibnamefont {Oka}},\ and\ \bibinfo {author}
  {\bibfnamefont {T.~T.}\ \bibnamefont {Takahashi}},\ }\href
  {https://doi.org/10.1007/JHEP05(2014)125} {\bibfield  {journal} {\bibinfo
  {journal} {JHEP}\ }\textbf {\bibinfo {volume} {05}},\ \bibinfo {pages}
  {125}},\ \Eprint {https://arxiv.org/abs/1310.5915} {arXiv:1310.5915
  [hep-lat]} \BibitemShut {NoStop}%
\bibitem [{\citenamefont {Can}(2021)}]{Can:2021ehb}%
  \BibitemOpen
  \bibfield  {author} {\bibinfo {author} {\bibfnamefont {K.~U.}\ \bibnamefont
  {Can}},\ }\href {https://doi.org/10.1142/S0217751X21300131} {\bibfield
  {journal} {\bibinfo  {journal} {Int. J. Mod. Phys. A}\ }\textbf {\bibinfo
  {volume} {36}},\ \bibinfo {pages} {2130013} (\bibinfo {year} {2021})},\
  \Eprint {https://arxiv.org/abs/2107.13159} {arXiv:2107.13159 [hep-lat]}
  \BibitemShut {NoStop}%
\bibitem [{\citenamefont {Wagner}\ \emph {et~al.}(1998)\citenamefont {Wagner},
  \citenamefont {Buchmann},\ and\ \citenamefont {Faessler}}]{Wagner:1998fi}%
  \BibitemOpen
  \bibfield  {author} {\bibinfo {author} {\bibfnamefont {G.}~\bibnamefont
  {Wagner}}, \bibinfo {author} {\bibfnamefont {A.~J.}\ \bibnamefont
  {Buchmann}},\ and\ \bibinfo {author} {\bibfnamefont {A.}~\bibnamefont
  {Faessler}},\ }\href {https://doi.org/10.1103/PhysRevC.58.3666} {\bibfield
  {journal} {\bibinfo  {journal} {Phys. Rev. C}\ }\textbf {\bibinfo {volume}
  {58}},\ \bibinfo {pages} {3666} (\bibinfo {year} {1998})},\ \Eprint
  {https://arxiv.org/abs/nucl-th/9809015} {arXiv:nucl-th/9809015} \BibitemShut
  {NoStop}%
\bibitem [{\citenamefont {Wagner}\ \emph {et~al.}(2000)\citenamefont {Wagner},
  \citenamefont {Buchmann},\ and\ \citenamefont {Faessler}}]{Wagner:2000ii}%
  \BibitemOpen
  \bibfield  {author} {\bibinfo {author} {\bibfnamefont {G.}~\bibnamefont
  {Wagner}}, \bibinfo {author} {\bibfnamefont {A.~J.}\ \bibnamefont
  {Buchmann}},\ and\ \bibinfo {author} {\bibfnamefont {A.}~\bibnamefont
  {Faessler}},\ }\href {https://doi.org/10.1088/0954-3899/26/3/306} {\bibfield
  {journal} {\bibinfo  {journal} {J. Phys. G}\ }\textbf {\bibinfo {volume}
  {26}},\ \bibinfo {pages} {267} (\bibinfo {year} {2000})}\BibitemShut
  {NoStop}%
\bibitem [{\citenamefont {Shen}\ \emph {et~al.}(2022)\citenamefont {Shen},
  \citenamefont {L\"ahde}, \citenamefont {Lee},\ and\ \citenamefont
  {Mei\ss{}ner}}]{Shen:2022bak}%
  \BibitemOpen
  \bibfield  {author} {\bibinfo {author} {\bibfnamefont {S.}~\bibnamefont
  {Shen}}, \bibinfo {author} {\bibfnamefont {T.~A.}\ \bibnamefont {L\"ahde}},
  \bibinfo {author} {\bibfnamefont {D.}~\bibnamefont {Lee}},\ and\ \bibinfo
  {author} {\bibfnamefont {U.-G.}\ \bibnamefont {Mei\ss{}ner}},\ }\href@noop {}
  {\  (\bibinfo {year} {2022})},\ \Eprint {https://arxiv.org/abs/2202.13596}
  {arXiv:2202.13596 [nucl-th]} \BibitemShut {NoStop}%
\bibitem [{\citenamefont {Hiyama}\ \emph {et~al.}(2003)\citenamefont {Hiyama},
  \citenamefont {Kino},\ and\ \citenamefont {Kamimura}}]{Hiyama:2003cu}%
  \BibitemOpen
  \bibfield  {author} {\bibinfo {author} {\bibfnamefont {E.}~\bibnamefont
  {Hiyama}}, \bibinfo {author} {\bibfnamefont {Y.}~\bibnamefont {Kino}},\ and\
  \bibinfo {author} {\bibfnamefont {M.}~\bibnamefont {Kamimura}},\ }\href
  {https://doi.org/10.1016/S0146-6410(03)90015-9} {\bibfield  {journal}
  {\bibinfo  {journal} {Prog. Part. Nucl. Phys.}\ }\textbf {\bibinfo {volume}
  {51}},\ \bibinfo {pages} {223} (\bibinfo {year} {2003})}\BibitemShut
  {NoStop}%
\bibitem [{\citenamefont {Jaffe}(2005)}]{Jaffe:2004ph}%
  \BibitemOpen
  \bibfield  {author} {\bibinfo {author} {\bibfnamefont {R.~L.}\ \bibnamefont
  {Jaffe}},\ }\href {https://doi.org/10.1016/j.physrep.2004.11.005} {\bibfield
  {journal} {\bibinfo  {journal} {Phys. Rept.}\ }\textbf {\bibinfo {volume}
  {409}},\ \bibinfo {pages} {1} (\bibinfo {year} {2005})},\ \Eprint
  {https://arxiv.org/abs/hep-ph/0409065} {arXiv:hep-ph/0409065} \BibitemShut
  {NoStop}%
\bibitem [{\citenamefont {Wang}\ \emph {et~al.}(2009)\citenamefont {Wang},
  \citenamefont {Leinweber}, \citenamefont {Thomas},\ and\ \citenamefont
  {Young}}]{Wang:2008vb}%
  \BibitemOpen
  \bibfield  {author} {\bibinfo {author} {\bibfnamefont {P.}~\bibnamefont
  {Wang}}, \bibinfo {author} {\bibfnamefont {D.~B.}\ \bibnamefont {Leinweber}},
  \bibinfo {author} {\bibfnamefont {A.~W.}\ \bibnamefont {Thomas}},\ and\
  \bibinfo {author} {\bibfnamefont {R.~D.}\ \bibnamefont {Young}},\ }\href
  {https://doi.org/10.1103/PhysRevD.79.094001} {\bibfield  {journal} {\bibinfo
  {journal} {Phys. Rev. D}\ }\textbf {\bibinfo {volume} {79}},\ \bibinfo
  {pages} {094001} (\bibinfo {year} {2009})},\ \Eprint
  {https://arxiv.org/abs/0810.1021} {arXiv:0810.1021 [hep-ph]} \BibitemShut
  {NoStop}%
\bibitem [{\citenamefont {Hackett-Jones}\ \emph {et~al.}(2000)\citenamefont
  {Hackett-Jones}, \citenamefont {Leinweber},\ and\ \citenamefont
  {Thomas}}]{Hackett-Jones:2000qxh}%
  \BibitemOpen
  \bibfield  {author} {\bibinfo {author} {\bibfnamefont {E.~J.}\ \bibnamefont
  {Hackett-Jones}}, \bibinfo {author} {\bibfnamefont {D.~B.}\ \bibnamefont
  {Leinweber}},\ and\ \bibinfo {author} {\bibfnamefont {A.~W.}\ \bibnamefont
  {Thomas}},\ }\href {https://doi.org/10.1016/S0370-2693(00)01115-1} {\bibfield
   {journal} {\bibinfo  {journal} {Phys. Lett. B}\ }\textbf {\bibinfo {volume}
  {494}},\ \bibinfo {pages} {89} (\bibinfo {year} {2000})},\ \Eprint
  {https://arxiv.org/abs/hep-lat/0008018} {arXiv:hep-lat/0008018} \BibitemShut
  {NoStop}%
\bibitem [{\citenamefont {Shanahan}\ \emph {et~al.}(2014)\citenamefont
  {Shanahan}, \citenamefont {Thomas}, \citenamefont {Young}, \citenamefont
  {Zanotti}, \citenamefont {Horsley}, \citenamefont {Nakamura}, \citenamefont
  {Pleiter}, \citenamefont {Rakow}, \citenamefont {Schierholz},\ and\
  \citenamefont {St\"uben}}]{Shanahan:2014cga}%
  \BibitemOpen
  \bibfield  {author} {\bibinfo {author} {\bibfnamefont {P.~E.}\ \bibnamefont
  {Shanahan}}, \bibinfo {author} {\bibfnamefont {A.~W.}\ \bibnamefont
  {Thomas}}, \bibinfo {author} {\bibfnamefont {R.~D.}\ \bibnamefont {Young}},
  \bibinfo {author} {\bibfnamefont {J.~M.}\ \bibnamefont {Zanotti}}, \bibinfo
  {author} {\bibfnamefont {R.}~\bibnamefont {Horsley}}, \bibinfo {author}
  {\bibfnamefont {Y.}~\bibnamefont {Nakamura}}, \bibinfo {author}
  {\bibfnamefont {D.}~\bibnamefont {Pleiter}}, \bibinfo {author} {\bibfnamefont
  {P.~E.~L.}\ \bibnamefont {Rakow}}, \bibinfo {author} {\bibfnamefont
  {G.}~\bibnamefont {Schierholz}},\ and\ \bibinfo {author} {\bibfnamefont
  {H.}~\bibnamefont {St\"uben}},\ }\href
  {https://doi.org/10.1103/PhysRevD.90.034502} {\bibfield  {journal} {\bibinfo
  {journal} {Phys. Rev. D}\ }\textbf {\bibinfo {volume} {90}},\ \bibinfo
  {pages} {034502} (\bibinfo {year} {2014})},\ \Eprint
  {https://arxiv.org/abs/1403.1965} {arXiv:1403.1965 [hep-lat]} \BibitemShut
  {NoStop}%
\bibitem [{\citenamefont {Barnea}\ \emph {et~al.}(2006)\citenamefont {Barnea},
  \citenamefont {Vijande},\ and\ \citenamefont {Valcarce}}]{Barnea:2006sd}%
  \BibitemOpen
  \bibfield  {author} {\bibinfo {author} {\bibfnamefont {N.}~\bibnamefont
  {Barnea}}, \bibinfo {author} {\bibfnamefont {J.}~\bibnamefont {Vijande}},\
  and\ \bibinfo {author} {\bibfnamefont {A.}~\bibnamefont {Valcarce}},\ }\href
  {https://doi.org/10.1103/PhysRevD.73.054004} {\bibfield  {journal} {\bibinfo
  {journal} {Phys. Rev. D}\ }\textbf {\bibinfo {volume} {73}},\ \bibinfo
  {pages} {054004} (\bibinfo {year} {2006})},\ \Eprint
  {https://arxiv.org/abs/hep-ph/0604010} {arXiv:hep-ph/0604010} \BibitemShut
  {NoStop}%
\bibitem [{\citenamefont {Anderson}(1976)}]{anderson1976quantum}%
  \BibitemOpen
  \bibfield  {author} {\bibinfo {author} {\bibfnamefont {J.~B.}\ \bibnamefont
  {Anderson}},\ }\href {https://aip.scitation.org/doi/10.1063/1.432868}
  {\bibfield  {journal} {\bibinfo  {journal} {The Journal of Chemical Physics}\
  }\textbf {\bibinfo {volume} {65}},\ \bibinfo {pages} {4121} (\bibinfo {year}
  {1976})}\BibitemShut {NoStop}%
\bibitem [{\citenamefont {Gandolfi}(2007)}]{Gandolfi:2007ed}%
  \BibitemOpen
  \bibfield  {author} {\bibinfo {author} {\bibfnamefont {S.}~\bibnamefont
  {Gandolfi}},\ }\emph {\bibinfo {title} {{The Auxiliary Field Diffusion Monte
  Carlo Method for Nuclear Physics and Nuclear Astrophysics}}},\ \href@noop {}
  {\bibinfo {type} {Other thesis}} (\bibinfo {year} {2007}),\ \Eprint
  {https://arxiv.org/abs/0712.1364} {arXiv:0712.1364 [nucl-th]} \BibitemShut
  {NoStop}%
\end{thebibliography}%

\end{document}